\documentclass[onecolumn,amsmath,amssymb,nofootinbib,letterpaper,11pt]{article}
\pdfoutput=1
\usepackage{jheppub}
\usepackage{ifpdf}
\usepackage[utf8]{inputenc}
\usepackage[T1]{fontenc}
\usepackage{graphicx}
\input{epsf}
\usepackage{epsfig}
\usepackage{epstopdf}
\usepackage{arcs}
\usepackage{subfigure}
\usepackage{tensor}
\usepackage{braket}
\usepackage[dvipsnames]{xcolor}
\usepackage{amsmath}
\newcommand{\be}{\begin{equation}}
\newcommand{\ee}{\end{equation}}

\DeclareMathOperator{\Tr}{Tr}

\usepackage{tikz}
\usetikzlibrary{calc}   % this is needed in order to be able to do coordinate calculations
\usetikzlibrary{topaths}

\usepackage{hyperref}
\hypersetup{
	colorlinks =WildStrawberry,
	linkcolor =red,
	pdftitle={Holographic studies of Einsteinian cubic gravity},
	citecolor=NavyBlue,
	filecolor=WildStrawberry,
	urlcolor=WildStrawberry,
}

\usepackage[toc]{appendix}
\usepackage{color,soul} 
\usepackage{datetime}
\newcommand{\labell}[1]{\label{#1}}

\newcommand{\bea}{\begin{eqnarray}}
\newcommand{\eea}{\end{eqnarray}}
\newcommand{\ba}{\begin{eqnarray}}
\newcommand{\ea}{\end{eqnarray}}

\newcommand{\beq}{\begin{equation}}
\newcommand{\eeq}{\end{equation}}
\newcommand{\beqa}{\begin{eqnarray}}
\newcommand{\eeqa}{\end{eqnarray}}
\newcommand{\beqar}{\begin{eqnarray*}}
\newcommand{\eeqar}{\end{eqnarray*}}

\newcommand{\ssc}{\scriptscriptstyle}
\newcommand{\eg}{{\it e.g.,}\ }
\newcommand{\ie}{{\it i.e.,}\ }

\newcommand{\req}[1]{(\ref{#1})}

\newcommand{\C}{\mathcal{C}}

\newcommand{\E}{\mathcal{E}}

\newcommand{\ct}{C_{T}} %{C_\mt{T}}

\newcommand{\rh}{{r_{\rm \ssc H}}}

\newcommand{\cs}{C_{\ssc S}}
\newcommand{\cse}{C_{ \ssc{S } }^{ \rm E}}
\newcommand{\ctt}{C_{\ssc T}}
\newcommand{\ctte}{C_{\ssc T}^{{\rm E}}}

\renewcommand{\href}[2]{#2}

\title{Holographic studies of Einsteinian cubic gravity}
\author[a]{Pablo Bueno,}
\author[b,c]{Pablo A. Cano}
\author[b]{and Alejandro Ruip\'erez}
\affiliation[a]{\vspace{0.1cm}Instituut voor Theoretische Fysica, KU Leuven,\\ Celestijnenlaan 200D, B-3001 Leuven, Belgium\vspace{0.1cm}}
\affiliation[b]{Instituto de F\'isica Te\'orica UAM/CSIC,\\C/ Nicol\'as Cabrera, 13-15, C.U. Cantoblanco, 28049 Madrid, Spain\vspace{0.1cm}}
\affiliation[c]{Perimeter Institute for Theoretical Physics,\\ Waterloo, ON N2L 2Y5, Canada}
\vspace{0.1cm}
\abstract{Einsteinian cubic gravity provides a holographic toy model of a nonsupersymmetric CFT in three dimensions, analogous to the one defined by Quasi-topological gravity in four. The theory admits explicit non-hairy AdS$_4$ black holes and allows for numerous exact  calculations, fully nonperturbative in the new coupling.  
	We identify several entries of the AdS/CFT dictionary for this theory, and study its thermodynamic phase space, finding interesting new phenomena. We also analyze the dependence of R\'enyi entropies for disk regions on universal quantities characterizing the CFT.  In addition, we show that $\eta/s$ is given by a non-analytic function of the ECG coupling, and that the existence of positive-energy black holes strictly forbids violations of the KSS bound.
Along the way, we introduce a new method for evaluating Euclidean on-shell actions for general higher-order gravities possessing second-order linearized equations on AdS$_{(d+1)}$. Our generalized action involves the very same Gibbons-Hawking boundary term and counterterms valid for Einstein gravity, which now appear weighted by the universal charge $a^*$ controlling the entanglement entropy across a spherical region in the CFT dual to the corresponding higher-order theory.
	
	  }

\preprint{
}

\begin{document}
\maketitle

\section{Introduction}\labell{Introduction}
Higher-order gravities play an important role in AdS/CFT \cite{Maldacena,Gubser,Witten}. Perturbative corrections to the large-$N$ and strong-coupling limits of holographic CFTs are encoded, from the bulk perspective, in higher-curvature interactions which modify the semiclassical Einstein (super)gravity action --- see \eg \cite{Grisaru:1986px,Gross:1986iv,Gubser:1998nz,Buchel:2004di}.
%When understood as perturbative modifications of semiclassical Einstein (super)gravity, they encode corrections to the large-$N$ and strong-coupling limits of the corresponding CFT. 
The introduction of such terms, which is in principle fully controlled by String Theory, gives rise to
%In that framework, the structure of such terms is fully controlled and constrained by the underlying String Theory.
%From a different perspective, the addition of simple higher-curvature terms in the bulk action
 %can give access to 
 holographic theories belonging to universality classes different from the one defined by Einstein gravity \cite{Buchel:2008vz,Hofman:2008ar,Hofman:2009ug} --- \eg one can construct CFTs with $a\neq c$ in $d=4$ \cite{Nojiri:1999mh,Blau:1999vz}.  Some care must be taken, however. As shown in \cite{Camanho:2014apa}, higher-curvature terms making finite contributions to physical quantities in the dual CFT can become acausal unless new higher-spin ($J> 2$) modes appear at the scale controlling the couplings of such terms.
 
 %On the one hand, if one of these terms makes finite contributions to the dual theory, it means that the background scale has approached stringy or Planckian regimes, where infinitely many other higher-curvature terms would become relevant as well, suggesting the breakdown of the local field-theory description in the bulk. On the other, if we assume the new couplings to be controlled by scales much larger than 
 
  In spite of this, a great deal of non-trivial information can be still obtained by considering particular higher-curvature interactions at finite coupling --- \ie beyond a perturbative approach. The idea is to select theories whose special properties make them amenable to calculations --- something highly nontrivial in general. The approach turns out to be very rewarding and, in some cases, it has led to the discovery of universal properties valid for completely general CFTs  \cite{Myers:2010tj,Myers:2010xs,Mezei:2014zla,Bueno1,Bueno2}. In other cases, higher-order gravities have served as a proof of concept, \eg providing counterexamples \cite{Buchel:2004di,Kats:2007mq,Brigante:2007nu,Myers:2008yi,Cai:2008ph,Ge:2008ni} to the Kovtun-Son-Starinets bound for the shear viscosity over entropy density ratio \cite{Kovtun:2004de} --- see discussion below. 
  Just like free-field theories, these holographic higher-order gravities should be regarded as toy models for which many calculations can be explicitly performed, hence providing important insights on physical quantities otherwise practically inaccesible for most CFTs --- see \eg \cite{HoloRen,Hung:2014npa,deBoer:2011wk,Bianchi:2016xvf} for additional examples.

A key property one usually demands from a putative holographic model of this kind is that it admits explicit AdS black-hole solutions. In $d\geq 4$, this canonically selects Gauss-Bonnet or, more generally, Lovelock gravities \cite{Lovelock1,Lovelock2}, for which numerous holographic studies have been performed in different contexts --- see \eg \cite{Camanho:2009vw,deBoer:2009pn,Buchel:2009sk,deBoer:2009gx,Camanho:2009hu,Grozdanov:2014kva,Grozdanov:2016fkt,Andrade:2016rln,Konoplya:2017zwo} and references therein. The next-to-simplest example in $d=4$ is Quasi-topological gravity (QTG) \cite{Quasi,Quasi2}, a theory which includes, in addition to the Einstein gravity and Gauss-Bonnet terms, an extra density, cubic in the Riemann tensor. Besides admitting simple generalizations of the Einstein gravity AdS black holes, and having second-order linearized equations of motion on maximally symmetric backgrounds, this theory contains three dimensionless parameters: the ratio of the cosmological constant scale over the Newton constant, $L^2/G$, and the new gravitational couplings, $\lambda$ and $\mu$. These can be translated into the three parameters characterizing the three-point function of the boundary stress tensor. As opposed to Lovelock theories, for which one of such parameters, customarily denoted $t_4$ \cite{Hofman:2008ar}, is always zero \cite{Buchel:2009sk,deBoer:2009gx,Camanho:2009hu,Camanho:2013pda}, the new QTG coupling gives rise to a nonvanishing $t_4$ \cite{Myers:2010jv}. For supersymmetric theories one also has $t_4=0$ \cite{Hofman:2008ar,Kulaxizi:2009pz}, so QTG provides a toy model of a non-supersymmetric CFT in four dimensions.

All studies performed so far involving finite higher-curvature couplings have been restricted to $d\geq 4$ --- observe that all theories mentioned in the previous paragraph reduce to Einstein gravity for $d=3$. Obviously, from the CFT side, there is no fundamental reason to exclude holographic three-dimensional theories. In fact, there exist many interesting CFTs in $d=3$ with known holographic duals, \eg \cite{Maldacena,Aharony:2008ug,Aharony:2008gk,Klebanov:2002ja,Leigh:2003gk,Aharony:2011jz}.
The actual reason for the absence of holographic studies involving higher-curvature terms in $d=3$ has been the lack of examples admitting generalizations of Einstein gravity black holes in four bulk dimensions. The situation has recently changed thanks to the discovery of Einsteinian cubic gravity (ECG) \cite{PabloPablo}, for which such generalizations are possible \cite{Hennigar:2016gkm,PabloPablo2} --- see section \ref{ECGG} for a detailed review. As we show here, ECG provides a holographic toy model of a nonsupersymmetric CFT in three dimensions, analogous to the one defined by QTG in four. The main purpose of this paper is to study the behavior of several physical quantities in this new model. Just like it occurs for Lovelock and QTG in $d\geq 4$, all results can be obtained fully nonperturbatively in the new gravitational coupling, which provides a much better handle on the corresponding quantities than any possible perturbative calculation.

On a more general front, we propose a new method for computing Euclidean on-shell actions for asymptotically AdS$_{(d+1)}$ solutions of an important class of general higher-order gravities --- those for which the linearized equations become second-order on maximally symmetric backgrounds. Our generalized action represents a drastic simplification with respect to standard approaches, as it utilizes the same boundary term and counterterms as for Einstein gravity, but weighted by a universal quantity related to the entanglement entropy across a spherical region in the boundary theory. 

A more precise summary of our findings can be found next.

\subsection{Summary of results}\label{summ}
The paper is somewhat divided into two main parts. In the first, which includes sections \ref{ECGG}, \ref{BHs} and \ref{osa}, we develop some preliminary results and techniques which are necessary for the holographic computations which we perform in sections \ref{tt} to \ref{shear}.
\begin{itemize}
\item{In section \ref{ECGG}, we start with a review of ECG and recent developments. Then, we characterize the AdS$_4$ vacua of the theory, and identify the range of (in principle) allowed values of the new coupling and its relation to the existence of a critical limit for which the effective Newton constant blows up.} 
\item{In section \ref{BHs}, we construct the AdS$_4$ black holes of ECG with general horizon topology.}
\item{In section \ref{osa}, we propose a new method for computing on-shell actions of asymptotically-AdS solutions of general higher-order gravities whose linearized spectrum on AdS$_{(d+1)}$ matches that of Einstein gravity. %We argue that the usual Gibbons-Hawking-York boundary term of general relativity can be still used in that case we argue should be valid for any theory with Einstein-like spectrum in an asymptotically AdS space. 
	We claim that the corresponding boundary term and counterterms can be chosen to be proportional to the usual Einstein gravity ones. Amusingly, we find that the proportionality factor is controlled by the charge $a^*$ characterizing the entanglement entropy across a spherical region $\mathbb{S}^{d-2}$ in the dual CFT. As a first consistency check of our proposal, we use our generalized action to prove the relation between $a^*$ and the on-shell gravitational Lagrangian $\mathcal{L}|_{\rm AdS}$ for odd-dimensional holographic CFTs with  higher-curvature duals. }
\item{In section \ref{tt}, we compute the charge $\ct$ controlling the correlator of the boundary stress-tensor from an explicit holographic computation and show that the result agrees with the (not so) naive expectation obtained from the effective Newton constant. We argue that the detailed cancellations between bulk and boundary contributions giving rise to the correct answer constitute a strong check of the generalized action proposed in the previous section.}
\item{In section \ref{therr}, we start with another check of our generalized action, consisting in an explicit calculation of the free energy of ECG AdS$_4$ black holes, which we show to agree with the one obtained using Wald's entropy approach. Then we compute the thermal entropy charge $\cs$, and we note that it presents notable differences with respect to previous results for other higher-curvature holographic models in $d\geq 4$. Then, we study the thermal phase space of holographic ECG with toroidal and spherical boundaries, respectively. In the latter case, we find that the standard Hawking-Page transition also occurs in ECG. However, the transition temperature increases with the ECG coupling, and actually diverges in the critical limit (for which thermal AdS always dominates). The phase diagram presents new phenomena, like the presence of `intermediate-size' black holes, a new phase of small and stable black holes, as well as the existence of a new critical point.}
\item{In section \ref{renyie}, we compute the Renyi entropy of disk regions in holographic ECG. In particular, we study the dependence of $S_q/S_1$ on the CFT-charges ratio $\ct/a^*$. Although the functional dependence is very complicated, we observe that the behavior is approximately linear for most values in the allowed range. We also obtain an exact result for the scaling dimension of twist operators, from which we are able to extract the value of the stress-tensor three-point function charge $t_4$, which is non-vanishing in general.}
\item{In section \ref{shear}, we compute the shear viscosity to entropy density ratio in ECG. Unlike all previous exact results ($d\geq 4$), the result turns out to be highly nonperturbative in the ECG coupling, as it involves a non-analytic function. Several approximations as well as a precise numerical evaluation are accesible. We find that violations of the KSS bound are strictly forbidden in ECG by the requirement that black holes have positive energy. On the other hand, we show that energy-flux bounds on $t_4$ impose a maximum value for the ratio, given by $(\eta/s)|_{\rm max.}\simeq 1.253/(4\pi)$.}
\item{In section \ref{discu}, we make a quick summary of the different universal charges computed throughout the paper and how they compare with the analogous ones for QTG in $d=4$. Here, we also speculate on the possible implications of the generalized on-shell action introduced in section \ref{osa} for holographic complexity.}
\item{In appendix \ref{ttt}, we show that the scaling dimension of twist operators can be used to obtain the exact results for the stress-tensor three-point function parameters $t_2$ and $t_4$ for holographic theories in which explicit calculations of such quantities had been performed before. Appendix \ref{BTcheck} provides an additional check of our generalized action, in this case for a theory for which the generalized version of the Gibbons-Hawking-York term is explicitly known, namely, Gauss-Bonnet. We show that our method gives rise to exactly the same divergent and finite terms as the standard prescription. Appendix \ref{2pbdy} contains some intermediate calculations omitted in section \ref{tt}.  }
\end{itemize}

\subsubsection*{Note on conventions}
We set $c=\hbar=1$ throughout the paper. $D$ stands for the number of spacetime dimensions of the bulk theory, and $d\equiv D-1$ for those of the boundary one. We use signature $(-,+,+,\dots)$, latin indices from the beginning of the alphabet for bulk tensors, $a,b,\dots=0,1,\dots,D$, Greek indices for boundary tensors, $\mu,\nu,\dots=0,1,\dots,d$ and $i,j,\dots=1,\dots,d$ for spatial indices on the boundary.
Our conventions for $\ctt$, $t_4$, $\cs$ and $a^*$ are the same as in \cite{Buchel:2009sk,Myers:2010jv,Myers:2010tj,Bueno2}. Superscripts `E' and `ECG' mean that the corresponding quantities are computed for Einstein and Einsteinian cubic gravities respectively, whereas we use the subscript `$E$' for Euclidean actions. $L$ is the cosmological constant length-scale ($-2\Lambda_0 \equiv (D-1)(D-2)/L^2$) whereas $\tilde{L}$ stands for the AdS$_{D}$ radius. We often use $L$ for intermediate calculations (including on-shell actions, etc.), but normally present final results in terms of $\tilde{L}$. It is then important to keep in mind that, when expressing our results in terms of the ECG coupling $\mu$, there is some additional dependence hidden in $\tilde{L}=L/\sqrt{f_{\infty}}$, as $f_{\infty}$ is also a function of $\mu$ --- see Fig. \ref{ffffi} and \req{finfs}. 
 
 \section{Einsteinian cubic gravity}\label{ECGG}
Let us start with a quick review of four-dimensional Einsteinian cubic gravity (ECG) and its most relevant properties. The $D$-dimensional version of the theory was presented in \cite{PabloPablo}, where it was shown to be the most general diffeomorphism-invariant metric theory of gravity which, up to cubic order in curvature, shares the linearized spectrum of Einstein gravity on general maximally symmetric backgrounds in general dimensions\footnote{More concretely, the theory is selected by asking it to be the `same' for arbitrary $D$, in the sense that the coefficients relating the various cubic invariants entering its definition do not depend on $D$.}. This criterion selects the Lovelock densities --- cosmological constant, Einstein-Hilbert, Gauss-Bonnet and cubic Lovelock densities --- plus a new invariant, which reads
\begin{equation}
\mathcal{P}=12 R_{a\ b}^{\ c \ d}R_{c\ d}^{\ e \ f}R_{e\ f}^{\ a \ b}+R_{ab}^{cd}R_{cd}^{ef}R_{ef}^{ab}-12R_{abcd}R^{ac}R^{bd}+8R_{a}^{b}R_{b}^{c}R_{c}^{a}\, .
\end{equation}
This invariant is neither trivial nor topological in $D=4$, so the action of the theory becomes
\begin{equation}\label{ECG}
I^{\rm ECG}=\frac{1}{16\pi G}\int d^4x \sqrt{|g|}\left[\frac{6}{L^2}+R-\frac{\mu L^4}{8} \mathcal{P} \right]\, ,
\end{equation}
in such a number of dimensions\footnote{From now on, we will always be referring to the four-dimensional version of the theory when referring to `ECG', unless otherwise stated.}. Here, $\mu$ is a dimensionless coupling. Note also that, for later convenience, in \req{ECG} we have chosen the cosmological constant to be negative, $-2\Lambda_0\equiv 6/L^2$, where $L$ is a length scale which will coincide with the corresponding AdS$_4$ radius for $\mu=0$. 

It was subsequently shown \cite{Hennigar:2016gkm,PabloPablo2} that \req{ECG} admits non-trivial generalizations of Einstein gravity's Schwarzschild black hole characterized by a single function $f(r)$ --- see next section. It was also observed \cite{Hennigar:2017ego,PabloPablo3,Ahmed:2017jod,PabloPablo4} that, in fact, ECG belongs to a broader class of theories --- coined \emph{Generalized Quasi-topological gravities} in \cite{Hennigar:2017ego} --- which also includes Lovelock \cite{Lovelock1,Lovelock2} and Quasi-topological \cite{Quasi2,Quasi,Myers:2010jv,Dehghani:2011vu,Dehghani:2013ldu,Cisterna:2017umf} gravities as particular examples, and which are characterized by: having a well-defined Einstein gravity limit; sharing the linearized spectrum of Einstein gravity on general maximally symmetric backgrounds; admitting non-hairy single-function generalizations of Schwarzschild's black hole. If the action does not include derivatives of the Riemann tensor, the full non-linear equations of a given theory belonging to this class reduce, on a general static and spherically symmetric ansatz, to a single (at most second-order) differential or algebraic --- depending on the case \cite{PabloPablo2} --- equation for $f(r)$, which indeed can be seen to correspond to a unique non-hairy black hole whose thermodynamic properties can be exactly obtained by solving a system of algebraic equations without free parameters.
 
The thermodynamic properties of the asymptotically flat ECG black holes and its higher-curvature generalizations are very different from their Einstein gravity counterparts, as they become stable below certain mass, which results in infinite evaporation times \cite{PabloPablo2,PabloPablo4}. The asymptotically-AdS black brane solutions of ECG, and generalizations above mentioned, have also been considered in \cite{PabloPablo3,Ahmed:2017jod,PabloPablo4} and, specially, in \cite{Hennigar:2017umz}. There, it was shown that, as opposed to  all previously considered higher-order gravities, the charged black brane solutions of the Generalized QTG class in $D\geq 4$ generically present nontrivial thermodynamic phase spaces, containing phase transitions and critical points.
 
Another relevant development entailed the identification of a critical limit of ECG (for which the effective Newton constant diverges) \cite{Feng:2017tev}, corresponding to $\mu=4/27$. In that particular case, the black holes --- as well as other interesting solutions, such as bounce universes --- can be constructed analytically. 

More recently, some of the possible observational implications of the theory were studied in \cite{Hennigar:2018hza}. There, an observational bound on the ECG coupling was found using Shapiro time delay, and the effects of ECG on black-hole shadows were discussed, including possible measurable differences with respect to Einstein gravity predictions.  Comparisons between general relativity and other theories of gravity regarding black-hole observables are highly limited by the lack of explicit four-dimensional alternatives, which makes ECG particularly appealing for this purpose.

Finally, from the holographic front, let us mention that a study of R\'enyi entropies for spherical regions, similar to the one we perform in section \ref{renyie}, was carried out in \cite{Dey:2016pei} for ECG in $D=5$. However, it should be stressed that in dimensions greater than four, ECG does not belong to the Generalized QTG class, in the sense that --- even though it shares the linearized spectrum of Einstein gravity --- simple black hole solutions satisfying the properties explained above do not exist for the theory and, as opposed to the $D=4$ case, one is restricted to perturbative calculations in the gravitational couplings, which makes them less interesting.

\subsection{AdS$_4$ vacua and linearized spectrum}
The AdS$_{4}$ vacua of \req{ECG} have a curvature scale $\tilde{L}$ related to the action length scale $L$ through
\begin{equation}\label{adsc}
\frac{1}{\tilde{L}^2}=\frac{f_{\infty}}{L^2}\, ,
\end{equation}
where $f_{\infty}$ is a solution to the algebraic equation
\begin{equation}\label{roo}
1-f_{\infty}+\mu f_{\infty}^3=0\, .
\end{equation}	
For negative values of $\mu$, two of the roots are imaginary, and one is  real and positive. For $0<\mu<4/27$, the three roots are real, one of them being negative and the other two positive. Finally, for $\mu>4/27$, two of the roots are imaginary, and the remaining one is negative. Hence, imposing $f_{\infty}>0$, constrains $\mu$ as
\begin{equation}\label{sis}
\mu<\frac{4}{27}\simeq 0.148\,.
\end{equation}
For larger values of $\mu$, no positive roots exist, which means that no AdS$_4$ vacuum exists in that case\footnote{This analysis is analogous to the one corresponding to QTG in $D\geq 5$ \cite{Quasi,Myers:2010jv}, with the difference that, in that case, the Gauss-Bonnet term is present, and the identification of the allowed stable vacua becomes more involved.}. However, not all real roots of \req{roo} satisfying \req{sis} give rise to stable vacua.

In order to see this, we can consider the linearized equations of motion of \req{ECG} on a general maximally symmetric background (in particular, one of these AdS$_4$), in the presence of minimally coupled fields. As already mentioned, these always reduce to the linearized equations of Einstein gravity, up to a normalization of the  Newton constant \cite{PabloPablo,Aspects}, namely 
\begin{equation}
G_{ab}^{ \ssc L}=8 \pi G_{\rm eff}^{\rm ECG} T_{ab}\, ,
\end{equation}
where $G_{ab}^{\ssc L}$ is the linearized Einstein tensor, $T_{ab}$ is the stress tensor of the extra fields, and $G_{\rm eff}^{\rm ECG}$ is the effective Newton's constant, which is given by
\begin{equation}\label{Geff}
G_{\rm eff}^{\rm ECG}=\frac{G}{1-3\mu f_{\infty}^2}\, .
\end{equation}
The sign of $G_{\rm eff}$ determines the sign of the graviton propagator. Whenever the denominator in the right-hand side --- which is nothing but (minus) the slope of \req{roo} --- is negative, the graviton becomes a ghost, and the corresponding vacuum is unstable. This imposes $\mu<0$ or $f_{\infty}^2< 1/(3\mu)$ for positive values of $\mu$. The condition kills one of the two positive roots of \req{roo} available for $0<\mu<4/27$, which would then correspond to unstable vacua. Hence, we conclude that, whenever \req{sis} is satisfied, there exists a single stable vacuum. No additional vacua exist for $\mu<0$, whereas an additional unstable vacuum exists for $0<\mu<4/27$. Special comment deserves the $f_{\infty}^2=1/(3\mu)$ case, corresponding to $\mu=4/27$, and for which $G_{\rm eff}\rightarrow +\infty$. This `critical' limit of the theory was identified in \cite{Feng:2017tev}, and gives rise to a considerable simplification of most calculations, as we further illustrate below.
\begin{figure}[t]
	\centering 
	\includegraphics[scale=0.65]{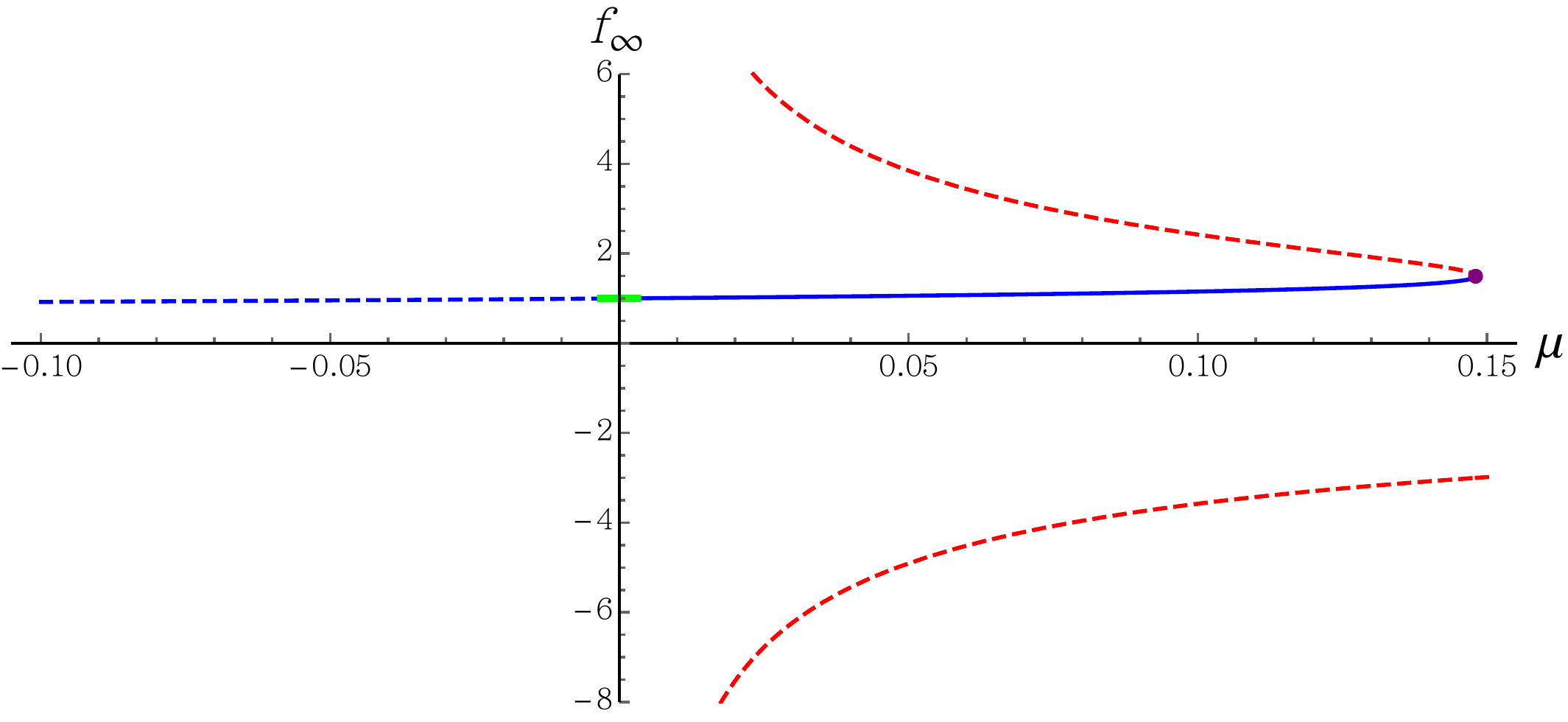}
	\caption{Real roots of \req{roo} for different values of $\mu$. The lower red dashed line corresponds to $f_{\infty}<0$, whereas the upper one corresponds to unstable vacua; the blue dashed line ($\mu<0$) corresponds to stable vacua which do not allow for black hole solutions --- see discussion under \req{hh}; the purple dot corresponds to the critical case, $\mu=4/27$; finally, the small green region corresponds to the set of parameters allowed by the positive-energy constraint $|t_4|\leq 4$ in \req{fifif}. }\label{ffffi}
\end{figure}

We summarize these observations in Fig. \ref{ffffi}, were we also include two additional constraints which we derive in sections \ref{BHs} and \ref{t44}, respectively. The first comes from imposing the existence of black holes solutions, which restricts the allowed values to $0\leq \mu \leq 4/27$. The second follows from the positivity of energy fluxes at null infinity which, as we can see from the figure, produces the very stringent constraint, $-0.00322\leq \mu \leq 0.00312$.

Throughout the paper, we will assume $\mu$ to lie in the range $0\leq \mu \leq 4/27$. From the two positive roots of \req{roo} in that range, we will be implicitly choosing the one corresponding to a stable vacuum, which is also the one connecting to the Einstein gravity one for $\mu\rightarrow 0$. While the positive-energy condition further limits this range, we find it convenient to also consider values close to $\mu=4/27$, for which many exact results can be obtained. Let us finally point out that the solution of \req{roo} corresponding to the relevant root (blue in Fig. \ref{ffffi}) can be written explicitly as
\begin{equation}\label{finfs}
f_{\infty}=\frac{2}{\sqrt{3\mu}}\sin \left[\frac{1}{3}\arcsin \left(\sqrt{\frac{27\mu}{4}} \right) \right]\, .
\end{equation}

%As we explain in the next subsection, additional bounds on $\mu$ can be obtained from 

\section{AdS$_4$ black holes}
\label{BHs}
ECG admits static asymptotically AdS$_4$ black holes of the form
\begin{equation}\label{bhss}
ds^2=-N^2 V_k(r) dt^2+\frac{dr^2}{V_k(r)}+\frac{r^2}{L^2} d\Sigma_k^2\, ,\quad \text{where} \quad d\Sigma_k^2= \begin{cases}
L^2 d\Omega^2_2\, ,\quad &\text{for}\quad k=+1\, , \\
d\vec{x}_2^2\, ,\quad &\text{for}\quad k=0\, ,\\
L^2 d\Xi^2\, ,\quad &\text{for}\quad k=-1\, ,\\
\end{cases}
\end{equation}
corresponding to spherical, planar and hyperbolic horizons, respectively, and where $V_k(r)$ is determined from the second-order differential equation
\begin{equation}\label{eqVs}
1-\frac{L^2 (V_k-k)}{r^2}-\frac{3L^6\mu}{4r^3} \left[\frac{V_k'^3}{3}+\frac{k V_k'^2}{r}-\frac{2 V_k (V_k-k)V'_k}{r^2}-\frac{V_k V_k''(rV'_k  -2 (V_k-k))}{r} \right]=\frac{\omega^3}{r^3}\,,
\end{equation}
where $\omega^3$ is an integration constant related to the ADM energy \cite{Abbott:1981ff,Deser:2002jk} of the solution --- see \req{adm}. Also, $N^2$ is a constant that we fix in different ways depending on the horizon geometry, \eg \cite{Quasi,Myers:2010jv,HoloRen}. In particular, we will choose $N^2=1$ for spherical horizons, $N^2=1/f_{\infty}$ for planar horizons, which sets the speed of light in the dual theory to one, and $N^2=L^2/(f_{\infty}R^2)$ for hyperbolic horizons, so that the boundary metric is conformally equivalent to that of $\mathbb{R}\times \mathbb{H}^2$, where $R$ is the curvature scale of the hyperbolic slices. 

The fact that ECG admits static solutions of the form \req{bhss}, characterized by a single function $V_k(r)$, such that the full nonlinear equations\footnote{These can be found explicitly \eg in \cite{Hennigar:2016gkm}.}  of the theory reduce to a single third-order differential equation, which can in turn be integrated once to yield \req{eqVs}, is a highly non-trivial property of ECG \cite{Hennigar:2016gkm,PabloPablo2}. Such property is shared by the higher-dimensional Lovelock \cite{Wheeler:1985nh,Wheeler:1985qd,Boulware:1985wk,Cai:2001dz,Dehghani:2009zzb}, QTG \cite{Quasi2,Quasi,Dehghani:2011vu,Cisterna:2017umf} (for these, the equation for $V_k(r)$ is algebraic instead) and Generalized Quasi-topological \cite{Hennigar:2017ego} gravities, as well as by other higher-curvature theories of the same class, recently discovered and characterized \cite{PabloPablo4,Ahmed:2017jod}. As mentioned before, this property is related to the absence of extra modes in the linearized spectrum of the theory, and can be shown to lead to non-hairy black holes whose thermodynamic properties can be computed analytically on general grounds \cite{PabloPablo3}.

In \req{bhss}, it is customary to make the redefinition 
\begin{equation}
V_k(r)=k+\frac{r^2}{L^2}f(r)\, ,
\end{equation}
specially when dealing with the planar and hyperbolic cases. In terms of $f(r)$, \req{eqVs} reads
\begin{equation}\label{eqsf}
1-f+\mu \left[f^3+\frac{3}{2}r^2 f f'^2-\frac{r^3}{4}f' (f'^2-3f f'')+\frac{3}{4}k L^2 f' (rf''+3 f')\right]=\frac{\omega^3}{r^3}\, .
\end{equation}
Observe that this reduces to \req{roo} for constant $f(r)$ and $\omega^3=0$. In particular, asymptotically, we require $\lim_{r\rightarrow+\infty}f(r)=f_{\infty}$, which then makes \req{bhss} become the metric of pure AdS$_4$ with radius $\tilde{L}$ given by \req{adsc}, and a different boundary geometry for each value of $k$ \cite{Emparan:1999pm}.

%There are two cases in which we can solve \req{eqsf} analitically. These correspond to Einstein gravity ($\mu=0$), and to the critical theory ($\mu=4/27$). For those, one finds
%\begin{equation}
%f(r)=\begin{cases}
%1-\frac{k L^2\rh+\rh^3}{r^3}\quad {\rm \, \, if}\quad \mu=0\, ,\\
%\frac{3}{2}-\frac{k L^2+3/2\rh^2}{r^2}\quad {\rm if}\quad \mu=4/27\, .\\
%\end{cases}
%\end{equation}
%For intermediate values of $\mu$, the solutions can be thought of as interpolating between these two limits. 

\subsection{Asymptotic expansion}
For general values of $\mu$, finding analytic black hole solutions of \req{eqsf} looks extremely challenging (if not impossible). Let us then start by exploring the asymptotic and near horizon expansions, from which we can gain a lot of relevant information (and, in fact, argue that non-hairy black hole solutions do really exist for general values of $\mu$).

The first terms in the asymptotic expansion of $f(r)$ read
%In order to solve \req{eqsf}, we need to impose appropriate boundary conditions. Let us first consider the limit $r\rightarrow \infty$, where we impose that the solution is asymptotically AdS$_4$, this is, $\lim_{r\rightarrow\infty}f(r)=f_{\infty}$. In order to study this limit, we can work out an asymptotic expansion of the function $f$ in powers of $1/r$. For the first terms we get
\begin{equation}\label{Asympt}
f_{1/r}(r)=f_{\infty}-\frac{\omega^3}{(1-3\mu f_{\infty}^2)r^3}-\frac{21\mu f_{\infty}\omega^6}{2(1-3\mu f_{\infty}^2)^3r^6}+\mathcal{O}(r^{-8})\, .
\end{equation}
Note that \req{eqsf} is a second-order differential equation, which therefore possesses a two-parameter family of solutions. In order to capture the asymptotic behavior of the most general one, we write $f(r)=f_{1/r}(r)+h(r)$ and then expand \req{eqsf} linearly in $h$. Keeping only leading terms in $1/r$, we get the following equation for $h$\footnote{For instance, we assume that the term $h' L^2r^{-4}$ is negligible compared to $h'' r^{-1}$ when $r\rightarrow +\infty$.}:
\begin{equation}\label{heq}
h''(r)-\frac{4(1-3\mu f_{\infty}^2)^2}{9f_{\infty}\mu \omega^3} r h(r)=0\, .
\end{equation}
Leaving aside the limiting cases, corresponding to $\mu=0$ and $\mu=4/27$, we see that there are two possibilities, depending on the sign of $\mu \cdot \omega^3$.
If $\mu\cdot \omega^3>0$, \req{heq} has the following approximate solutions as $r\rightarrow+ \infty$\footnote{The exact solution of \req{heq} is given by the Airy functions,
\begin{equation*}
h(r)=A \textrm{AiryAi}\left[\left(\frac{4(1-3\mu f_{\infty}^2)^2}{9f_{\infty}\mu \omega^3} \right)^{1/3}r\right]+ B\textrm{AiryBi}\left[\left(\frac{4(1-3\mu f_{\infty}^2)^2}{9f_{\infty}\mu \omega^3} \right)^{1/3}r\right]\, ,
\end{equation*}
but we only need the asymptotic behavior for the discussion.}:
\begin{equation}\label{expA}
h(r)\sim A \exp\left[\frac{4|1-3\mu f_{\infty}^2|}{9\sqrt{f_{\infty}\mu\cdot \omega^3}}r^{3/2}\right]+B\exp\left[-\frac{4|1-3\mu f_{\infty}^2|}{9\sqrt{f_{\infty}\mu\cdot \omega^3}}r^{3/2}\right]\, .
\end{equation}
In order to obtain an asymptotically AdS$_4$ solution, we need to kill the growing mode, which forces us to set $A=0$. Therefore, this boundary condition fixes one of the integration constants required by \req{eqsf}. Now, even though the remaining exponentially decaying term is extremely subleading, in general we will have $B\neq 0$. In fact, this constant ends up being fixed by the horizon-regularity condition. In particular, this implies that the solutions show a strongly nonperturbative character, as   $\sim e^{-1/\sqrt{\mu}}$ terms generically appear. Indeed, it is possible to show that a series expansion of the full solution in powers of $\mu$ is always divergent.  

The second possibility corresponds to $\mu\cdot \omega^3<0$. An approximate solution of \req{heq} for large $r$ is then given by
\begin{equation}\label{hh}
h(r)\sim \frac{A}{r} \cos\left[\frac{4|1-3\mu f_{\infty}^2|}{9\sqrt{f_{\infty}|\mu\cdot \omega^3|}}r^{3/2}\right]+\frac{B}{r}\sin\left[\frac{4|1-3\mu f_{\infty}^2|}{9\sqrt{f_{\infty}|\mu\cdot \omega^3|}}r^{3/2}\right]\, .
\end{equation}
This solution is sick. Although $h(r)\rightarrow 0$ as $r\rightarrow +\infty$, the derivatives of $h$ diverge wildly in this limit, which would force us to set $A=B=0$ in order to get an asymptotically AdS$_4$ solution. However, this leaves us with no additional free parameters, and regularity at the (would-be) horizon cannot be imposed. Therefore, no regular black hole solution exists for $\mu\cdot \omega^3<0$: the solution is always sick, either at the horizon or at infinity. 

As shown later in \req{adm}, $\omega^3$ is proportional to the total energy $E$ (or mass) of the black hole, which leads us to impose $\mu\ge 0$. Hence, interestingly, the range of values of $\mu$ which allows for positive-energy solutions, forbids the negative-energy ones, which simply do not exist for $\mu\ge 0$.
%, in order for $f_{\infty}\mu \omega^3$ to be positive in \req{expA}. 
%\comment{This will imply that solutions with negative energy will not exist. It is interesting that in this theory negative energy configurations are forbidden by dynamics.} 

\subsection{Near-horizon expansion}
Let us now consider the near-horizon behavior. For that, we assume that there is a value $\rh$ of the radial coordinate for which the function $V_k$ vanishes and is analytic. Analyticity ensures that the solution can be maximally extended beyond the horizon using Kruskal-Szekeres-like coordinates. 

The derivative of $V_k$ at the horizon is related to the temperature through: $V_k'(\rh)=4\pi T/N$ so, in terms of $f$, the near-horizon expansion can be written as
\begin{equation}\label{nH}
k+\frac{r^2}{L^2}f(r)=\frac{4\pi T}{N}(r-\rh)+\sum_{n=2}^{\infty} a_n (r-\rh)^n\, ,
\end{equation}
where the relation between $f'(r)$ and the temperature reads in turn
\begin{equation}
T=\frac{N}{4\pi}\left[\frac{\rh^2}{L^2}f'(\rh) -\frac{2k}{\rh}\right]\,.
\end{equation}
Note also that $f(\rh)=-k L^2/\rh^2$. Now, if we plug \req{nH} into \req{eqsf} and we expand it in powers of $(r-\rh)$, we are led to the equation 
\begin{align}
0= 1+\frac{k L^2}{\rh^2}-\frac{\omega^3}{\rh^3}-\frac{4L^6\pi^2 T^2 \mu}{N^2 \rh^3}\left(\frac{3k}{\rh}+\frac{4\pi T}{N}\right)+\, \\
\left[-\frac{2kL^2}{\rh^2}+\frac{3\omega^3}{\rh^3}-\frac{4L^2\pi T}{N \rh}+\frac{24L^6\pi^2 T^2 \mu}{N^2 \rh^3}\left(\frac{k}{\rh}+\frac{2\pi T}{N}\right)\right](r-\rh)
+\mathcal{O}\left( (r-\rh)^2\right)&\, .
\end{align}
Since every coefficient must vanish independently, we get an infinite number of equations relating the parameters in the near-horizon expansion \req{nH}.
From the first two equations, we can obtain $T^{\rm ECG}$ and ${\omega}^{\rm ECG}$ as functions of $\rh$, the result being (in order to minimize the clutter, we often omit the `ECG' superscripts throughout the text)
\begin{align}\label{T}
T^{\rm ECG}&=\frac{N}{2\pi \rh} \left(k+\frac{3\rh^2}{L^2}\right)\left[1+\sqrt{1+\frac{3k L^4\mu}{\rh^4}\left(k +3\frac{\rh^2}{L^2}\right)} \right]^{-1} \, ,\\
{(\omega^{\rm ECG})}^3&=kL^2\rh+\rh^3-\frac{\mu L^6}{4}\left[\frac{3k}{\rh}\left(\frac{4\pi T^{\rm ECG}}{N}\right)^2+\left(\frac{4\pi T^{\rm ECG}}{N}\right)^3\right]\,.
\end{align}
These reduce to the usual Einstein gravity results for $\mu=0$, namely
\begin{align}
T^{\rm  E}=\frac{N}{4\pi \rh} \left(k+\frac{3\rh^2}{L^2}\right)\, ,\quad
{(\omega^{\rm  E})}^3=\rh^3+k \rh L^2\,.
\end{align}
The rest of equations, which we do not show here, fix all coefficients $a_{n>2}$ in terms of $a_2$. Hence, for a fixed $\rh$, the series \req{nH} contains a single free parameter, which is nothing but the value of $f''$ at the horizon. This  must be carefully chosen so that the solution has the appropriate asymptotic behavior, \ie so that $A=0$ in \req{expA}.

\subsection{Full solutions}\label{fullsol}
Equation \req{eqsf} can be solved analytically in two cases, namely: for Einstein gravity, $\mu=0$, and in the critical limit, $\mu=4/27$ \cite{Feng:2017tev}. For those, one finds\footnote{\label{BTZ}A curious property of the critical-theory solutions is that they look identical to three-dimensional BTZ black holes \cite{Banados:1992wn}, with an additional `angular' direction:
\begin{eqnarray}
ds^2_{\rm  ECG,\, crit}&=&-\frac{3(r^2-\rh^2)}{2L^2}dt^2-\frac{2 L^2 dr^2}{3(r^2-\rh^2)}+\frac{r^2}{L^2} d\Sigma_{(k)}^2\, ,\\
ds^2_{\rm BTZ}&=&-\frac{(r^2-\rh^2)}{L^2}dt^2-\frac{L^2 dr^2}{(r^2-\rh^2)}+r^2d\phi^2\, .
\end{eqnarray}
 We point out that an analogous behavior has been observed to occur for critical Gauss-Bonnet gravity ($\lambda_{\rm \ssc GB}=1/4$), see \eg \cite{Grozdanov:2016fkt} as well as for Einstein gravity coupled to an axionic field in a particular limit \cite{Davison:2014lua}. The connection of this phenomenon to other instaces of background-symemtry enhancement --- \eg \cite{Compere:2012jk} --- deserves further attention.}
\begin{equation}\label{limits}
f(r)=\begin{cases}
1-\frac{\rh^3+k L^2\rh}{r^3}\quad {\rm\,\, if}\quad \mu=0\, ,\\
\frac{3}{2}-\frac{3\rh^2+2k L^2}{2r^2}\quad {\rm if}\quad \mu=4/27\, .\\
\end{cases}
\end{equation}
For intermediate values of $\mu$, the solutions can be constructed numerically. In order to do so, we solve \req{eqsf} setting the initial condition at the horizon, and then applying the shooting method to obtain the value of $a_2$ for which $f(r)\rightarrow f_{\infty}$. The differential equation \req{eqsf} is very stiff when $r$ is large but, by choosing $a_2$ accurately, it is always possible to extend the numerical solution well into the region in which the asymptotic expression \req{Asympt} applies. In all cases, there is a unique value of $a_2$ for which this happens. Hence, for each value of $\mu$ and each horizon geometry, there exists a unique regular  black fully characterized by $\rh$ (or, more physically, by $\omega^{\rm ECG}$).

\begin{figure}[t]
	\centering 
	\includegraphics[scale=0.485]{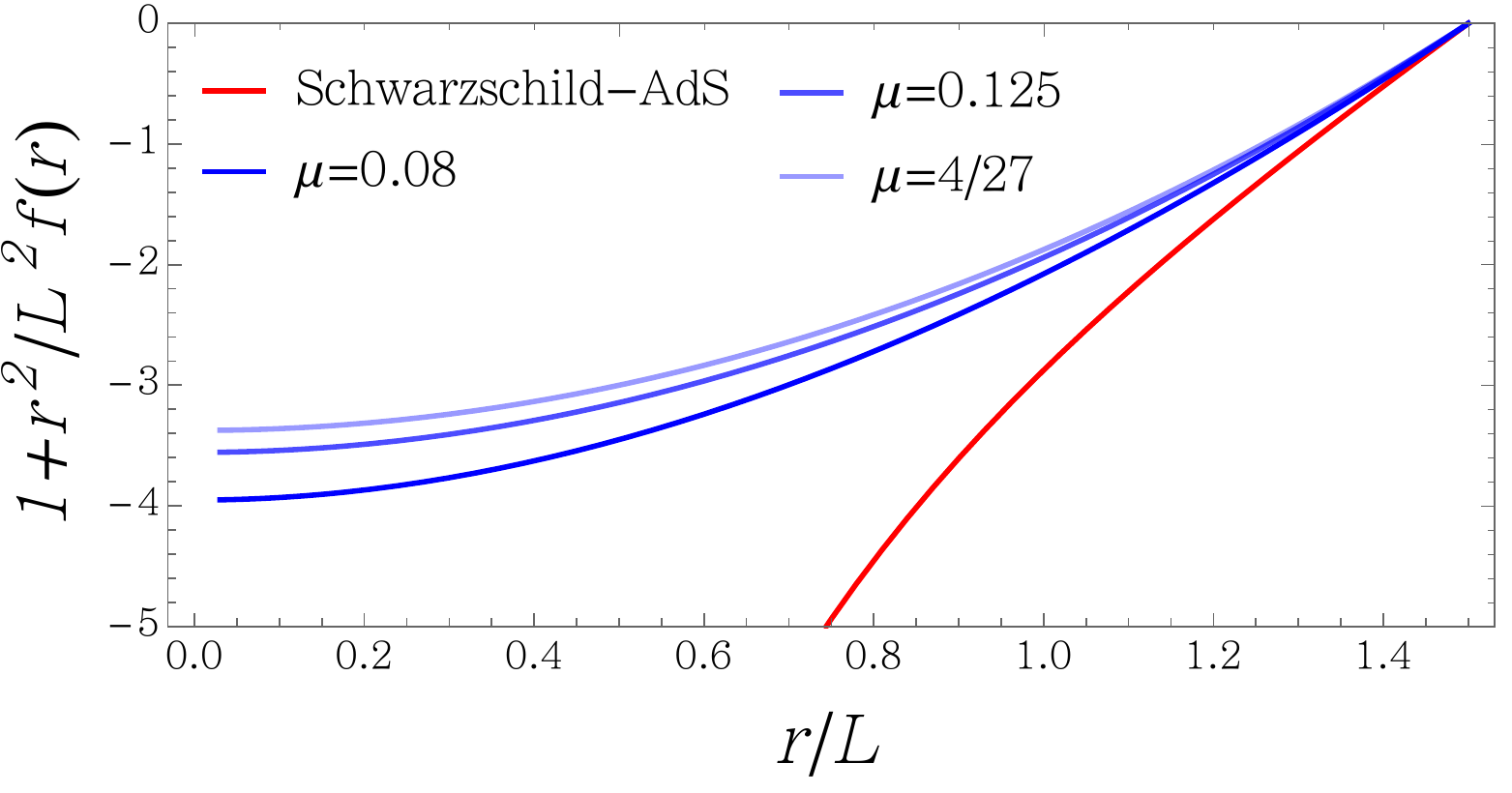}
	\includegraphics[scale=0.485]{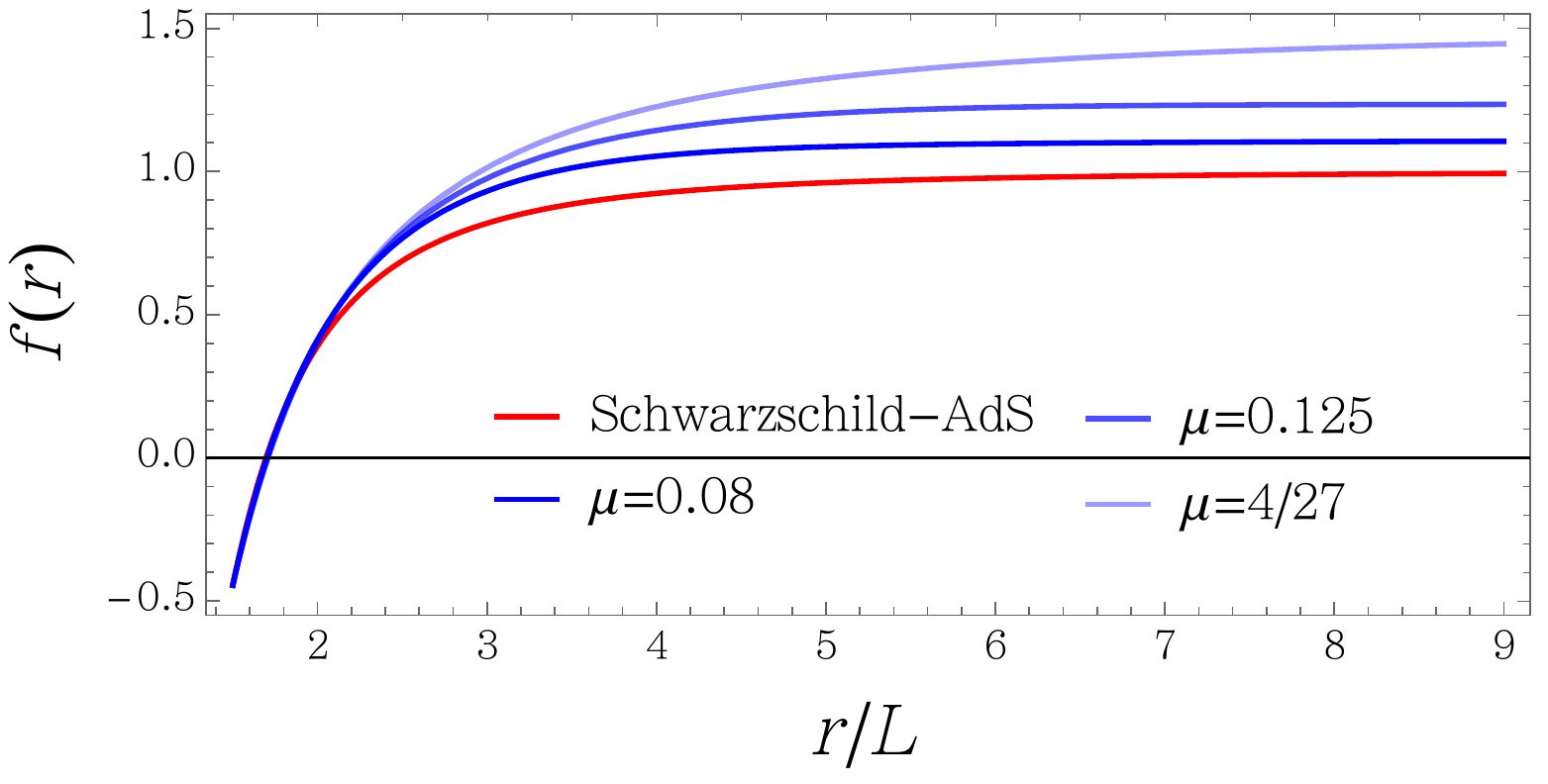}
	\includegraphics[scale=0.485]{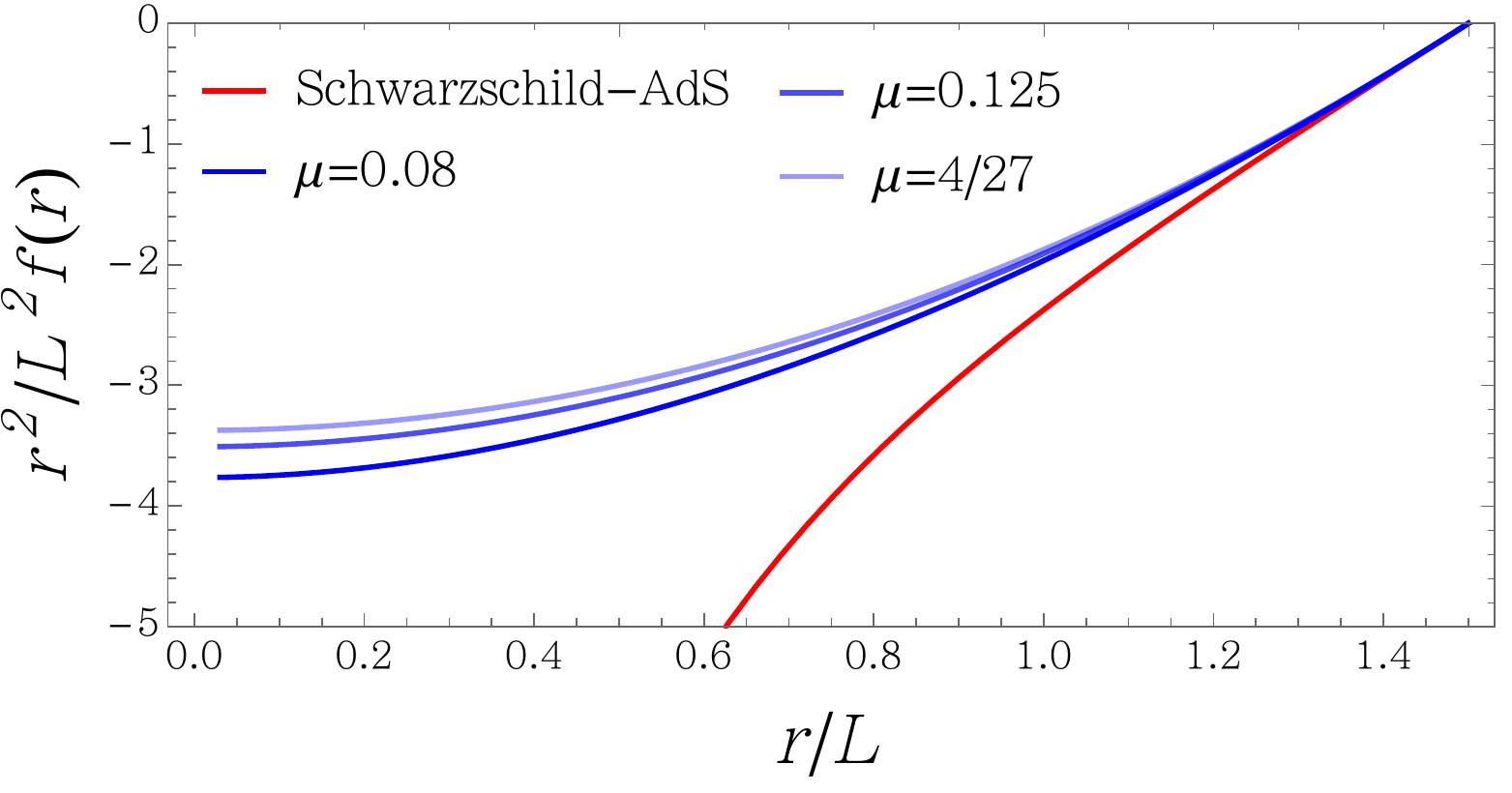}
	\includegraphics[scale=0.485]{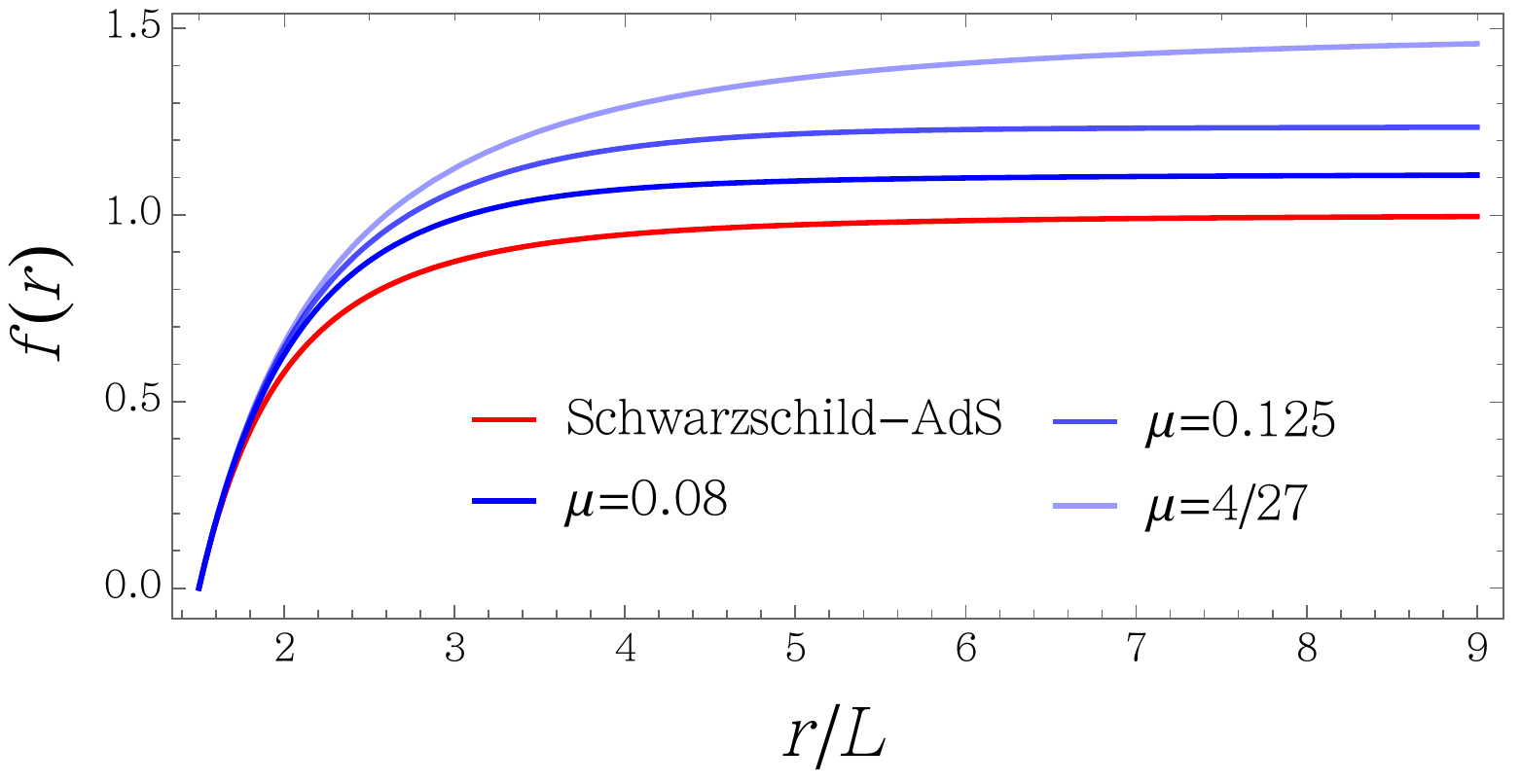}
	\includegraphics[scale=0.485]{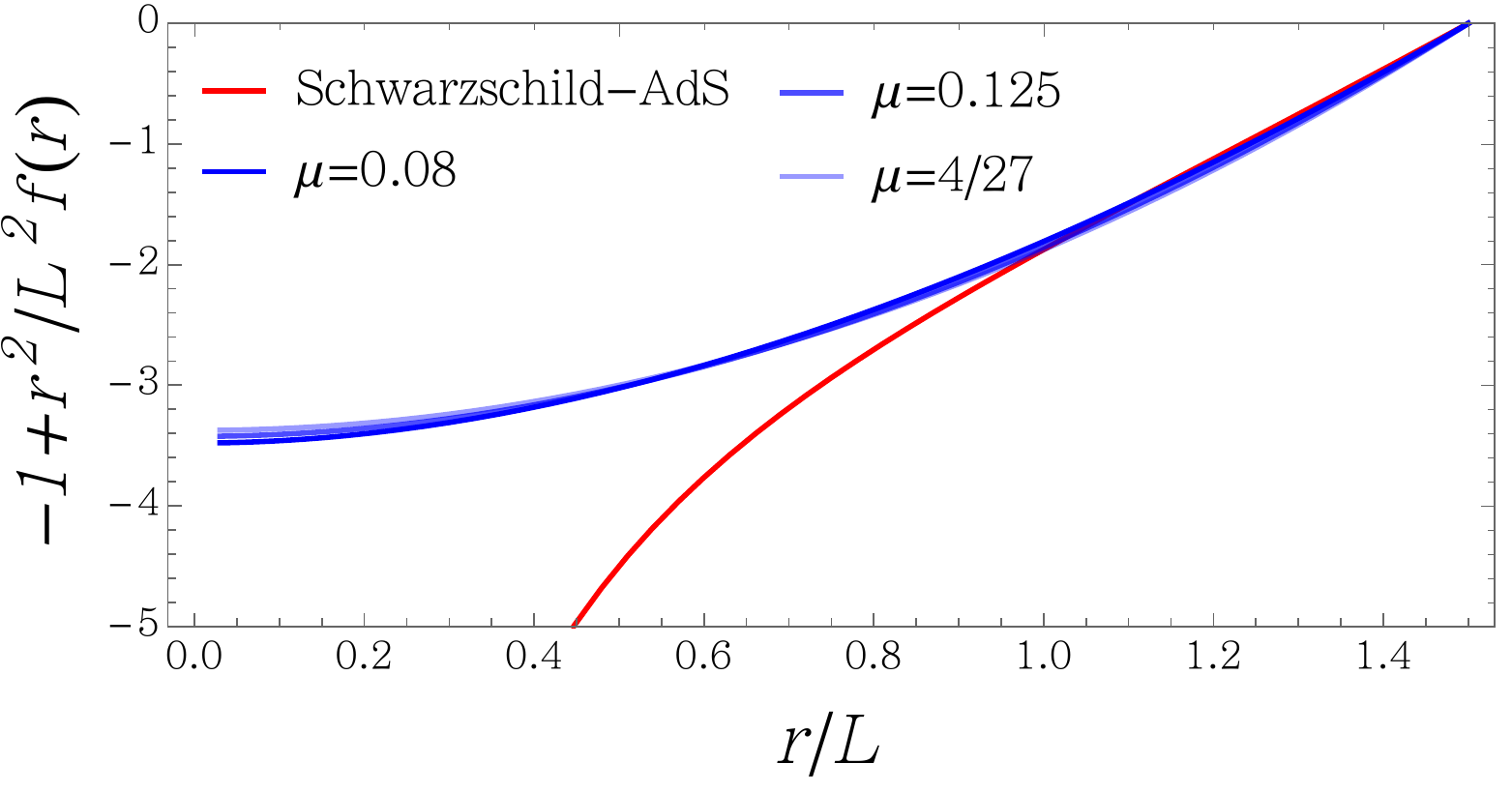}
	\includegraphics[scale=0.485]{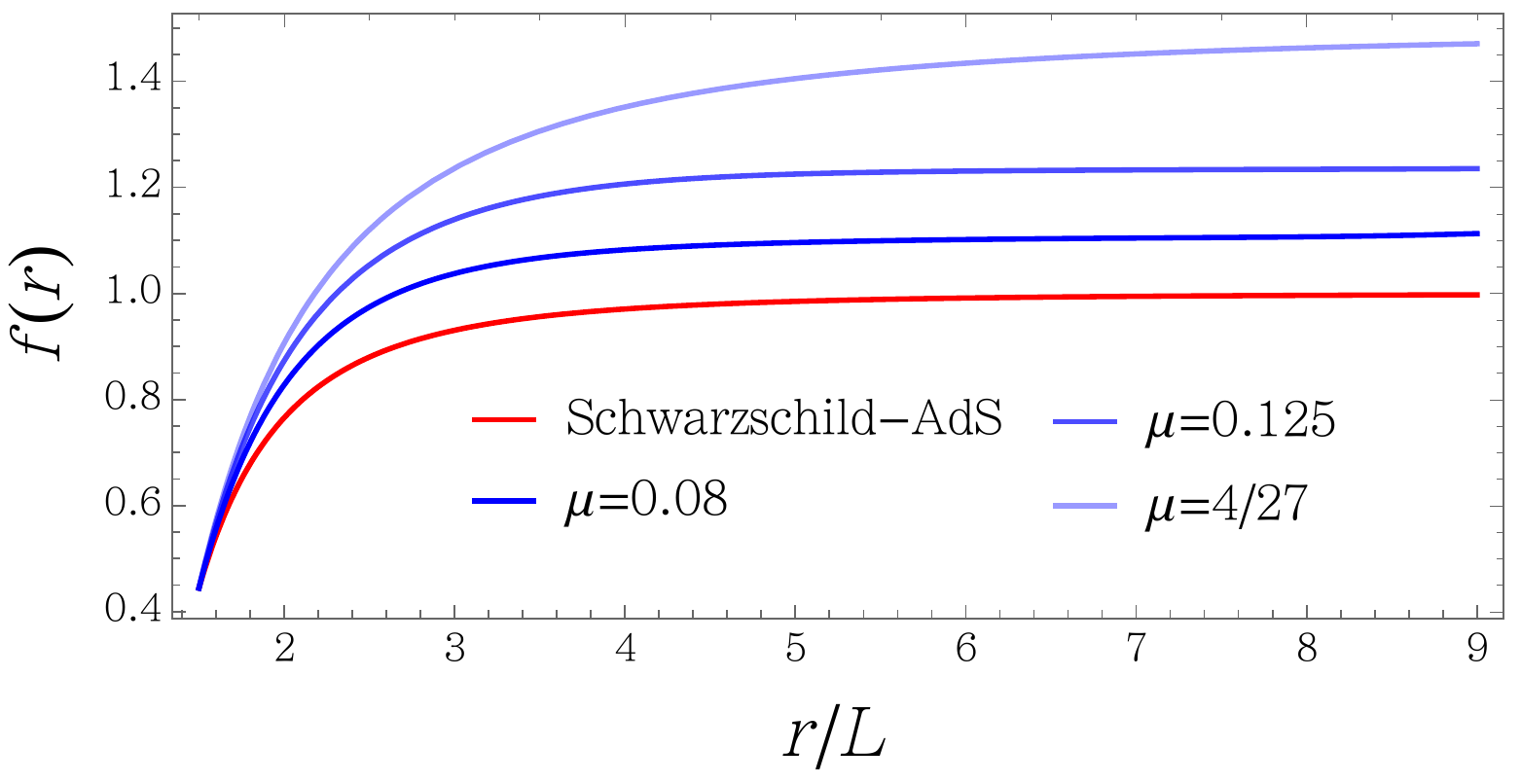}
	\caption{Black hole solutions for several values of $\mu$ (we take $\rh=3L/2$), including the Einstein gravity ($\mu=0$) and critical ($\mu=4/27$) cases. From top to bottom $k=1,0,-1$. For the sake of clarity, we have made separate plots for the interior and exterior solutions. In the left column, we plot $V_k(r)=k+r^2/L^2 f(r)$ for the black-hole interior range. In the right column we plot $f(r)$ instead for the exterior solutions. }\label{fPlot1}
\end{figure}

 In Fig. \ref{fPlot1} we show a couple of these numerical solutions
for $\rh=1.5 L$. As we can see, the corresponding curves lie between the analytic limiting solutions in \req{limits}.  Far from the horizon, the functions $f(r)$ tend to the constant values $f_{\infty}$ which, as explained above, are different for each value of $\mu$ --- see Fig. \ref{ffffi}. Besides the exterior solutions, we also show plots of the black hole interior profiles\footnote{For the sake of visual clarity, we present the interior and exterior solutions in different figures, plotting $V_k(r)$ for the former, and $f(r)$ for the latter.}, which present the curious feature of having regular metrics at $r=0$. However, as observed in \cite{PabloPablo2} for the asymptotically flat case, curvature invariants still diverge. For example, in the critical case, one finds
\begin{equation}
R_{abcd}R^{abcd}=\frac{4k^2L^4+54r^4-6\rh^2 r^2+\rh^4-4kL^2(3r^2-\rh ^2)}{L^4r^4}\sim \mathcal{O}\left(r^{-4} \right)\, , %\quad R^2=\frac{(r_h^2+2kL^2-18r^2)^2}{L^4 r^4}\, ,
\end{equation}
 which is two powers of $r$ softer than in the usual Schwarzschild case. Such behavior is common to all solutions with $\mu\neq 0$. This singularity-softening phenomenon appears to be generic for higher-curvature generalizations of Einstein gravity black holes. For example, for the Gauss-Bonnet black hole \cite{Cai:2001dz}, one finds \cite{Ohta:2010ae} $R_{abcd}R^{abcd}\sim \mathcal{O}(r^{-{(D-1)}})$, which is in turn $(D-1)$ powers of $r$ softer than the Kretschmann invariant of the $D$-dimensional Schwarzschild black hole.

\section{Generalized action for higher-order gravities}\label{osa}

%$a^*$ The  charge appearing in the universal contribution to the entanglement entropy across a $S^{D-3}$ spherical region both in odd and even-dimensional theories\footnote{$a^*$ is an important quantity, as it reduces to the trace anomaly charge $a$ in even dimensions, and \comment{F theorem, bla, bla in odd $(D-1)$}. }. In particular, it was been argued \comment{it is proven for $(D-1)$ even and arguments for $(D-1)$ odd and more to come in the holoAspects of ECG paper} that for a general holographic higher-order gravity theory, this is given by\comment{check signs}
%\begin{equation}
%	a^*=\frac{\pi^{\frac{(D+1)}{2}} \tilde{L}^D}{2\pi \Gamma\left[\frac{D+1}{2}\right]}  \mathcal{L}|_{\rm AdS}\, .
%\end{equation}

When performing holographic calculations with higher-curvature bulk duals, one is faced with the challenge of identifying appropriate boundary terms which render the action differentiable, as well as counterterms which, along with those, give rise to finite and well-defined on-shell actions, when evaluated on stationary points of the functional. In this section, we propose a novel prescription for computing the on-shell action of arbitrary asymptotically AdS solutions of any $D$-dimensional higher-order gravity whose linearized spectrum on a maximally symmetric background matches that of Einstein gravity\footnote{\label{ELC}This property defines the `Einstein-like' class in the classification of \cite{Aspects}, and includes, in particular: Lovelock, QTG, ECG in general $D$ and, more generally, all theories of the Generalized QTG type. Additional examples of theories of this type can be found \eg in \cite{Karasu:2016ifk,Love,Li:2017ncu,Li:2017txk}.}. The procedure represents an important simplification with respect to previous methods, as it only makes use of the usual Gibbons-Hawking-York boundary term and the counterterms of Einstein gravity. As we argue here --- and illustrate throughout the rest of the paper and in appendix \ref{BTcheck} with various non-trivial checks of the proposal --- such contributions can be also used to produce the correct on-shell actions for this class of higher-order theories. Interestingly, for those, the only modification with respect to the Einstein gravity case is that such contributions appear weighted by the Lagrangian of the corresponding theory evaluated on the AdS background, \ie $\mathcal{L}
|_{{\rm AdS}}$. This quantity has been argued to be proportional to the charge $a^*$ appearing in the universal contribution to the entanglement entropy of the dual theory across a $\mathbb{S}^{d-2}$, and our prescription can be used to actually prove such a connection explicitly for this class of theories, as we show below. %In the discussion section, we also speculate on the possible implications of our result in the context of the complexity$=$action paradigm. 
%\comment{Allows for calculations of Quasi-topological on-shell actions for non-planar horizons}

%In holographic calculations, one often needs to compute on-shell actions

% \comment{often needs to compute on-shell actions. Prescription for Einstein gravity includes  }

Let us start considering a general higher-curvature theory of the form
\begin{equation}\label{geng}
I=\int_{\mathcal{M}}d^Dx\sqrt{|g|}\mathcal{L}(g^{ef},R_{abcd})\, ,
\end{equation}
where the Lagrangian density $\mathcal{L}(g^{ef},R_{abcd})$ is assumed to be constructed from arbitrary contractions of the Riemann and metric tensors. The variation of the action with respect to the metric yields
\begin{equation}
\delta I=\int_{\mathcal{M}}d^Dx\sqrt{|g|}\E_{ab}\delta g^{ab}+\epsilon \int_{\partial\mathcal{M}}d^{D-1}x\sqrt{|h|}n_{a}\delta v^{a}\, .
\label{var}
\end{equation}
In this expression we defined
%The equations of motion of the theory take the form $\mathcal{E}_{ab}=0$, where
\begin{equation}
\mathcal{E}_{ab}\equiv P_{a}\,^{c d e}R_{b c d e}-\frac{1}{2}g_{ab}\mathcal{L}-2\nabla^{c}\nabla^{ d}P_{a c d b}\, , %=\frac{1}{2}T_{ab}\, .
\label{fieldequations}
\end{equation}
the equations of motion reading $\mathcal{E}_{ab}=0$, and
\begin{equation}
\delta v^{a}=2g^{dc}P_{e d}^{\ \ ab}\nabla^{e}\delta g_{bc}\, , \quad \text{where}\quad P^{abcd}\equiv \left[\frac{\partial \mathcal{L}}{\partial R_{abcd}}\right]_{g^{ef}}\, .\label{Ptensor}
\end{equation}
In addition, $n^{a}$ is the unit normal to $\partial \mathcal{M}$, normalized as $n^{a}n_{a}\equiv \epsilon=\pm 1$, and  $h_{ab}=g_{ab}-\epsilon n_{a}n_{b}$ is the induced metric. In order to have a well-posed variational problem, the action must be differentiable, in the sense that $\delta I\propto \delta g^{ab}$, so that $\delta I=0$ whenever the field equations --- and the boundary conditions --- are satisfied. %This requirement is essential if we want to define a path integral for quantum gravity, since the action must be stationary for solutions of the classical equations of motion. 
This is not the case of \req{var}, due to the presence of the boundary contribution. In the case of Einstein gravity, $\mathcal{L}^{\rm E}=\left[R+(D-1)(D-2)/L^2\right]/(16\pi G)$, this problem is solved by the addition of the Gibbons-Hawking-York term \cite{York:1972sj,Gibbons:1976ue},
\begin{equation}
I_{\rm GHY}=\frac{\epsilon}{8\pi G}\int_{\partial\mathcal{M}}d^{D-1}x\sqrt{|h|}K\, ,
\end{equation}
where $K=K_{ab}g^{ab}$ is the trace of the second fundamental form of the boundary, $K_{ab}=h_{a}^{\ c} \nabla_{c} n_{b}$. When this term is included, the variation of the action, when we keep $g_{ab}$ fixed at the boundary, reads
\begin{equation}
\delta(I^{\rm E}+I_{\rm GHY})\Big|_{\delta g_{ab}|_{\partial\mathcal{M}}=0}=\frac{1}{16 \pi G}\int_{\mathcal{M}}d^4x\sqrt{|g|}\left[R_{ab}-\frac{1}{2}g_{ab}\mathcal{L}_{\rm EH}\right]\delta g^{ab}\, ,
\end{equation} 
 and so the action is stationary whenever the metric satisfies Einstein's field equations.

For higher-order gravities, the situation is much more involved in general.
 One of the main issues arises from the fact that these theories generally posses fourth-order equations of motion. This implies that the boundary-value problem is not fully specified by the induced metric on $\partial\mathcal{M}$, and one needs to impose additional boundary conditions on derivatives of the metric. Furthermore, even if we know which components of the metric and its derivatives to fix, determining what boundary term needs to be added to yield a differentiable action for such variations is a far from trivial task. Some notable examples for which differentiable actions have been constructed are: quadratic gravities (perturbatively in the couplings) \cite{Smolic:2013gz}, Lovelock gravities \cite{Teitelboim:1987zz,Myers:1987yn}, which are the most general theories with second-order covariantly-conserved field equations \cite{Lovelock1,Lovelock2} (and for which one only needs to fix $g_{ab}$ at the boundary), $f(R)$ \cite{Madsen:1989rz,Dyer:2008hb,Guarnizo:2010xr} and, more generally, $f($Lovelock$)$ gravities \cite{Love}. In these cases, it is also necessary to fix the value of some of the densities on the boundary --- \eg $\delta R\big|_{\partial\mathcal{M}}=0$ for $f(R)$ --- which is related to the fact that these theories propagate additional scalar modes. With the goal of providing a canonical formulation for arbitrary $f$(Riemann$)$ gravities, an interesting proposal for constructing satisfactory boundary terms for such general class of theories was presented in \cite{Deruelle:2009zk} --- see also \cite{Teimouri:2016ulk}.
 %In \cite{Deruelle:2009zk}, an interesting proposal for constructing satisfactory boundary terms for general $f$(Riemann$)$ gravity was presented --- see also \cite{Teimouri:2016ulk}. 
 Unfortunately, the procedure involves the introduction of auxiliary fields and it is quite implicit in general, which seems to limit its practical applicability in the holographic framework.
 % it is not clear how to understand it in a pure metric formalism.

The problem can be simplified if we specify the boundary structure in advance, \eg by restricting the analysis to spacetimes which are maximally symmetric asymptotically. %This approach was succesfully followed, \eg in \cite{} for \comment{bla bla Loevloek...}. 
%he problem of finding the boundary term which yields the action stationary is a difficult task, but we can relax the problem by specifying the boundary conditions on advance, and by finding a boundary term which yields the action stationary for those conditions. In particular, the most natural assumption is that the spacetime is asymptotically maximally symmetric, \textit{i.e.}, Minkowski, de Sitter or anti de Sitter. 
Let us, in particular, assume that the space is asymptotically AdS$_D$, so that the Riemann tensor behaves as $R_{abcd}\rightarrow -\tilde L^{-2} (g_{ac}g_{bd}-g_{ad}g_{bc})$ asymptotically. Then, on general grounds, the tensor $P_{cd}^{\ \  ab}$ appearing in the boundary term in \req{var} takes the simple form 
\begin{equation}\label{pp}
P_{cd}^{\ \  ab}\rightarrow C(\tilde L^2)\delta_{[c}^{\ a}\delta_{d]}^{\ b}+\text{subleading}\, ,
\end{equation}
where $C(\tilde L^2)$ is a constant which depends on the background curvature, and is in general given by\footnote{As shown in \cite{Aspects}, this quantity can be equivalently written as \begin{equation}
	C(\tilde L^2)=\frac{\tilde L^4}{D(D-1)}\frac{d\mathcal{L}
		|_{{\rm AdS}}}{d\tilde L^2}\, ,
	\end{equation}
	the relation between both expressions being nothing but the embedding equation of AdS$_D$ in the corresponding theory --- \eg \req{roo} for ECG.
	 } \cite{Aspects}
\begin{equation}\label{cdd}
C(\tilde L^2)=-\frac{\tilde L^2}{2(D-1)}\mathcal{L}
|_{{\rm AdS}}\, ,
\end{equation}
%\begin{equation}
%C(\tilde L^2)=\frac{\tilde L^4}{D(D-1)}\frac{d\mathcal{L}(\tilde L^2)}{d\tilde L^2}\, ,
%\end{equation}
%\comment{$C(\Lambda)$ es proporcional al coeficiente universal de la EE para una superficie esf'erica en cualquier n'umero de dimensiones (en dimensiones impares, este coincide con la free energy de la teor'ia puesta en una esfera, y en dimensiones pares coincide con el $a$-term de la Weyl anomaly)! Funciona para ECG y tambi'en para GB en dimensiones generales, en ambos casos, $C(\Lambda)$ y $F$ ambos $\propto (1-2(D-2)/(D-4))\lambda_{\ssc \rm GB} f_{\infty}$}
where $\mathcal{L}|_{{\rm AdS}}$ is the Lagrangian of the corresponding theory evaluated on the AdS$_D$ background with curvature scale $\tilde{L}$. 
%In \req{cdd}, the subleading terms will arise from 

For Einstein gravity, we simply have $C^{\rm E}=1/(16\pi G)$ and, in fact, there are no subleading terms in \req{pp} for any spacetime --- simply because $P_{cd}^{\ \  ab}$ only involves products of deltas in that case. Now, asymptotically AdS$_D$ solutions of higher-order gravities will in general produce subleading contributions in \req{pp} as we move away from the asymptotic region. However, the leading term can still be canceled out by adding a generalized GHY term of the form
\begin{equation}\label{GGHY}
I_{\rm GGHY}=2 	C(\tilde L^2)\epsilon \int_{\partial\mathcal{M}}d^{D-1}x\sqrt{|h|}K\, .
\end{equation}
The question is, of course, whether or not the subleading terms for a given theory will give additional non-vanishing contributions asymptotically, forcing us to add extra terms. We expect this to be the case in general. In addition, one generally needs to specify extra boundary conditions, which is related to the metric propagating additional degrees of freedom. However, as we have mentioned, some theories --- see footnote \ref{ELC} ---
do not propagate additional modes on general maximally symmetric backgrounds. For those, the asymptotic dynamics is the same as for Einstein gravity, so it is reasonable to expect the only data that we need to fix on $\partial\mathcal{M}$ to be $g_{ab}$, and also that \req{GGHY} will be enough to make the action stationary for solutions of the field equations. 

% The dynamics of these theories is asymptotically the same as in Einstein gravity, up to rescaling of the gravitational constant, so we can expect that the only boundary data that we need to specify is $g_{\mu\nu}$ on $\partial\mathcal{M}$, and nothing else. In this case, we claim that this boundary term will be enough to make the action stationary for solutions of the equations of motion. \\
In order to obtain finite on-shell actions, one also needs to include counterterms, which only depend on the boundary induced metric. For asymptotically AdS$_D$ spacetimes, there is a generic way of finding them \cite{Emparan:1999pm}. Let us focus on Euclidean signature. In that case, we always have $\epsilon=+1$, and an additional global $(-)$ with respect to Lorentzian signature arises, \eg \cite{Myers:2010tj}, so we have
%
%We will work in Euclidean signature, so $\epsilon=+1$, but there is also a global change of sign in the action.
% Let us first recall the embedding equation for the AdS radius $\tilde L$ \cite{Aspects}:
%\begin{equation}
% \frac{d}{d\tilde L^2}\left(\tilde L^{D}\mathcal{L}(\tilde L^2)\right)=0
% \end{equation}
%Using this, we may rewrite the boundary term \ref{GGHY} as
%\begin{equation}
%I_{\rm GGHY}=\frac{\tilde L^2\mathcal{L}(\tilde L^2)}{(D-1)}\int_{\partial \mathcal{M}}d^{D-1}x\sqrt{h}K\, ,
%\end{equation}
%and the complete Euclidean action reads
\begin{equation}\label{SEcomplete}
I_E=-\int_{\mathcal{M}}d^Dx\sqrt{g}\mathcal{L}(g^{ef},R_{abcd})-2 	C(\tilde L^2) \int_{\partial\mathcal{M}}d^{D-1}x\sqrt{|h|}K+I_{\rm GCT}\, ,
\end{equation}
where we seek to construct the generalized counterterms, $I_{\rm GCT}$. In order to identify all possible divergences, one possibility consists in evaluating the action on pure AdS$_D$ spaces with different boundary geometries \cite{Yale:2011dq}. Observe however that, whenever we evaluate the bulk term on pure AdS$_D$, this will produce an overall constant $\mathcal{L}
|_{{\rm AdS}}$, which is precisely proportional to $C(\tilde L^2)$. This already appears in front of the boundary term, and the result is that the combination of the bulk and boundary contributions reduce to those of Einstein gravity, up to a common overall $C(\tilde L^2)$. Hence, the divergences are exactly the same as for Einstein gravity, and we can use the same counterterms.
% But, anytime we evaluate the action on AdS$_D$ we get 
%\begin{equation}
%I_E=-C(\tilde L^2)\left[-\frac{2(D-1)}{\tilde L^2}\int_{\mathcal{M}}d^4x\sqrt{g}+2\int_{\partial \mathcal{M}}d^{D-1}x\sqrt{h}K\right]+I_{\rm GCT}\, .
%\end{equation}
%Now, the term between parenthesis is the same as in Einstein gravity when the AdS scale is $\tilde L$. 
%
%Therefore, the divergences of this action are the same as those of Einstein gravity times $C(\tilde L^2)$. Then, we can use the same counterterms as for EG, times this constant. 
For example, up to $D=5$ we find \cite{Emparan:1999pm,Yale:2011dq}
\begin{equation}\label{gass}
\begin{aligned}
I_{\rm GCT}&=-2C(\tilde L^2)\int_{\partial \mathcal{M}}d^{D-1}x\sqrt{h}\bigg[-\frac{D-2}{\tilde L}-\frac{\tilde L \Theta[D-4]}{2(D-3)}\mathcal{R}
%&-\frac{\tilde L^3\Theta[D-6]}{2(D-3)^2(D-5)}\left(\mathcal{R}_{ab}\mathcal{R}^{ab}-\frac{D-1}{4(D-2)}\mathcal{R}^2\right)
+\ldots\bigg]\, ,
\end{aligned}
\end{equation}
where $\Theta[x]=1$ if $x\geq 0$, and zero otherwise, and the dots refer to additional counterterms arising for $D\geq 6$. Combining \req{gass} with \req{SEcomplete}, we obtain the final form of the action.

Below, we show that \req{SEcomplete} successfully yields the right answers for ECG in various highly non-trivial situations in which the corresponding on-shell actions can be deduced from alternative considerations --- \eg it correctly computes the free energy of black holes, in agreement with the result obtained using Wald's entropy, as well as the holographic stress tensor two-point charge, $\ctt$, which can be alternatively deduced from the effective Newton constant. Besides, in appendix \ref{BTcheck} we consider arbitrary radial perturbations of AdS$_5$ in Gauss-Bonnet gravity, and show that \req{SEcomplete} produces exactly the same finite and divergent contributions as those obtained using the standard Gauss-Bonnet boundary term and counterterms, \eg \cite{Teitelboim:1987zz,Myers:1987yn,Emparan:1999pm, Mann:1999pc, Balasubramanian:1999re, Brihaye:2008xu,Astefanesei:2008wz}. 
%In the next subsection, we perform another check of \req{SEcomplete}.
%HEREEEE
%\comment{Y el GB de la induced metric? Clarificar qu'e pasa con los CT en general} \comment{counterterm related to conformal gravity}
%In appendix \ref{BTcheck} we compute the complete boundary contribution in \ref{SEcomplete} for Gauss-Bonnet gravity by using this method and we compare it to the actual Gauss-Bonnet boundary term. Of course, both terms coincide (by construction) when evaluated on AdS, so we consider radial perturbations and we find that both boundary terms give the same result for finite and divergent parts, meaning that they are equivalent, at least in the kind of situations considered. 

 %Boundary terms are introduced so that the variational principle is well defined 

\subsection{$a^*$ and generalized action}
Let us momentarily switch to $d\equiv D-1$ notation. As we have seen, both the boundary term and the counterterms appearing in \req{SEcomplete} have the property of being identical to those of Einstein gravity up to an overall constant $C(\tilde{L}^2)$ proportional to the Lagrangian of the corresponding theory evaluated on the AdS background \req{cdd}. Now, an interesting quantity that one would like to compute holographically is the charge $a^*$  appearing in the universal contribution to the entanglement entropy (EE) across a radius-$R$ spherical region $\mathbb{S}^{d-2}$ which, for a general CFT$_d$, is given by \cite{Myers:2010tj,Myers:2010xs,CHM}
\begin{equation}\label{asta}
S_{\rm \ssc EE\, univ.}=\begin{cases}
(-)^{\frac{d-2}{2}} 4a^* \log(R/\delta) \quad &\text{for even } d \, , \\
 (-)^{\frac{d-1}{2}}2\pi a^* \quad &\text{for odd } d\, .
\end{cases}
\end{equation}
$a^*$ coincides with the $a$-type trace-anomaly charge in even dimensional theories. In odd dimensions, $a^*$ is proportional to the free energy, $F=-\log Z$, of the corresponding theory evaluated on $\mathbb{S}^{d}$ \cite{CHM}, namely 
\begin{equation}\label{fffs}
F_{\mathbb{S}^{d}}= (-)^{\frac{d+1}{2}} 2\pi a^* \, , \quad \text{for odd } d\, .
\end{equation}
For even-dimensional holographic theories dual to any higher-order gravity of the form \req{geng} in the bulk, $a^*$ is given by \cite{Imbimbo:1999bj,Schwimmer:2008yh}
\begin{equation}\label{astar}
a^*=-\frac{\pi^{d/2}\tilde{L}^{d+1}}{d \Gamma(d/2)}\mathcal{L}
|_{{\rm AdS}}\, ,
\end{equation}
\ie it is precisely proportional to the charge $C(\tilde L^2)$ defined in \req{cdd}, namely
\begin{equation}
C(\tilde{L}^2)=\frac{a^* }{ \Omega_{(d-1)}\tilde{L}^{d-1}} \, ,
\end{equation}
where $\Omega_{(d-1)}\equiv 2\pi^{d/2}/ \Gamma(d/2)$ is the area of the unit sphere $\mathbb{S}^{d-1}$.
For odd-dimensional theories, it was argued in \cite{Myers:2010tj,Myers:2010xs} that \req{astar} also yields the right $a^*$ for general cubic theories. We can readily extend this result to all theories for which \req{SEcomplete} and \req{gass} hold. From \req{fffs}, it follows that $(-)^{\frac{d+1}{2}} 2\pi a^*$ can be obtained from the on-shell action of pure Euclidean AdS$_{(d+1)}$ with boundary geometry $\mathbb{S}^d$. Since $C(\tilde{L}^2)$ appears as an overall factor in \req{SEcomplete} when evaluated in pure AdS, it follows that $F_{\mathbb{S}^{d}}$ matches the Einstein gravity result up to an overall factor $16 \pi G\cdot  C(\tilde L^2)$. Then, using the result for the free energy in Einstein gravity,
\begin{equation}
F^{\rm E}_{\mathbb{S}^{d}}= (-)^{\frac{d+1}{2}}\frac{\pi^{d/2}\tilde{L}^{d-1} }{4\Gamma(d/2)G}\, ,
\end{equation}
it follows immediately that for any theory of the form \req{geng}, for which our generalized on-shell action can be used,
\begin{equation}
F_{\mathbb{S}^{d}}=16 \pi G\cdot  C(\tilde L^2)F^{\rm E}_{\mathbb{S}^{d}}= (-)^{\frac{d-1}{2}}\frac{2\pi^{d/2+1}\tilde{L}^{d+1}}{d \Gamma(d/2)}\mathcal{L}|_{{\rm AdS}}\, ,
\end{equation}
which takes the expected general form \req{fffs}, with $a^*$ precisely given by \req{astar}. Hence, we have obtained the expected form of the charge $a^*$ from an explicit holographic calculation of the free energy on $\mathbb{S}^{d}$ using our generalized action. The consistency between \req{SEcomplete} and \req{astar} provides support for both expressions.

Reversing the logic, we can rewrite our generalized action in terms of $a^*$, which is way more charismatic than $C(\tilde L^2)$. The result reads
\begin{equation}\label{SEcomplete2}
I_E=-\int_{\mathcal{M}}d^Dx\sqrt{g}\mathcal{L}(g^{ef},R_{abcd})-\frac{2a^*}{\Omega_{(d-1)}\tilde{L}^{d-1}} \int_{\partial\mathcal{M}}d^{D-1}x\sqrt{|h|} \left[ K-\frac{d-1}{\tilde{L}}+\cdots \right]\, ,
\end{equation}
where we have omitted most of the counterterms in \req{gass}. The explicit appearance of $a^*$ in the boundary terms is rather suggestive, and somewhat striking. In section \ref{discu} we comment on the possible implications of \req{SEcomplete2} for holographic complexity.

\subsection{Generalized action for Quasi-topological gravity}
The QTG density in five bulk dimensions is given by \cite{Quasi2,Quasi}
\begin{equation}
\begin{aligned}
\mathcal{Z}_5&= \tensor{R}{_{a}^{c}_{b}^{d}}\tensor{R}{_{c}^{e}_{d}^{f}}\tensor{R}{_{e}^{a}_{f}^{b}}+\frac{1}{56}\Big(-72\tensor{R}{_{abcd}}\tensor{R}{^{abc}_{e}}R^{de}+21\tensor{R}{_{abcd}}\tensor{R}{^{abcd}}R+120\tensor{R}{_{abcd}}\tensor{R}{^{ac}}\tensor{R}{^{bd}}\\
&+144 R^{b}_{a} R_{b}^{c} R_{c}^{a}- 132R_{ab}R^{ab}R+15 R^3\Big)\, .
\end{aligned}
\end{equation}
Just like ECG in $D=4$, the linearized equations of this theory on constant-curvature backgrounds are Einstein-like \cite{Quasi}. Hence, the method developed in the previous subsection should be valid for computing Euclidean on-shell actions of AdS$_5$ solutions of the theory. In this case, the full generalized action \req{SEcomplete2} is given by
\begin{equation}\label{QTBT}
\begin{aligned}
I_E^{\rm QTG}&=-\frac{1}{16\pi G}\int_{\mathcal{M}}d^5x\sqrt{g}\left[\frac{12}{L^2}+R+\frac{L^2\lambda}{2}\mathcal{X}_4+\frac{7\mu L^4}{4}  \mathcal{Z}_5\right]\\
&-\frac{1-6\lambda f_{\infty}+9\mu f_{\infty}^2}{8 \pi G}\int_{\partial \mathcal{M}}d^{4}x\sqrt{h}\bigg[K-\frac{3\sqrt{f_{\infty}}}{L}-\frac{L}{4\sqrt{f_{\infty}}}\mathcal{R}\Bigg]\, ,
\end{aligned}
\end{equation}
where we also included the Gauss-Bonnet density $\mathcal{X}_4=R^2-4\tensor{R}{_{ab}}\tensor{R}{^{ab}}+\tensor{R}{_{abcd}}\tensor{R}{^{abcd}}$. In this case, the charge $a^*$ reads \cite{Myers:2010jv}
\begin{equation}
a^{*{\rm QTG}}=\left(1-6\lambda f_{\infty}+9\mu f_{\infty}^2\right)\frac{\pi \tilde L^{3}}{8G}\, ,
\end{equation}
while $f_{\infty}$ is determined by the equation \cite{Quasi}
\begin{equation}
1-f_{\infty}+\lambda f_{\infty}^2+\mu f_{\infty}^3=0\, .
\end{equation}
A generalized boundary term for QTG was proposed in  \cite{Dehghani:2011hm}. It would be interesting to check whether \req{QTBT} provides the same results as those obtained using such term. As we mentioned above, in appendix \ref{BTcheck} we perform an explicit check of that kind for Gauss-Bonnet gravity.
%It would be interesting to compare the results produced using \req{QTBT} with those obtained from the generalized boundary term constructed in  \cite{Dehghani:2011hm}.
%As we mentioned before, the exact generalization of the Gibbons-Hawking-York term is known for Gauss-Bonnet. In that case, we can compare both prescriptions, which we show to yield identical results for asymptotically AdS spaces in appendix \ref{BTcheck}. As for QTG, a boundary term was constructed in \cite{Dehghani:2011hm}
%On the other hand, no exact boundary term is known for QTG when $\mu\neq 0$, so \req{QTBT} should allow for new calculations.

\subsection{Generalized action for Einsteinian cubic gravity}
Let us now return to ECG. In that case, the full generalized Euclidean action \req{SEcomplete2} becomes
\begin{equation}\label{EuclideanECG}
\begin{aligned}
I_E^{\rm ECG}=&-\frac{1}{16\pi G}\int d^4x \sqrt{|g|}\left[\frac{6}{L^2}+R-\frac{\mu L^4}{8} \mathcal{P} \right]\\
&-\frac{1+3\mu f_{\infty}^2}{8 \pi G}\int_{\partial \mathcal{M}}d^3x\sqrt{h}\left[K-\frac{2\sqrt{f_{\infty}}}{L}-\frac{L}{2\sqrt{f_{\infty}}}\mathcal{R}\right]\, ,
\end{aligned}
\end{equation}
where recall that $f_{\infty}$ can be obtained as a function of $\mu$ from \req{roo}. Observe also that the charge $a^*$ reads in this case
\begin{equation}\label{aae}
a^{* \rm ECG}=(1+3\mu f_{\infty}^2)\frac{\tilde L^2}{4 G}\, .
\end{equation}
We use \req{EuclideanECG} in several occasions in the remainder of the paper, finding exact agreement with the expected results in all cases for which alternative methods can be used.

\section{Stress tensor two-point function charge $\ctt$}\label{tt}
In order to characterize the holographic dual of ECG, we must translate the two available dimensionless parameters in \req{ECG}, namely: $L^2/G$ and $\mu$, into universal defining quantities of the boundary theory. Since we are only considering the gravitational sector of the bulk theory, the most relevant `charges' to be identified in the CFT are those characterizing the boundary stress tensor. Conformal symmetry highly constrains the structure of stress-tensor two- and three-point functions \cite{Osborn:1993cr}. We will deal with the three-point function charges in section \ref{t44}. Let us start here with the stress-tensor correlator which, for an arbitrary CFT$_3$, is given by \cite{Osborn:1993cr}
\begin{equation}\label{2pf}
\braket{ T_{\mu\nu}(x)T_{\rho\sigma}(x')} =\frac{\ctt}{|x-x'|^6}\mathcal{I}_{\mu\nu,\rho\sigma}(x-x')\, ,
\end{equation}
where
\begin{equation}
\mathcal{I}_{\mu\nu,\rho\sigma}(x)\equiv\frac{1}{2}\left(I_{\mu\rho}(x)I_{\nu\sigma}(x)+I_{\mu\sigma}(x)I_{\nu\rho}(x)\right)-\frac{1}{4}\delta_{\mu\nu}\delta_{\rho\sigma}\, ,\quad \text{and} \quad I_{\mu\nu}(x)\equiv\delta_{\mu\nu}-2\frac{x_{\mu}x_{\nu}}{x^2}\ ,
\end{equation}
are fixed tensorial structures. This correlator is then fully characterized by a single theory-dependent parameter, customarily denoted $\ctt$. This quantity, which in even dimensions is proportional to the trace anomaly charge $c$, also plays a relevant role in three-dimensional CFTs --- see \eg \cite{Huh:2014eea,Diab:2016spb,Giombi:2016fct} for recent studies. As opposed to the $d=2$ case \cite{Zamolodchikov:1986gt}, $\ctt$ is not monotonous under general RG flows in three dimensional CFTs \cite{Nishioka:2013gza}. However, it universally shows up in various contexts, including relevant quantities in entanglement and R\'enyi entropies \cite{HoloRen,Hung:2014npa,Perlmutter:2013gua,Mezei:2014zla,Bueno1};  quantum critical transport --- see \eg \cite{Witczak-Krempa:2015pia,Lucas:2017dqa} and references therein; or partition functions on deformed curved manifolds \cite{Closset:2012ru,Bobev:2017asb,Fischetti:2017sut}.

In AdS/CFT, the dual of $T_{\mu\nu}(x)$ is the normalizable mode of the metric \cite{Witten,Gubser}. Hence, evaluating \req{2pf} entails determining the two-point  boundary  correlator  of  gravitons  in  the corresponding AdS  vacuum. For Einstein gravity in $d=3$, the result \cite{Buchel:2009sk,Liu:1998bu} reads
\begin{equation}
\ctte=\frac{3}{\pi^3}\frac{\tilde{L}^2}{G}\, .
\end{equation}
Naturally, the introduction of higher curvature terms in the bulk modifies this result, \eg \cite{Buchel:2009sk,Myers:2010jv,Bueno2}. In general, higher order gravities give rise to equations of motion involving more than two derivatives of the metric. In those cases, the metric generically contains additional degrees of freedom besides the usual massless graviton. From the holographic perspective, this means that the metric couples to additional operators which are typically nonunitary\footnote{See \eg \cite{Myers:2010tj,Bueno2} for more detailed discussions of this issue.}. This is not always the case, however. In fact, there exist families of higher order gravities whose linearized equations around maximally symmetric backgrounds are identical to those of Einstein gravity, up to a normalization of the Newton constant --- see footnote \ref{ELC} and  \eg \cite{Aspects} for details. For those, the only mode is the usual spin-2 graviton, the metric only couples to the stress tensor, and $\ctt$ can be straightforwardly extracted from the effective Newton constant. This generically reads $G_{\rm eff}=G/\alpha$, where $\alpha$ is a constant which depends on the new couplings. The appearance of $\alpha$ can be alternatively understood as changing the normalization of the graviton kinetic term which, holographically, gets translated into a modification of the stress-tensor correlator charge, which then becomes $\alpha \cdot \ctte$.

For ECG, using \req{Geff}, we find then
\begin{equation}\label{cttecg}
\ctt^{\rm ECG}= (1-3\mu f_{\infty}^2)\frac{3}{\pi^3}\frac{\tilde{L}^2}{G}\, .
\end{equation}
Observe that unitarity imposes $\ctt$ to be positive, which translates into $1-3\mu f_{\infty}^2 >0$. This is of course equivalent to asking the effective bulk gravitational constant to be positive.
It can be seen that this constraint is automatically satisfied whenever \req{sis} holds.

While we have been able to compute $\ctt$ for ECG using $G_{\rm eff}^{\rm ECG}$, it is instructive to obtain it from an explicit holographic calculation. This will also serve as a highly-nontrivial consistency check for the new on-shell action method introduced in the previous section.

Let us then consider a metric perturbation: $g_{ab}=\bar g_{ab}+h_{ab}$, on the Euclidean AdS$_4$ vacuum
\begin{equation}\label{EAdS}
ds^2=\frac{r^2}{L^2}\left[d\tau^2+dx^2+dy^2\right]+\frac{L^2}{r^2f_{\infty}}dr^2\, .
\end{equation}
Since all components of the two-point function will be determined by $\ctt$, computing one of them will be enough. It is then sufficient to consider a perturbation of the form $h_{xy}=\frac{r^2}{L^2}\phi(r,\tau)$. Plugging this into the Euclidean version of \req{ECG} and expanding up to quadratic order in $\phi$, we find
\begin{equation}\label{ttw}
I^{\rm ECG}_{E\, \rm \small Bulk}=\frac{(1-3\mu f_{\infty}^2)}{32\pi G}\int d^3xdr\left[\frac{1}{\sqrt{f_{\infty}}}(\partial_{\tau}\phi)^2+\sqrt{f_{\infty}}\frac{r^4}{L^4}(\partial_r\phi)^2\right]-\frac{1}{16\pi G}\int d^3x\, \Gamma_r\Big|_{r=r_{\infty}}\, ,
\end{equation}
where $\Gamma_r$ is a boundary term which appears after integration by parts --- see \req{gaga}. Recall also that, in this coordinates, the boundary corresponds to $\lim_{r\rightarrow \infty} r \equiv L^2/\delta$, where we introduce the UV cutoff $\delta\ll 1$. 
% Before considering the boundary terms in \req{EuclideanECG}, let us consider
%In addition to this, we have to take into account the generalized Gibbons-Hawking term as well as the counterterms, which are shown in \ref{EuclideanECG}. For now, let us consider only the bulk action. From it we can read 
The equation of motion for $\phi$ follows from \req{ttw}, and reads
\begin{equation}
\frac{\partial}{\partial r}\left(\frac{r^4}{L^4}\frac{\partial \phi}{\partial r}\right)+\frac{1}{f_{\infty}}\frac{\partial ^2\phi}{\partial \tau^2}=0\, .
\end{equation}
In order to solve it, we Fourier-transform the dependence on the coordinate $\tau$,
\begin{equation}
\phi(r,\tau)=\frac{1}{2\pi}\int dp \phi_0(p)e^{ip\tau}H_p(r)\, .
\end{equation}
$H_p$ satisfies the equation
\begin{equation}
\frac{d}{d r}\left(\frac{r^4}{L^4}\frac{d H_p}{d r}\right)-\frac{p^2}{f_{\infty}}H_p=0\, ,
\end{equation}
whose general solution reads
\begin{equation}
H_p(r)=c_1 e^{-\frac{L^2 |p|}{\sqrt{f_{\infty}}r}}\left(1+\frac{L^2 |p|}{\sqrt{f_{\infty}}r}\right)+c_2 e^{\frac{L^2 |p|}{\sqrt{f_{\infty}}r}}\left(1-\frac{L^2 |p|}{\sqrt{f_{\infty}}r}\right)\, .
\end{equation}
In order to get a regular solution, we set $c_2=0$, and we also fix $c_1=1$ so that $H_p(r \rightarrow L^2/\delta)=1$. With this solution, we evaluate the Lagrangian, which can be expressed as a total derivative. Further integrating over the $r$ coordinate and substituting the solution in Fourier space, we get
%, so the on-shell action reads
%\begin{equation}
%I_{\rm ECG}^{\rm \small Bulk}=\frac{(1-3\mu f_{\infty}^2)\sqrt{f_{\infty}}}{32\pi G}\int d^3xdr\partial_r\left(\frac{r^4}{L^4}\phi\partial_r\phi\right)-\frac{1}{16\pi G}\int d^3x %\Gamma_r\Big|_{r=r_0}\, .
%\end{equation}
%Then, we integrate over the $r$ coordinate and we substitute the solution in Fourier space. This way we get
\begin{equation}\label{kd}
I^{\rm ECG}_{E\, \rm \small Bulk}=\frac{\sqrt{f_{\infty}}V_{\mathbb{R}^2}}{64\pi^2 G_{\rm eff}^{\rm ECG} }\int dpdq\phi_0(p)\phi_0(q)\delta(p+q)\frac{L^4}{\delta^4}H_p\partial_r H_p\Big|_{r=L^2/\delta}-\frac{1}{16\pi G}\int d^3x \Gamma_r\Big|_{r =L^2/\delta}\, ,
\end{equation}
where $V_{\mathbb{R}^2}=\int dxdy$, and where we used $\int d\tau e^{i(p+\tau)}=2\pi \delta(p+q)$.

Let us now turn to the boundary contributions in the generalized action \req{EuclideanECG}. As we explain in appendix \ref{2pbdy}, when these terms are added to \req{kd}, most divergences in $\Gamma_r\Big|_{r=L^2/\delta}$ disappear, and we are left with the following result for the full action:
%HEREEEE
 %Now we must consider the boundary contribution in \req{EuclideanECG}. When we include these terms we can see that they cancel most of the divergences which appear in $\Gamma_r$ \comment{Appendix with this} and at the end only one term survives. This way, the complete action $I_{\rm ECG}=I_{\rm ECG}^{\rm \small Bulk}+I_{\rm ECG}^{\rm \small Bdry}$ reads
\begin{align}\label{ICT}
I^{\rm ECG}_E&=I^{\rm ECG}_{E\, \rm \small Bulk}+I^{\rm ECG}_{E\, \rm \small GGHY+GCT}\\ \notag&=\frac{V_{\mathbb{R}^2}}{64\pi^2 G_{\rm eff}^{\rm ECG} \sqrt{f_{\infty}}}\int dpdq\phi_0(p)\phi_0(q)\delta(p+q)\left[f_{\infty}\frac{L^4}{\delta^4}H_p\partial_r H_p\Big|_{r=L^2/\delta }-\frac{L^2 p^2}{\delta} H_p^2\right]\, .
\end{align}
Observe that, even though $1/G_{\rm eff}^{\rm ECG}$ and $a^{* \rm ECG}$ have  a different dependence on $\mu$ --- see \req{Geff} and \req{aae} respectively --- and that it is $a^{* \rm ECG}$ the one  appearing as an overall constant in the generalized GHY term and the counterterms \req{EuclideanECG}, everything conspires to produce a single finite contribution which is instead proportional to $1/G_{\rm eff}^{\rm ECG}$, as it must.

If we take the limit $\delta \rightarrow 0$ explicitly in \req{ICT}, we get the simple result
\begin{equation}
I^{\rm ECG}_E[\phi_0]=-\frac{V_{\mathbb{R}^2}\tilde{L}^2}{64\pi^2 G_{\rm eff} }\int dpdq\phi_0(p)\phi_0(q)\delta(p+q)|p|^3\, .
\end{equation}
Using the holographic dictionary \cite{Witten}, we can compute one of the components of the boundary stress tensor two-point function in momentum space as
%
%Then we have expressed the on-shell action in terms of the boundary perturbations. Hence, according to the holographic dictionary we can compute the following component of the two-point function in momentum space:
\begin{equation}\label{TTECG}
\langle T_{xy}(0,0,p)T_{xy}(0,0,q)\rangle=-(2\pi)^2\frac{\delta^2 I^{\rm ECG}_E[\phi_0]}{\delta \phi_0(-p)\delta \phi_0(-q)}=\frac{\tilde{L}^2V_{\mathbb{R}^2}}{8 G_{\rm eff} } \delta(p+q)|p|^3\, .
\end{equation}
Now, from the CFT side, this is given by
\begin{equation}\label{df}
\langle T_{xy}(0,0,p)T_{xy}(0,0,q)\rangle=\int d^3x\int d^3x'e^{-i p\tau}e^{-iq\tau'}\langle T_{xy}(x)T_{xy}(x')\rangle \, ,
\end{equation}
where
\begin{equation}
\langle T_{xy}(x)T_{xy}(x')\rangle=\frac{\ctt}{2|x-x'|^6}\left[-1+2\frac{(\tau-\tau')^2}{|x-x'|^2}+8\frac{(x-x')^2(y-y')^2}{|x-x'|^4}\right]\, .
\end{equation}
The integration in \req{df} can be performed without further complications and we obtain the result
\begin{equation}
\langle T_{xy}(0,0,p)T_{xy}(0,0,q)\rangle=\frac{\pi^3 \ctt V_{\mathbb{R}^2}}{24}\delta(p+q)|p|^3\, .
\end{equation}
Comparing this expression with \req{TTECG}, we obtain the result for $\ctt$, which agrees with the one in \req{cttecg}, as it should. The fact that our generalized action \req{EuclideanECG} succeeds in providing the right answer for this quantity, including various non-trivial cancellations  between $I^{\rm ECG}_{E\, \rm \small Bulk}$ and $I^{\rm ECG}_{E\, \rm \small GGHY+GCT}$ --- see appendix \ref{2pbdy} --- provides strong evidence for the validity of the method developed in section \ref{osa}. 
%Therefore, comparing with \ref{TTECG} we get the central charge $\ctt$:
%\begin{equation}
%\ctt=\frac{3(1-3\mu f_{\infty}^2)L^2}{\pi^3 G f_{\infty}}\, .
%\end{equation}

Note finally that, as explained at the beginning of this section, $\ctt$ provides information about many different physical quantities appearing in numerous contexts. Hence, by the same price we computed \req{cttecg}, we gain access to all such quantities for the CFT$_3$ dual to ECG.

\section{Thermodynamics}\label{therr}
In this section we study the thermodynamic properties of the ECG black holes constructed in section \ref{BHs}. First, we compute the Wald entropy, ADM energy and free energy of the solutions, and compare the result with the one obtained from an explicit on-shell action calculation, which serves as a further check of the method proposed in section \ref{osa}. Then, focusing on the flat boundary case, $k=0$, we identify the quantity $\cs$ which relates the thermal entropy density to the temperature, and show that, in contradistinction to Einstein gravity, it defines an independent charge with respect to $\ct$. In subsections \ref{TT2} and \ref{SS2}, we study the phase space of holographic ECG on $\mathbb{S}^1_{\beta}\times \mathbb{T}^2$ and $\mathbb{S}^1_{\beta}\times \mathbb{S}^2$, respectively. In the first case, we show that the standard phase transition between the ECG AdS soliton and black brane keeps occurring at the same temperature as for Einstein gravity. In the second, we show that depending on the value of $\mu$, one, two or three black hole solutions can coexist at the same temperature. The dominating phases are still thermal AdS at small temperatures and large black holes at large temperatures, but the Hawking-Page-transition temperature becomes arbitrarily large as we approach the critical limit $\mu=4/27$. Besides, small black holes become thermodynamically stable for $\mu\neq0$, although their contribution to the partition function is always subleading with respect to thermal AdS.

\subsection{Entropy, energy and free energy}
Let us start by computing the Wald entropy of the solutions which, for any covariant theory of gravity is given by \cite{Wald:1993nt,Iyer:1994ys}
\begin{equation}
S=-2\pi \int_{\rm \ssc H}  d^{d-1}x\, \sqrt{h} \frac{\partial \mathcal{L}}{\partial R^{ab}\,_{cd}}\varepsilon^{ab}\varepsilon_{cd} \, ,
\end{equation}
where $\varepsilon_{ab}$ is the binormal to the horizon. Now, for metrics of the form \req{bhss}, the integration can be performed straightforwardly, yielding
\begin{equation}
S=-\frac{2\pi \rh^2}{L^2} V_{\Sigma} \left.\frac{\partial \mathcal{L}^{\rm ECG}}{\partial R^{ab}\,_{cd}}\varepsilon^{ab}\varepsilon_{cd} \right|_{r=\rh}\, ,
\end{equation}
where $V_{\Sigma}$ is the regularized volume of $\mathbb{S}^2$, $\mathbb{R}^2$ or $\mathbb{H}^2$ for $k=1,0,-1$ respectively. Explicitly, the final result for the ECG black holes reads
\begin{equation}\label{sth}
S^{\rm ECG}=\frac{\rh^2  V_{\Sigma}}{4G L^2}  \left[1-\frac{3 \mu L^4  \left(k+\frac{3\rh^2}{L^2}\right)\left[ \left(k+\frac{3\rh^2}{L^2}\right)+2k \left[1+\sqrt{1+\frac{3k L^4\mu}{\rh^4}\left(k +3\frac{\rh^2}{L^2}\right)} \right] \right]}{\rh^4 \left[1+\sqrt{1+\frac{3k L^4\mu}{\rh^4}\left(k +3\frac{\rh^2}{L^2}\right)} \right]^2}\right]   \, .
\end{equation}
Again, this reduces to the Einstein gravity result
\begin{equation}\label{E}
S^{\rm E}=\frac{\rh^2 V_{\Sigma} }{4GL^2}    \, ,
\end{equation}
when we set $\mu=0$.
Once we have $S(T)$ (defined implicitly), we can use the first law, $dE=TdS$, to find the energy. The result is\begin{equation}\label{adm}
E^{\rm ECG}=\frac{{(\omega^{\rm ECG})}^3 V_{\Sigma} N}{8\pi G L^4}\, .
\end{equation}
As expected, this coincides with the result one would obtain for the generalized ADM energy from the asymptotic expansion \req{Asympt}.

The entropy of the solutions can be alternatively computed from the free energy as $S=-\partial F/\partial T$. Hence, we can perform an additional check of our generalized action \req{EuclideanECG}, which evaluated on the Euclidean version of the solutions --- for which we identify $t_E\sim t_E+\beta$ --- should yield the free energy as  $F^{\rm ECG}= I^{\rm ECG}_{E}/\beta$.
%Even though we have completely determined the thermodynamic properties of these black holes, it is important for the applications that we are going to consider, to motivate the thermodynamics from the point of view of Euclidean path integral. For that, we first need to define a well-posed action, in the sense that it is stationary for solutions of the equations of motion. In Appendix A we propose a way to achieve this in ECG by adding a generalized York-Gibbons-Hawking term. In Euclidean signature, we propose the following action functional
%\begin{equation}\label{EECG1}
%\begin{aligned}
%I^{\rm ECG}_{E}=&-\frac{1}{16\pi G}\int d^4x \sqrt{|g|}\left[\frac{6}{L^2}+R-\frac{\mu L^4}{8} \mathcal{P} \right]\\
%&-\frac{1+3\mu f_{\infty}^2}{8 \pi G}\int_{\partial \mathcal{M}}d^3x\sqrt{h}\left(K-\frac{2\sqrt{f_{\infty}}}{L}-\frac{L}{2\sqrt{f_{\infty}}}\mathcal{R}\right)\, ,
%\end{aligned}
%\end{equation}
%The free-energy of the solution should be obtained from the Euclidean on-shell action as $F= I^{\rm ECG}_{E}/\beta$. 
Plugging \req{bhss} in \req{EuclideanECG}, we find that the bulk term is a total derivative that can be integrated straightforwardly, namely
\begin{equation}
I^{\rm ECG}_{E\, {\rm Bulk}}=\frac{\beta N V_{\Sigma}}{16\pi G L^2 }\left[H(\rh)-H(L^2/\delta) \right]\, ,
\end{equation}
where
\begin{equation}
H(r)\equiv \frac{r^3}{L^2}\left[(2-4 f-r f')-\frac{\mu }{4}\left(2f+r f'\right)^2 \left(4f-r f'\right)\right]\, . 
\end{equation}
Using the asymptotic expansion \req{Asympt}, we get
\begin{equation}
H(L^2/ \delta)=\frac{2L^4}{\delta^3}(1-2f_{\infty}-2\mu f_{\infty}^3)+\frac{{(\omega^{\rm ECG})}^3}{L^2}\frac{(1+3\mu f_{\infty}^2)}{(1-3\mu f_{\infty}^2)}+\mathcal{O}(\delta)\, . 
\end{equation}
We can also evaluate the boundary contributions in \req{EuclideanECG}. For these, we use 
%the following results for the induced metric determinant, trace of 
%On the other hand we can also determine the boundary contribution by using again the asymptotic expansion. The extrinsic curvature and the the Ricci scalar of the boundary $r=L^2/\delta$ read
\begin{equation}
\begin{aligned}
d^3x\sqrt{h}&=N dt\wedge d\Sigma_{k} \left(\frac{\sqrt{f_{\infty}}L^3}{\delta^3}+\frac{k L}{2\delta\sqrt{f_{\infty}}}-\frac{{(\omega^{\rm ECG})}^3}{2\sqrt{f_{\infty}}L^3(1-3\mu f_{\infty}^2)}\right)+\mathcal{O}(\delta)\, ,\\
K&=\frac{3\sqrt{f_{\infty}}}{L}+\frac{k \delta^2}{2L^3\sqrt{f_{\infty}}}+\mathcal{O}(\delta^{4})\, ,\quad \mathcal{R}=\frac{2k\delta^2}{L^4}\, .
\end{aligned}
\end{equation}
Then, we find
\begin{equation}
I^{\rm ECG}_{E\, \rm \small GGHY+GCT}=-\frac{\beta N V_{\Sigma}(1+3\mu f_{\infty}^2)}{8\pi G L^4 }\left[\frac{L^6 f_{\infty}}{\delta^3} -\frac{{(\omega^{\rm ECG})}^3}{2(1-3\mu f_{\infty}^2)}\right]+\mathcal{O}(\delta)\, .
\end{equation}
Now, if we add up both contributions we obtain the finite result
\begin{equation}
I^{\rm ECG}_{E}=\frac{\beta N V_{\Sigma}}{16\pi G L^2 }H(\rh)\, ,
\end{equation}
where we made use of the AdS$_4$ embedding equation \req{roo}. Hence, all boundary contributions cancel out and the on-shell action is reduced to the evaluation of the function $H(r)$ at the horizon. Using the near-horizon expansion \req{nH}, we can finally write the free energy as
%HEEEEEREEEEE
%Therefore, after using the near-horizon expansion \req{nH} we get the free energy $F= I_{\rm ECG}^{E}/\beta$, which can be written as
\begin{equation}\label{fecg}
F^{\rm ECG}=\frac{N V_{\Sigma}}{8\pi G L^2 }\left[k\rh +\frac{\rh^3}{L^2}-\frac{2\pi T\rh^2}{N}+\mu L^4 \left(\frac{3k}{\rh}\left(\frac{2\pi T}{N}\right)^2+\left(\frac{2\pi T}{N}\right)^3\right)\right]\, .
\end{equation}
Note that this can be also written fully in terms of $\rh$ using \req{T}.
%By using \req{T} we can also write explicitly the free energy as a function of the radius $\rh$.
%\begin{equation}
%H(\rh)=2\rh^3 \left[ 1+\frac{L^2}{\rh^2}\left(k-P_k(\mu,r)\right)+\frac{\mu L^6}{\rh^6}\left(3k+P_k(\mu,r)\right) P_k(\mu,r)^2 \right]\, ,
%\end{equation}
%where
%\begin{equation}
%P_k(\mu,r)=\left(k+\frac{3\rh^2}{L^2} \right) \left[1+\sqrt{1+\frac{3k L^4\mu}{\rh^4}\left(k +3\frac{\rh^2}{L^2}\right)} \right]^{-1}\,.
%\end{equation}
%The final result for the free energy reads
%\begin{equation}\label{fecg}
%F_{\rm ECG}=\frac{N V_{\Sigma}}{16\pi G L^2 }H(\rh)\, .
%\end{equation}
When $\mu=0$, \req{fecg} reduces to the Einstein gravity result
\begin{equation}
F^{\rm E}=\frac{N V_{\Sigma} \rh}{16\pi G L^2 }\left(k-\frac{\rh^2}{L^2}\right)\, .
\end{equation}
Using \req{fecg} and the thermodynamic identity $S=-\partial F/\partial T$, we can recompute the entropy of the solutions. The result precisely matches \req{sth}, computed using Wald's formula, which provides another check for our generalized action.

\subsection{Thermal entropy charge $\cs$}\label{cssex}
%The planar case corresponds to $k=0$ and, for that, it is customary to make the redefinition
%\begin{equation}
%V_{k=0}(r)=\frac{r^2}{L^2}f(r)\, , 
%\end{equation}
When the boundary geometry is flat, $k=0$, it is convenient to set $N^2=1/f_{\infty}$, a choice which fixes the speed of light to one in the dual CFT \cite{Buchel:2009sk}. In that case, the thermodynamic expressions simplify considerably. In particular, we find
\begin{eqnarray}\label{Trh}
T=\frac{3 \rh}{4\pi L^2\sqrt{f_{\infty}}}\, , \quad {\omega}^3= \rh^3 \left(1-\frac{27}{4}\mu \right)\, , \\  \label{srh}
s=\frac{ \rh^2}{4GL^2}\left(1-\frac{27}{4}\mu\right)\, ,\quad \varepsilon=\frac{ \rh^3}{8\pi G L^4\sqrt{f_{\infty}}}\left(1-\frac{27}{4}\mu \right)\, ,
\end{eqnarray}
where we defined the entropy and energy densities $s\equiv S/V_{\mathbb{R}^2}$, $\varepsilon \equiv E/V_{\mathbb{R}^2}$. %Assuming $\omega^3>0$, the second equation also imposes the constraint \ref{sis}. 
%\comment{comparison with GB Quasitopo on the depend of the grav coupling}
We can explicitly write these quantities in terms of the temperature, the result being
%\comment{Is it possible to have black holes with $\omega^3<0$. I think the condition at asymptotic infinity reads $\omega^3\mu<0$, so if it is the case, one could also have black holes with positive values of $\mu$}
%It is convenient to define the entropy and energy densities as $s=S/V_2$ and $\rho=E/V_2$, respectively, where $V_2=\int dx_1dy_1$. From the general expressions \req{sth} and \req{E} we find the following result
\begin{equation}\label{entropy}
s=\frac{4\pi^2 \tilde{L}^2f_{\infty}^2}{9G}\left(1-\frac{27}{4}\mu\right) T^2\, ,\quad \varepsilon=\frac{8\pi^2 \tilde{L}^2f_{\infty}^2}{27G}\left(1-\frac{27}{4}\mu\right) T^3.
\end{equation}
%\begin{equation}
%s=\frac{16\pi^2 L^2 f_{\infty}}{9G}\left(1-\frac{27}{4}\mu \right)T^2
%\end{equation}
From \req{entropy}, it immediately follows that ECG black branes satisfy
%Note that the following relation holds,
\begin{equation}
\varepsilon=\frac{2}{3}T s\, ,
\end{equation}
as expected for a thermal plasma in a general three-dimensional CFT.

The dependence on the temperature of the thermal entropy density is also fixed for any CFT$_3$ to take the form
\begin{equation}
s=\cs T^2\, , 
\end{equation}
where $\cs$ is a theory-dependent quantity. From, \req{entropy}, it follows that
\begin{equation}\label{csss}
\cs^{\rm ECG}=\left(1-\frac{27}{4}\mu \right)f_{\infty}^2\, \cse\, ,\quad \text{where} \quad \cse=\frac{4\pi^2}{9}\frac{\tilde{L}^2}{G}\, ,
\end{equation}
is the Einstein gravity result --- see \eg \cite{Buchel:2009sk}. As we can see, in the holographic model defined by ECG, $\cs$ is no longer proportional to $\ctt$, and therefore defines an additional well-defined  independent `charge' which characterizes the theory\footnote{Observe that $\cs$ can be rewritten as $\cs^{\rm ECG}=f_{\infty}^2(1-3\mu f_{\infty}^2/4)(1-3\mu f_{\infty}^2)^2\cse$, %$=\textcolor{red}{\frac{(f_{\infty}+3)(3-2 f_{\infty})}{27f_{\infty}^2}\ct^{\rm ECG}}$ 
which makes it more obvious that this charge is not proportional to $\ct^{\rm ECG}$.}. For growing values of $\mu$, $\cs$ monotonously decreases with respect to the Einstein gravity value and, funnily, it vanishes for the critical case\footnote{This would seem to suggest that the black brane has a unique microstate in that case, but it is probably just another evidence of the problematic properties of the critical theory.} , $\mu=4/27$. 

The fact that $\cs$ vanishes for certain value of the gravitational coupling is quite unusual, and does not occur for QTG or Lovelock black holes (in the Einstein gravity branch) in any number of dimensions --- see \eg \cite{Buchel:2009sk,Quasi,Myers:2010jv,Dehghani:2009zzb,Camanho:2011rj}. In fact, in those cases, the only modification in $\cs$ with respect to Einstein gravity is an overall $f_{\infty}^{(d-1)}$ factor, \ie the result reads  $\cs^{\rm QTG/Lovelock}=f_{\infty}^{(d-1)} \cse$, where $\cse$ is the Einstein gravity result written in terms of $\tilde{L}$. In fact, in view of the results for those theories, one would have naively expected all `$(1-27/4\mu)$' factors in \req{Trh}-\req{csss} not to appear for ECG. This seems to be a simple manifestation of the fact that the theories belonging to the Generalized QTG class (including ECG) for which $f(r)$ is determined through a second-order differential equation possess rather different properties from those for which $f(r)$ is determined from an algebraic equation --- see below and \cite{Hennigar:2017ego,PabloPablo3,Ahmed:2017jod,Hennigar:2017umz} for more evidence in this direction.

%\subsection{Disk entanglement entropy universal term $a^*$}	
%\begin{equation}
%s_{\rm EE}=a_1 \frac{R}{\delta}-2\pi a^*\, , 
%\end{equation}
%where $a^*$ is a universal contribution given by
%\begin{equation}\label{aae}
%a^*=\left(4-\frac{3}{f_{\infty}}\right)a^*_{{\rm \ssc E}}\, , \quad \text{where} \quad a^*_{{\rm \ssc E}}=\frac{1 }{4}\frac{\tilde{L}^2}{G}
%\end{equation}	

\subsection{Toroidal boundary: black brane vs AdS$_4$ soliton}\label{TT2}
In this subsection we study the phase space of thermal configurations when the spatial dimensions of the boundary CFT form a torus $\mathbb{T}^2$. The first obvious saddle corresponds to Euclidean AdS$_4$ with toroidal boundary conditions, given by
\begin{equation}\label{TAdS}
ds^2=\frac{r^2}{ L^2}\left[d\tau ^2+dx_1^2+dx_2^2\right]+\frac{L^2}{r^2f_{\infty}}dr^2\, ,
\end{equation}
where the coordinates $x_1$ and $x_2$ are assumed to be periodic, $x_{1,2}\sim x_{1,2}+l_{1,2}$, where $l_{1,2}$ is the period of each coordinate. Without loss of generality we assume $l_1\le l_2$. As before, $\tau \sim \tau+\beta$. The next candidate is the Euclidean black brane
\begin{equation}\label{BB}
ds^2=\frac{r^2}{L^2}\left[\frac{f(r)}{f_{\infty}}d\tau^2+dx_1^2+dx_2^2\right]+\frac{L^2}{r^2f(r)}dr^2\, ,
\end{equation}
for which the temperature is fixed in terms of the horizon radius through \req{Trh}. Finally, it should be evident that moving the $f(r)/f_{\infty}$ factor from $g_{\tau \tau}$ to $g_{11}$ or $g_{22}$ should also give rise to solutions of ECG, \eg
\begin{equation}\label{soliton}
ds^2=\frac{r^2}{L^2}\left[d\tau^2+\frac{f(r)}{f_{\infty}}dx_1^2+dx_2^2\right]+\frac{L^2}{r^2f(r)}dr^2\, .
\end{equation}
These are the so called AdS$_4$ `solitons' \cite{Witten:1998zw,Horowitz:1998ha}. The crucial difference with respect to the black brane is that, for these, regularity no longer imposes a relation between the temperature and the horizon radius. Instead, it fixes the periodicity of $x_1$ (or $x_2$ if $f(r)/f_{\infty}$ appears in $g_{22}$ instead) in terms of $\rh$ as
\begin{equation}
l_{1,2}=\frac{4\pi L^2\sqrt{f_{\infty}}}{3 \rh}\, .
\end{equation}
Of course, $\tau$ is still periodic with period $\beta$, but, as opposed to the black-brane case, the temperature can be now arbitrary for a given value of $\rh$.

Now, the Euclidean action vanishes for pure Euclidean AdS$_4$, whereas for the black brane and the solitons we find, respectively 
\begin{eqnarray}\label{bbe}
I_E^{\rm bb}=-\frac{4\pi f_{\infty}L^2}{27 G}\left(1-\frac{27}{4}\mu\right)T^2 l_1l_2\, ,\quad 
I_E^{\rm soliton\, 1,2}=-\frac{4\pi f_{\infty}L^2}{27 G}\left(1-\frac{27}{4}\mu\right)\frac{l_1l_2}{Tl_{1,2}^3}\, .
\end{eqnarray}
The solution which dominates the partition function is the one with the smaller on-shell action (or free energy, $ \beta F\equiv I_E$). As we can see from \req{bbe}, for the set of values of $\mu$ for which the ECG solutions exist, the free energies of the black brane and the AdS solitons are always negative, just like for Einstein gravity, which implies that pure AdS$_4$ never dominates. 
We observe that for (arbitrarily) small temperatures, the partition function is dominated by the soliton with the shortest periodicity, the other one being always subleading. For large temperatures, the black brane dominates instead. At $T=1/l_1$, (recall we are assuming $l_1<l_2$), there is a first-order phase transition which connects both phases.
Hence, the phase-transition temperature is not modified with respect to Einstein gravity. The latent heat, computed as the difference between the energy densities of both configurations at $T=1/l_1$, does change and is given by
\begin{equation}
\delta Q=\frac{4\pi f_{\infty}L^2}{9 G}\left(1-\frac{27}{4}\mu\right)\frac{l_2}{l_1^2}\, .
\end{equation}
Again, something unusual happens in the critical limit. In that case, the free energy of both the black brane and the soliton --- which have a simple metric function given by $f(r)=\frac{3}{2}(r^2-\rh^2)/L^2$ --- vanishes. Then, for $\mu=4/27$, the black brane, the two solitons and pure AdS$_4$ are all equally probable configurations.
%When $\mu\rightarrow 4/27$ the latent heat vanishes. Indeed, the black brane and soliton solution exist in the critical limit, and they are simply given by $f(r)=\frac{3}{2}(r^2-\rh^2)/L^2$, with the corresponding value of $\rh$. However, in the critical limit the free energy of both configurations vanish. This means that the black brane, the solitons and pure AdS are equally probable configurations.

\subsection{Spherical boundary: Hawking-Page transitions}\label{SS2}
Let us now consider the boundary theory on $\mathbb{S}^1_{\beta}\times \mathbb{S}^2$. In that case, apart from Euclidean AdS$_4$ foliated by spheres, the other candidate saddle of the semiclassical action corresponds to the Euclidean spherically symmetric black hole
\begin{equation}\label{Spheric}
ds^2=\left[1+\frac{r^2}{L^2}f(r)\right]d\tau^2+\frac{dr^2}{\left[1+\frac{r^2}{L^2}f(r)\right]}+r^2d\Omega^2_{(2)}\, ,
\end{equation}
where we have chosen $N^2=1$. Also, note that the `volume' of the transverse space is, in this case, $V_{\mathbb{S}^2}=4\pi L^2$. As a function of the horizon radius, the temperature of these solutions is given by \req{T}
\begin{equation}\label{eq:temperatureBHsECG}
T(\rh)=\frac{1}{2\pi \rh }\left(1+3\frac{\rh^2}{L^2}\right)\left[1+\sqrt{1+\frac{3\mu L^4}{\rh^4}\left(1+3\frac{\rh^2}{L^2}\right)}\right]^ {-1}\ .
\end{equation}
The contribution coming from the cubic term in the action becomes less and less relevant as we make $\rh$ larger, but its effect is highly nonperturbative for small radius. For example, a non-vanishing value of $\mu$ makes the temperature vanish, instead of blowing up, as  $\rh\rightarrow 0$. More precisely, one finds $T\approx \rh/(2\pi\sqrt{3\mu}L^2)$ in that regime. This is no different from the behavior observed for the asymptotically flat ECG black holes \cite{Hennigar:2016gkm,PabloPablo2,PabloPablo4} --- small black holes do not care whether they are inside AdS$_4$ or flat space.

\begin{figure}[t]
\centering
\includegraphics[scale=0.85]{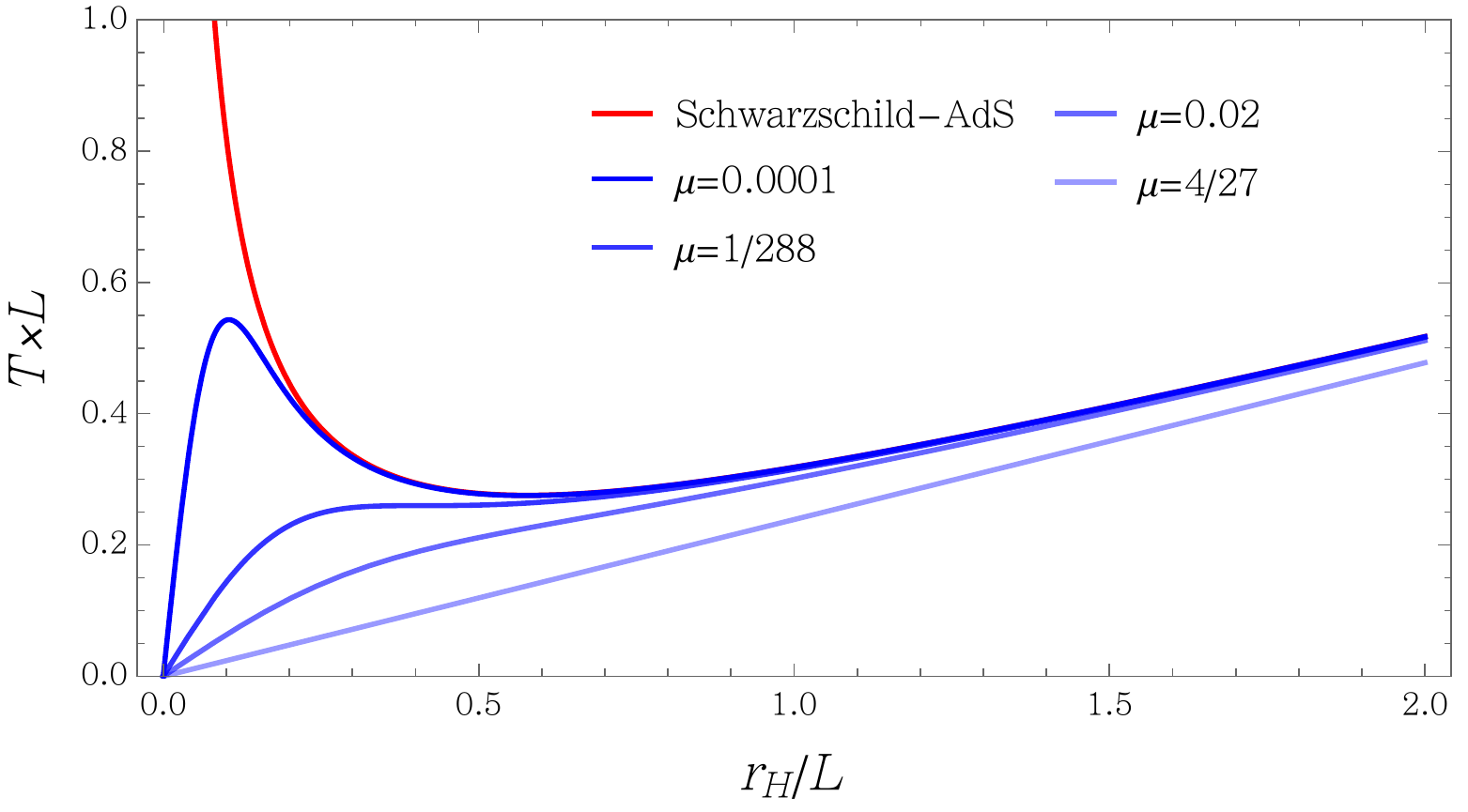}
\caption{Temperature as a function of the horizon radius for various values of $\mu \in [0,4/27]$. Depending on $\mu$, there exist one, two, or three black holes with the same temperature. }
\label{TempECG}
 \end{figure}

% For instance, one can see that, as a consequence of turning on the higher-order parameter $\mu$, the temperature does not diverge when $\rh\rightarrow 0$, even if $\mu$ is really small. Instead, one has $T\approx \frac{\rh}{2\pi\sqrt{3\mu}L^2}$, so that it vanishes when $\rh\rightarrow 0$.
Besides this, the introduction of the cubic term in the action leads to some additional differences with respect to Einstein gravity --- see Fig. \ref{TempECG}. For the usual Schwarzschild-AdS$_4$ Einstein gravity black hole, the temperature is always higher than a certain value,  $ T>T_{\rm min}\equiv\sqrt{3}/(2\pi L)$. In that case, for a given $T>T_{\rm min}$, there exist two black holes, one large, and one small. There are no solutions for which $T<T_{\rm min}$. For ECG the situation is quite different. On the one hand, one observes that there is no minimum temperature, this is, as long as $\mu\neq 0$, there always exists at least one black hole solution for a given $T$. We can distinguish two qualitatively different behaviors depending on $\mu$. For $0<\mu<\mu_{T}\equiv 1/288$, there is an interval %\footnote{The dependence of $T_{\rm min}$ and $T_{\rm max}$ on $\mu$ can be analytically obtained by finding the roots of the following cubic equation: $3 z^3 + (36 \mu- 1) z^2 + 12 \mu z + \mu =0$, which, in the interval under consideration, has three real roots, two of them positive and one negative. The minima and maxima are given by the square root of the two positive ones.} 
of temperatures $\left(T_{\rm min},T_{\rm max}\right)$ for which three black hole solutions with the same temperature exist. However, if $T\ge T_{\rm max}$ or $T<T_{\rm min}$. we just have one. On the other hand, if  $\mu>\mu_T$, there is always a single black hole solution for each temperature. In the critical limit, for which $f(r)=3(r^2-\rh^2)/(2L^2)$, the relation \req{eq:temperatureBHsECG} becomes linear \cite{Feng:2017tev}, and reads $T=3\rh/(4\pi L^2)$.

%\begin{itemize}
%\item When $0<\mu<\mu_{T}=1/288$, there is an interval\footnote{The dependence of $T_{\rm min}$ and $T_{\rm max}$ with $\mu$ can be analytically obtained by finding the roots of the following cubic equation: $3 z^3 + (36 \mu- 1) z^2 + 12 \mu z + \mu =0$, which, in the interval under consideration, has three real roots, two of them positive and one positive. The minima and the maxima are given by the square root of the two positive ones.}$\left(T_{\rm min},T_{\rm max}\right)$ where we have three black hole solutions with the same temperature. If $T\ge T_{max}$ or $T<T_{min}$ we just have one. 
%\item If $\mu>\mu_T$, we find that there is only one black hole solution at a given temperature. 
%\end{itemize}
%This is illustrated in fig. \ref{TempECG}.

% Besides, we see that for the critical vale $\mu=4/27$ the relation \req{eq:temperatureBHsECG} becomes linear: $T=\frac{3\rh}{4\pi L^2}$. In fact, it was found by Lu et al \comment{Cite here} that in the critical limit the solution \req{Spheric} becomes 
%\begin{equation}
%ds^2_{\rm cr}=-\frac{3(r^2-\rh^2)}{2L^2}dt^2+\frac{2L^2dr^2}{3(r^2-\rh^2)}+r^2d\Omega^2_{(2)}\, ,
%\end{equation}
%whose temperature precisely agrees with the previous value. 
 
In sum, at a fixed temperature $T$, we have several solutions with $\mathbb{S}^1_{\beta}\times \mathbb{S}^2$ boundary geometry: thermal AdS$_4$, and one or three black holes depending on the value of $\mu$. In order to identify which phase dominates the holographic partition function at each temperature, let us again compare the on-shell actions of the solutions. For thermal AdS$_4$, one finds a vanishing result, whereas for the black holes, the result can be obtained from \req{fecg}, from which we can obtain $I_E(T)$ implicitly using \req{eq:temperatureBHsECG}. 
\begin{figure}[t]
	\centering 
	\includegraphics[scale=0.48]{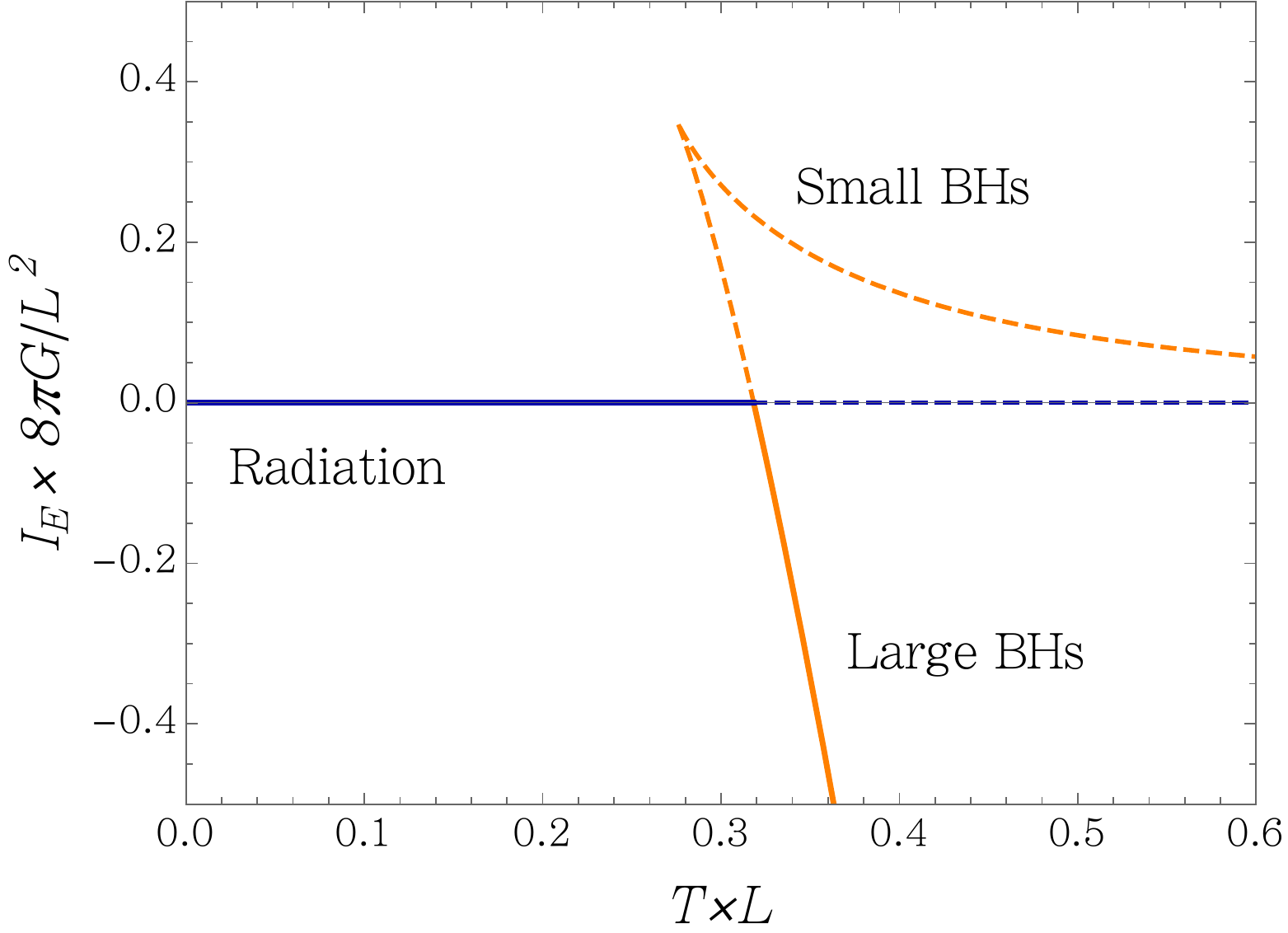}
	\includegraphics[scale=0.48]{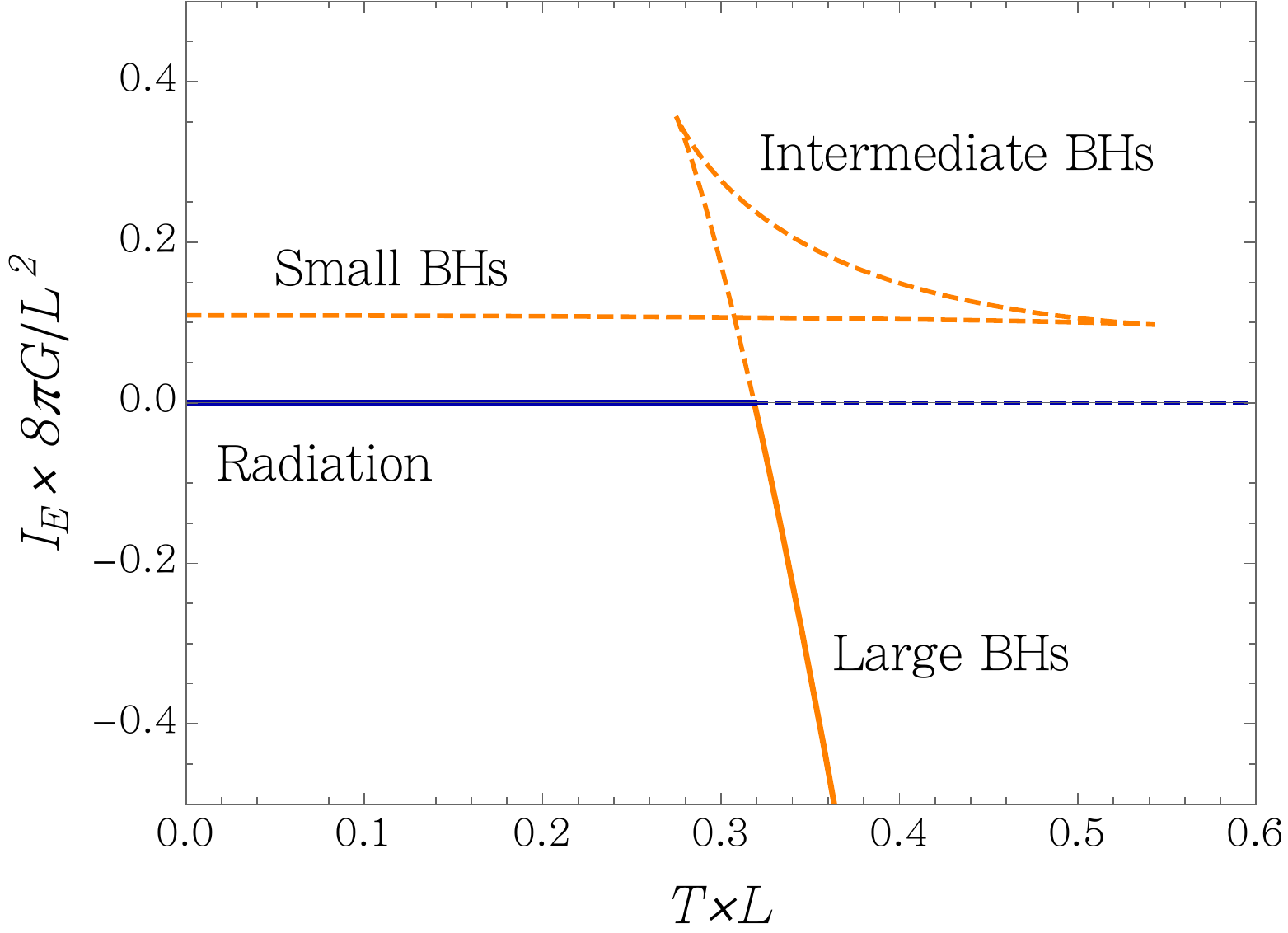}
	\includegraphics[scale=0.48]{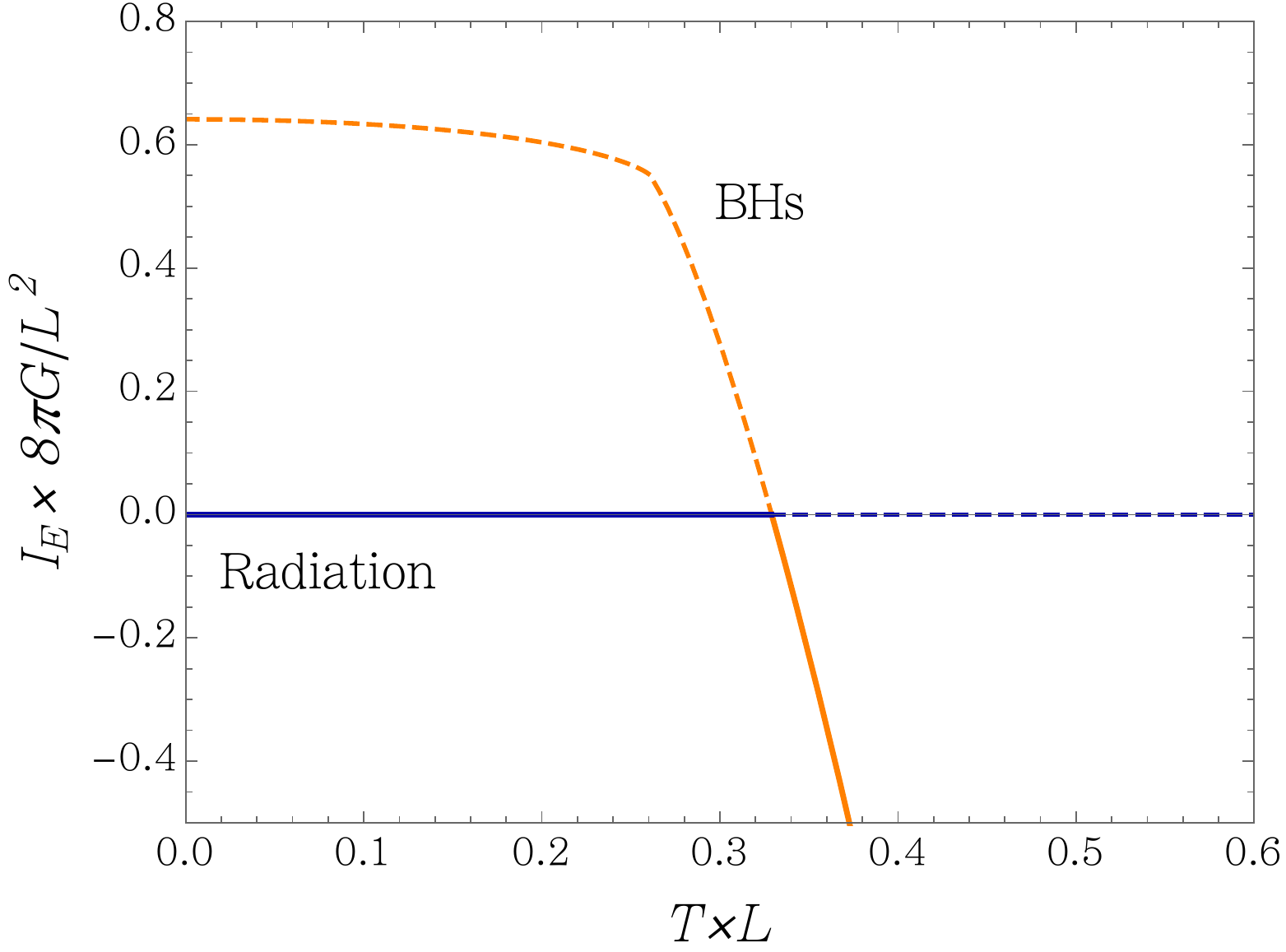}
	\includegraphics[scale=0.48]{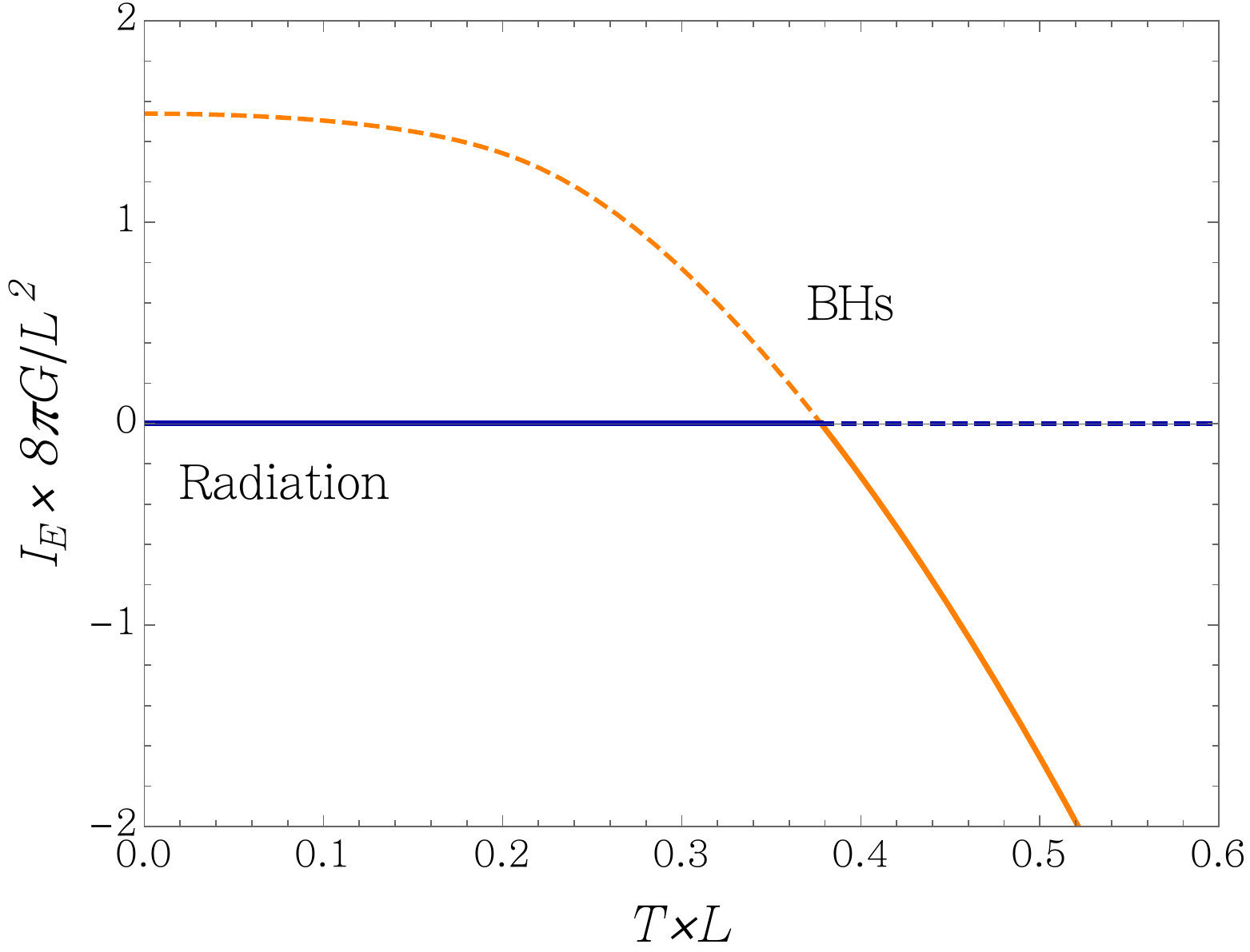}
	\caption{We plot $I_E$ as a function of the temperature for the different phases of holographic ECG in $\mathbb{S}^1_{\beta}\times \mathbb{S}^2$. Solid lines represent the dominant phase in each case. Blue lines correspond to thermal AdS$_4$, and orange lines to black holes. From left to right and top to bottom: $\mu=  0,\, 0.0001,\, 1/288,\, 0.02$. For $\mu=0$ we get the usual Einstein gravity result, with two orange branches corresponding to small and large black holes, and a Hawking-Page transition at $T_{\rm HP}=1/(\pi L)$. For $0<\mu<1/288$, there exist either one or three black-hole branches, depending on the temperature, while for $\mu>1/288$ there is a single black hole for every temperature. As $\mu$ approaches the critical value, the Hawking-Page transition temperature grows as $T_{\rm HP}\sim 3/(2\pi L \sqrt{1-27\mu/4})$. In the limit $\mu=4/27$, the on-shell action is constant (not shown in the figure), $I_E=4\pi L^2/(3G)$, so thermal AdS$_4$ always dominates, and there is no Hawking-Page transition.}\label{HP}
\end{figure}
In Fig.  \ref{HP}, we plot $I_E$ for various values of $\mu$. At a given temperature, we always have several possible phases: a pure thermal vacuum (radiation), and one or several black holes. The dominating phase (shown in solid line) is the one with smaller on-shell action. Regardless of the value of $\mu$, the qualitative behavior is always the same: for small temperatures, the partition function is dominated by radiation, while for large enough temperatures there is a Hawking-Page phase transition \cite{Hawking:1982dh,Witten:1998zw} to a large black hole. The temperature at which the transition occurs depends on $\mu$. For Einstein gravity, one finds $T_{\rm HP}=1/(\pi L)$, while for $\mu \ll 1$, this result gets corrected as 
\begin{equation}
T_{\rm HP}=\frac{1+10 \mu}{\pi L}+\mathcal{O}(\mu^2)\, .
\end{equation}
Hence, the introduction of the ECG density increases the temperature at which the transition occurs. The black-hole radius for which the phase transition takes place also grows if we turn on $\mu$, and is given by $\rh=L(1+26\mu +\mathcal O(\mu^2))$, and the same happens with the latent heat, $\delta Q=L/G\cdot \left(1+38\mu+\mathcal{O}(\mu^2)\right)$. As we increase $\mu$, the Hawking-Page transition temperature grows. In fact, it diverges in the critical limit $\mu=4/27$, which means that no transition at all occurs in that case. If we define $\epsilon\equiv 1-27/4\mu$, the transition temperature for $\epsilon \ll 1$ can be seen to be given by
\begin{equation}
T_{\rm HP}= \frac{3}{2\pi L \sqrt{\epsilon}}\left[1-\frac{\epsilon}{4}+\mathcal{O}(\epsilon^2)\right] \, ,\quad \text{which occurs for}\quad \rh= \frac{2L }{\sqrt{\epsilon}}\left[1-\frac{1}{4}\epsilon+\mathcal{O}(\epsilon^2)\right]\, .
\end{equation}
The reason for the disappearance of the transition is that the critical black holes have a temperature-independent on-shell action, namely\footnote{The fact that the on-shell action of black holes does not depend on the horizon size is yet another unusual property of the critical theory.}

\begin{equation}
I_E=\frac{4\pi L^2}{3G}\left[1-\frac{9+8\pi^2 T^2L^2}{18}\epsilon + \mathcal O(\epsilon^2)\right] \, ,
\end{equation}
which in the $\epsilon=0$ limit is a positive constant, therefore greater than the thermal AdS$_4$ value\footnote{As $\epsilon \rightarrow 0$, the latent heat also diverges as $\delta Q=4 L/ (G\sqrt{\epsilon})\cdot \left(1-3\epsilon/4+\mathcal{O}(\epsilon^2)\right)$, although the entropy increase tends to a constant value, $\delta S=8L^2\pi/(3G)\cdot \left(1-\epsilon/2+\mathcal{O}(\epsilon^2)\right)$.}.  

Although we have seen that only radiation and large black holes can dominate the partition function, it is worth stressing certain new features that appear in the thermal phase space of ECG.
First, we observe that a low-temperature phase of small black holes becomes available as we turn on $\mu$. For small $T$, the corresponding on-shell action is given by
% other phases are never dominant, we observe that for any $\mu>0$ we have a phase of small black holes at low temperatures which is not present in Einstein gravity. The partition function of this phase in the limit $T\rightarrow 0$ reads
\begin{equation}
I_E^{\rm Small\,\, BHs}=\frac{2\pi L^2\sqrt{3\mu}}{G}\, .
\end{equation}
Hence, if $\mu$ is small enough (but not zero!), a spontaneous transition from radiation to small black holes is likely to occur at low temperatures. However, a too small value of $\mu$ could be outside the limits of validity of this approach. Indeed, if the cubic corrections came from string theory, one would expect something like $\sqrt{\mu}L^2\sim \alpha'$, which is assumed to be much larger than $G$ in the holographic setup. %In any case, these additional phases at low temperature are an interesting novelty which is not found in Einstein gravity.
On the other hand, the phase space has a critical point (not to be confused with the critical limit of the theory) where the three black-hole phases in Fig. \ref{HP} (top right) stop existing separately\footnote{We thank Robie Hennigar for pointing this out to us.}. This occurs for $\mu=\mu_T$ which separates the cases for which there are three phases, from those for which there is only one. The phase transition is second-order, and takes place at a temperature $T_{\rm c}=\frac{\sqrt{2/3}}{\pi L}$, corresponding to the non-smooth point on the dashed orange curve in Fig. \ref{HP} bottom left. The critical exponent of the specific heat at the transition turns out to be $-2/3$. More precisely, we find
	% As observed previously, this happens when $\mu\ge \mu_T$. For $\mu>\mu_T$ there are no different phases, but precisely at the critical point $\mu=1/288$ we find a second-order phase transition. This can be clearly observed as the non-smooth point in Fig. \ref{HP} bottom left. The phase transition occurs at a temperature $T_c=\frac{\sqrt{2/3}}{\pi L}$ and it can be shown that the critical exponent of the specific heat $(C=-T \partial^2F/\partial T^2)$ is $-2/3$. More exactly, we find
\begin{equation}
C\equiv -T\frac{\partial^2 F}{\partial T^2}=\frac{\pi 5^{4/3} L^2}{9\cdot 2^{7/3}G}\left(\frac{T}{T_{\rm c}}-1\right)^{-2/3}\, \quad \text{as} \quad T\rightarrow T_{\rm c}\, .
\end{equation}

Let us finally mention that the thermodynamic behavior of our black holes is qualitatively similar to the one observed for $D=5$ Gauss-Bonnet black holes \cite{Cai:2001dz}\footnote{See also \cite{Cho:2002hq} for the case of general quadratic gravity --- the analysis becomes perturbative in that case though.}. Just like for ECG, a new phase of  small stable black holes appears also in that case, as a consequence of the Gauss-Bonnet term. Again, thermal AdS$_5$ is always globally preferred over such solutions. Observe also that the fact that there is no phase transition for critical ECG seems to be related to the fact that, in that case, the solutions become `very similar' to three-dimensional BTZ black holes (see footnote \ref{BTZ}), for which no Hawking-Page transition exists either \cite{Myung:2005ee}. Finally, let us point out that more sophisticated phase transitions connecting different AdS vacua have been identified for Lovelock gravities in various dimensions \cite{Camanho:2013uda}. It would be interesting to explore their possible existence in ECG or, more generally, for the class of theories introduced in  \cite{Ahmed:2017jod,PabloPablo4,Hennigar:2017ego}.

\section{R\'enyi entropy}\label{renyie}
R\'enyi entropies \cite{renyi1961,renyi1} are useful probes of the entanglement structure of quantum systems --- see \eg \cite{HoloRen,Klebanov:2011uf,Laflorencie:2015eck}, and references therein. Roughly speaking, given a state $\rho$ and some spatial subregion $V$ in a QFT, R\'enyi entropies characterize `the degree of entanglement' between the degrees of freedom in $V$ and those in its complement (when such a bi-partition of the Hilbert space is possible). More precisely, they are defined as
\begin{equation}\label{rr}
S_q(V)=\frac{1}{1-q}\log \Tr \rho_V^q \, , \quad q\geq 0\, ,
\end{equation}
where $\rho_V$ is the partial-trace density matrix obtained integrating over the degrees of freedom in the complement of the entangling region. Whenever \req{rr} can be analytically continued to $q\in\mathbb{R}$, the corresponding EE can be recovered as the $q\rightarrow 1$ limit of $S_q$.

In this section we use the methods developed in \cite{CHM,HoloRen} to compute the R\'enyi entropy for disk regions in the ground state of holographic ECG. In subsection \ref{ere}, we study the dependence of $S_q/S_1$ on $\mu$, as well as on some of the charges characterizing the CFT. In subsection \ref{twist}, we compute the conformal scaling dimension of twist-operators $h_q$ for ECG --- see below for definitions --- as an intermediate step to obtain in subsection \ref{t44}, using the results in \cite{Chu:2016tps}, the charge $t_4$ characterizing the three-point function of the stress tensor.

\subsection{Holographic R\'enyi entropy}
\label{ere}
In \cite{CHM}, it was shown that the entanglement entropy across a radius-$R$ spherical region $\mathbb{S}^{d-2}$ for a generic $d$-dimensional CFT equals the thermal entropy of the theory at a temperature $T_0=1/(2\pi R)$ on the hyperbolic cylinder $\mathbb{R}\times \mathbb{H}^{d-1}$, where the curvature scale of the hyperbolic planes is given by $R$.
%\begin{equation}
%\see = S_{\rm thermal}
%\end{equation}
The result is particularly useful in the holographic context, where the latter can be computed as the Wald entropy of pure AdS$_{(d+1)}$ foliated by $\mathbb{R}\times \mathbb{H}^{d-1}$ slices\footnote{Observe that this means, in particular, that for odd-dimensional holographic CFTs, we can in principle access $a^*$ --- see \req{asta} --- in three different ways: 1) from an explicit EE calculation using the Ryu-Takayanagi functional \cite{Ryu:2006bv,Ryu:2006ef} or its generalizations, \eg \cite{Fursaev:2013fta,Dong:2013qoa,Camps:2013zua}, depending on the bulk theory; 2) from the Euclidean on-shell action of pure AdS$_{(d+1)}$ with $\mathbb{S}^{d}$ boundary \cite{CHM}; 3) from the Wald entropy of AdS$_{(d+1)}$ with $\mathbb{R}\times \mathbb{H}^{d-1}$ boundary \cite{CHM}.}. Later, in \cite{HoloRen}, it was argued that this result could be in fact extended to general R\'enyi entropies, the result being 
\begin{equation}\label{sqq}
S_q=\frac{q}{(1-q)T_0}\int_{T_0/q}^{T_0} S_{\rm \ssc thermal}(T) dT\, ,
\end{equation}
where $S_{\rm \ssc thermal}(T)$ is the corresponding thermal entropy on $\mathbb{R}\times \mathbb{H}^{d-1}$ at temperature $T$. While for $T=T_0$, general results for the EE across a spherical region can be obtained for arbitrary holographic higher-derivative theories as long as AdS$_{(d+1)}$ is a solution, the situation becomes more involved for general $q$. In that case, \req{sqq} requires that we know $S_{\rm \ssc thermal}(T)$ for arbitrary values of $T$. Holographically, the calculation can only be performed if the bulk theory admits hyperbolic black-hole solutions for which we are able to compute the corresponding thermal entropy. Examples of such theories for which R\'enyi entropies have been computed using this procedure include: Einstein gravity, Gauss-Bonnet, QTG \cite{HoloRen} and cubic Lovelock \cite{Puletti:2017gym}. Analogous studies for theories in which the corresponding black holes solutions were only accesible approximately --- typically at leading order in the corresponding gravitational couplings --- have also been performed, \eg in \cite{Galante:2013wta,Belin:2013dva,Dey:2016pei}. 
ECG allows us to perform the first exact calculation (fully nonperturbative in the gravitational couplings)  of the holographic R\'enyi entropy of a disk region in $d=3$ for a bulk theory different from Einstein gravity. 

Following \cite{HoloRen}, let us start by rewriting \req{sqq} as 
\begin{equation}\label{sex}
S_q=\frac{q}{(q-1)T_0}\left[\left. S(x)T(x) \right|^1_{x_q}-\int_{x_q}^1S'(x) T(x)dx \right]\, ,
\end{equation}
where we defined the variable $x\equiv \rh/\tilde{L}$, and where $S$ and $T$ stand for the thermal entropy and temperature of the hyperbolic AdS black hole of the corresponding theory. For $x=1$, one has, in general $T(1)=T_0$, whereas $x_q$ is defined as a solution to the equation $T(x_q)=T_0/q$. 
For ECG cubic gravity, the expressions for $S(x)$ and $T(x)$ can be extracted from \req{sth} and \req{T} respectively by setting $k=-1$,
\begin{align}\notag
S(x)&=\frac{x^2 \tilde{L}^2 V_{\mathbb{H}^2}}{4G}  \left[1-\frac{3 \mu f_{\infty}^2  \left(\frac{3x^2 }{f_{\infty}}-1\right)\left[  \left(\frac{3x^2 }{f_{\infty}}-1\right)-2 \left[1+\sqrt{1-\frac{3 f_{\infty}^2\mu}{x^4}\left(\frac{3x^2 }{f_{\infty}}-1\right)} \right] \right]}{x^4\left[1+\sqrt{1-\frac{3 f_{\infty}^2\mu}{x^4}\left(\frac{3x^2 }{f_{\infty}}-1\right)} \right] ^2}\right] \, , \\ \label{sx}
T(x)&=\frac{1}{2\pi R x} \left(\frac{3x^2 }{f_{\infty}}-1\right)\left[1+\sqrt{1-\frac{3 f_{\infty}^2 \mu}{x^4}\left(\frac{3x^2}{f_{\infty}}-1 \right)} \right]^{-1} \, ,
\end{align}
where, in addition, we have set $N^2=L^2/(f_{\infty}R^2)$. This makes the boundary metric conformally equivalent to 
\begin{equation}
ds^2_{\rm bdy}=-dt^2+R^2 d\Xi^2\, ,
\end{equation}
so that the boundary theory lives on $\mathbb{R}\times \mathbb{H}^{2}$, with the `radius' of the hyperbolic plane given by $R$, as required \cite{HoloRen}. From \req{sx}, it can be seen that $x_q$ corresponds to the real and positive solution of
%\begin{equation}\label{eq:x_q 1}
%\sqrt{1-\frac{3 f_{\infty}^2\mu}{x_q^4}\left(\frac{3x_q^2 }{f_{\infty}}-1\right)}=\frac{q}{x_q}\left(\frac{3x_q^2 }{f_{\infty}}-1\right)-1 \ ,
%\end{equation}
%which can also be written as 
%\begin{equation}\label{eq:x_q 2}
%x_q^2\left(\frac{3q^2x_q^2}{f_{\infty}}-q^2-2qx_q\right)=-3\mu f^2_{\infty}	\ ,
%\end{equation}
\begin{equation}\label{eq:x_q 2}
x_q^2\left(3q^2x_q^2-q^2-2qx_q\right)=3\mu f^2_{\infty}\left(q^2 x_q^4-1 \right)	\ ,
\end{equation}
which for Einstein gravity reduces to 
\begin{equation}
x^{\rm E}_{q}=\frac{1}{3q}\left(1+\sqrt{1+3q^2} \right)\ .
\end{equation}

Observe that we have not said anything yet about the divergent nature of $V_{\mathbb{H}^2}$. Of course, one expects the entanglement and R\'enyi entropies to contain (a particular set of) divergent terms, so one could have only expected some source of divergences to appear in the calculation. It is a remarkably feature of the procedure outlined above that all necessary divergent terms in the R\'enyi entropy  (and no others)  are produced by the volume of the hyperbolic plane. In the case of interest for us, corresponding to $d=3$, the regularized volume reads \cite{CHM}
\begin{equation}\label{reguV}
V_{\mathbb{H}^2}=2\pi \left[\frac{R}{\delta}-1 \right]\, ,
\end{equation}
	where we introduced a short-distance cut-off $\delta$. From this expression, we shall only retain the universal piece\footnote{As stressed in \cite{Casini:2015woa}, the universality of constant terms comes with a grain of salt. For example, in \req{reguV}, one could think of rescaling $R$ by an order-$\delta$ constant, which would pollute the constant term. In the case of EE, this issue was overcome in \cite{Casini:2015woa} using mutual information as a regulator. We will ignore this problem here.}, and hence we will replace $V_{\mathbb{H}^2} \rightarrow -2\pi$ from now on, keeping in mind that $S_q$  also contains a cut-off dependent `area' law piece.  
	Taking this into account, after some massaging, which includes using \req{roo}, we can check that 
\begin{equation}
T(1)=T_0\, \quad \text{and} \quad S(1)=-2\pi a^{*\rm ECG}\, ,
\end{equation}
where $a^{*\rm ECG}$ was defined in \req{aae}. Hence, we obtain the same result for the EE of a disk as the one found in section \ref{osa} from the free energy of holographic ECG on $\mathbb{S}^3$. This is another check of our proposed generalized action \req{EuclideanECG}.
%\begin{equation}
%S(x_q)=\frac{x_q^2 \tilde{L}^2}{4G}V_{\mathbb{H}^2}\left[1-\frac{3\mu f_{\infty}^2}{x_q^3 q^2}\left(x_q-2q\right)\right]
%\end{equation}

With all the above information together, we are ready to evaluate the R\'enyi entropy from \req{sex}. The result reads
\begin{equation}\label{reni}
S^{\rm ECG}_q=\frac{q}{(1-q)}\frac{\pi \tilde L^2}{2G}\left[1-x_q-\frac{x^2_q}{q}+x_q^3-\mu f_\infty^2 \left(\frac{3}{q^2 x_q}-3-\frac{1}{q^3}+x_q^3\right)\right]\, ,
\end{equation}
which reduces to the Einstein gravity one \cite{HoloRen} for $\mu=0$. In Fig. \ref{figr1}, we plot $S_q/S_1$ as a function of the R\'enyi index for various values of $\mu$.
\begin{figure}[t]
	\centering 
	\includegraphics[scale=0.83]{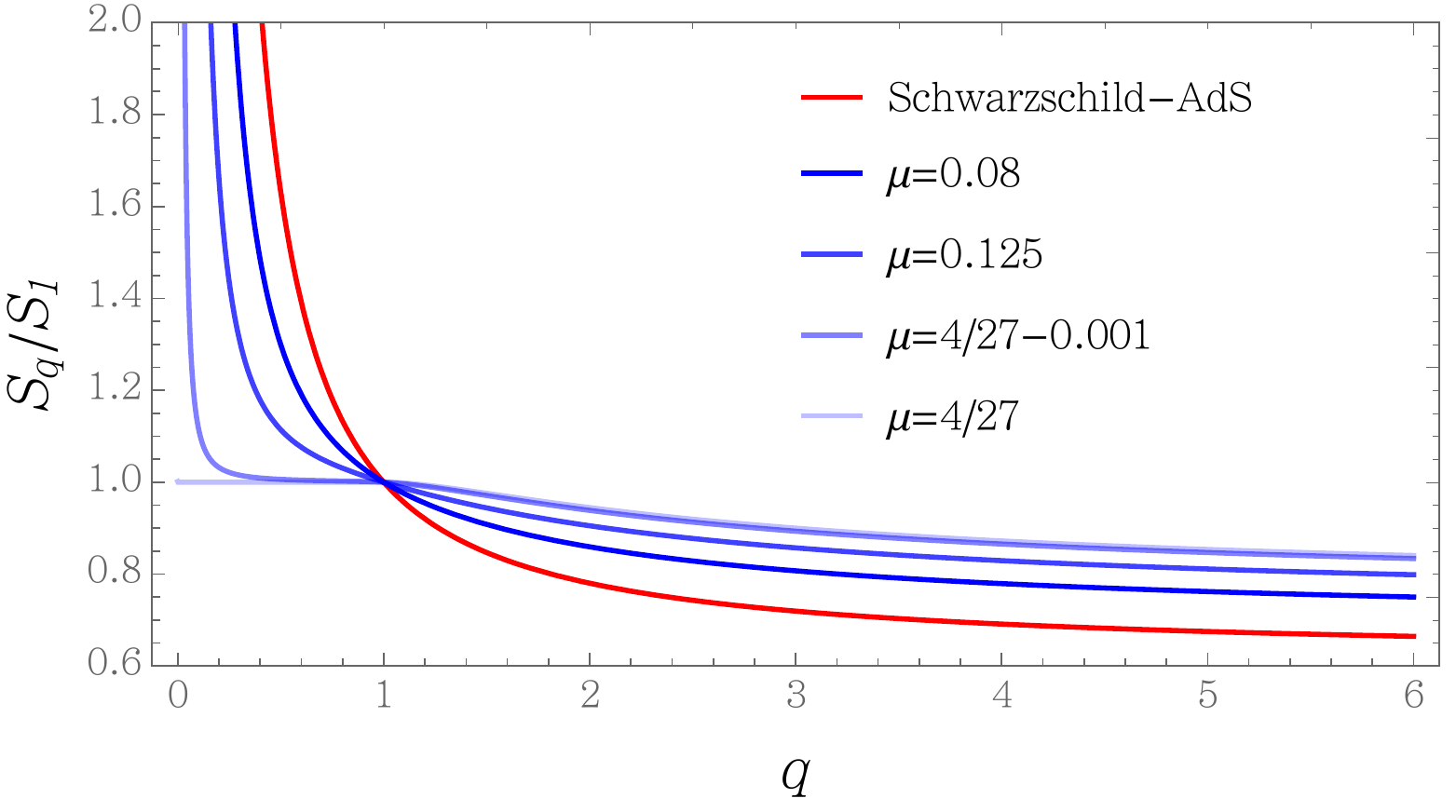}
	\caption{We plot the ratio of the R\'enyi entropy and the EE, $S_q/S_1$, as a function of the R\'enyi index $q$ for various values of the ECG coupling $\mu$.}
	\labell{figr1}
\end{figure}
As we increase $\mu$, $S_q/S_1$ becomes smaller in the $q<1$ region, but it remains larger than $1$ for all values of $\mu$. The opposite occurs for  $q>1$, where $S_q/S_1$ tends to grow as we increase $\mu$, but $S_q/S_1<1$ for all $\mu$.   In the critical limit, there is a jump, and $S_q/S_1$ no longer diverges near $q=0$. In fact, in that case, \req{reni} reduces to a $q$-independent constant for $q<1$, $S^{\rm crit.\, ECG}_q=-\pi \tilde{L}^2/G$. As we approach $\mu\rightarrow 4/27$, $S_{\infty}$ tends to another constant, $S_{\infty}\rightarrow -\pi \tilde{L}^2/G\times(1-1/(3\sqrt{2}))$. Note also that the curve is no longer concave for  $\mu \sim 0.135$ or larger.

 Explicit Taylor expansions of $S^{\rm ECG}_q$ around $q=\{0,1,\infty\}$ can be easily obtained. A few terms suffice in such expansions to provide excellent approximations to the exact curve for most values of $\mu$. At leading order we find, respectively,
\begin{eqnarray}
\lim_{q\rightarrow 1}S^{\rm ECG}_q&=&-2\pi a^{*\rm ECG}\ , \\ \label{peniss}
\lim_{q\rightarrow 0}S^{\rm ECG}_q&=&-\frac{1}{6\pi q^2}\cs^{\rm ECG}\ , \\
\lim_{q\rightarrow \infty}S^{\rm ECG}_q&=&-\frac{\pi \tilde L^2 }{2G}\left[1+3\mu f^2_\infty-\frac{2}{3\sqrt{3(1-\mu f_{\infty}^2)}}\right]\ .
\end{eqnarray}
The first result corresponds to the EE, and we have mentioned it already. As for the second, the appearance in the $q\rightarrow 0$ regime of the thermal entropy charge $\cs^{\rm ECG}$, identified in section \ref{cssex},  should not come as a surprise either. The reason is the following. As shown in \cite{HoloRen}, the R\'enyi entropy $S_q$ across a $\mathbb{S}^{d-2}$ in a general CFT$_d$ can be alternatively written as 
\begin{equation}\label{sqqs}
S_q=\frac{q}{(1-q)}\frac{R^{d-1}V_{\mathbb{H}^{d-1}}}{T_0}\left[\mathcal{F}(T_0)-\mathcal{F}(T_0/q)\right]\, ,
\end{equation}
where $\mathcal{F}(T)$ is the free energy density of the theory at temperature $T$ on $\mathbb{R}\times \mathbb{H}^{d-1}$. The point is that, as $q\rightarrow 0$, the second term in \req{sqqs} dominates over the first. Then, one can use the fact that, at high temperatures, the free energy density on $\mathbb{R}\times \mathbb{H}^{d-1}$ tends to the free energy density on $\mathbb{R}^d$ \cite{Swingle:2013hga}, $\mathcal{F}_{\mathbb{R}\times \mathbb{H}^{d-1}}(T)=\mathcal{F}_{\mathbb{R}^d}(T) \left[1+\mathcal{O}(1/(RT)^2)\right]$, since $1/R$ becomes irrelevant compared to $T$ in that regime. Using the general relation $\mathcal{F}_{\mathbb{R}^d}(T)=-\cs T^d/d$, valid for any CFT in flat space, it follows then that\footnote{See \cite{Bueno3,Galante:2013wta} for analogous arguments.}
\begin{equation}
\lim_{q\rightarrow 0}S_q=\frac{V_{\mathbb{H}^{d-1}}\cs }{d}\left(\frac{1}{2\pi q} \right)^{d-1} \, ,
\end{equation}
which should hold for any CFT$_d$ and, in particular, precisely agrees with \req{peniss} for ECG. Besides, we can readily check that
\begin{equation}
\left. \partial_q S^{\rm ECG}_q\right|_{q=1}=\frac{\pi^4}{12}\ctt^{\rm ECG}\, ,
\end{equation}
as expected from the general relation found in \cite{Perlmutter:2013gua}.

\begin{figure}[t]
	\centering 
	\includegraphics[scale=0.83]{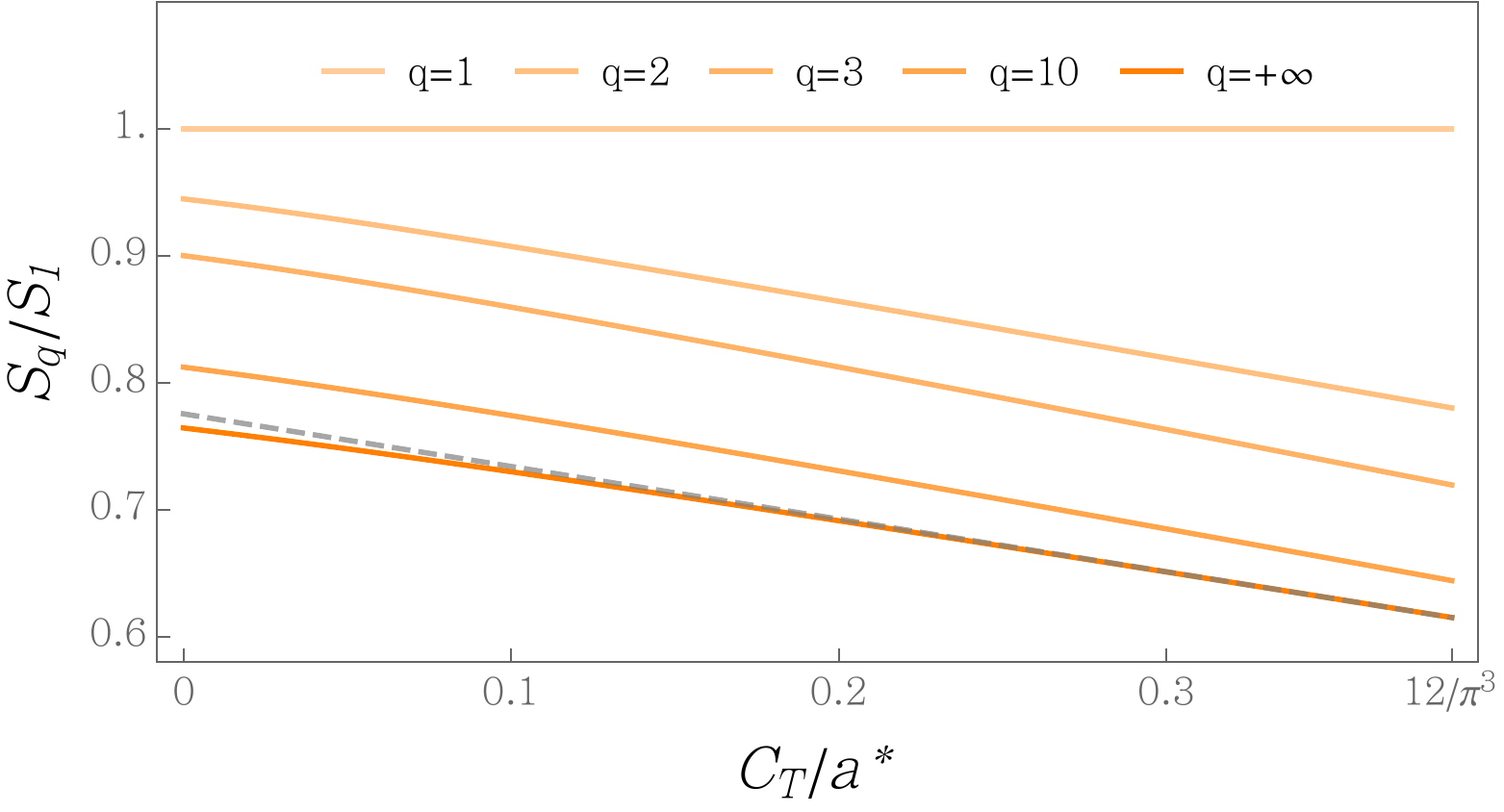}
	\caption{We plot the ratio of the R\'enyi entropy and the EE, $S_q/S_1$, as a function of the quotient $(\ctt/a^*)^{\rm ECG}$ for $q=\{1,2,3,10,\infty\}$. The limits of the range plotted correspond to the critical theory, $(\ctt/a^*)^{\rm ECG}=0$, and Einstein gravity $(\ctt/a^*)^{\rm ECG}=12/\pi^3$, respectively. The dashed line corresponds to the linear approximation to $S_{\infty}/S_1$ in \req{lin}.}
	\labell{figr2}
\end{figure}

Let us now gain some insight on the dependence of $S_q$ on quantities characterizing the CFT. In order to do that, we can use the relations
\begin{equation}
\frac{\tilde{L}^2}{G}= a^* \frac{\pi^3}{6}  \left[(\ctt/a^*)+\frac{12}{\pi^3} \right]\, ,\quad \mu f_{\infty}^2=-\frac{1}{3}\frac{\left[(\ctt/a^*)-\frac{12}{\pi^3} \right]}{\left[(\ctt/a^*)+\frac{12}{\pi^3}\right] }\, .
\end{equation}
It is then straightforward to substitute these in \req{eq:x_q 2} and \req{reni} to obtain $S_q$ as a function of $a^{*\rm ECG}$ and $(\ctt/a^*)^{\rm ECG}$.  Observe that $a^*$ appears as a global factor, so that $S_q/S_1$ is a function of $\ctt/a^*$ alone. We plot this ratio for several values of $q$ in Fig. \ref{figr2}. Observe that $\ctt/a^*$ takes values between $0$ and $12/\pi^3\simeq 0.3870$, corresponding to the critical value, $\mu=4/27$, and Einstein gravity respectively. Interestingly, even though the dependence of $S_q/S_1$ on  $\ctt/a^*$ is in principle highly non-linear, all curves seem to be approximately linear in the full range. In addition, we find that 
\begin{equation}
\left. \frac{S_q}{S_1}\right|_{({\ctt}_1/a_1^*)}<\left. \frac{S_q}{S_1}\right|_{({\ctt}_2/a_2^*)}\,  \quad \text{for} \quad ({\ctt}_1/a_1^*)> ({\ctt}_2/a_2^*)\, ,
\end{equation} 
\ie $S_q/S_1$ monotonously decreases as $\ctt/a^*$ grows, for all values of $q$. These features are very similar to the ones observed in \cite{HoloRen} for holographic Gauss-Bonnet in $d\geq 4$. 

We can gain some understanding on the approximately linear behavior of $S_q/S_1$ by expanding $S_{\infty}/S_1$ around the Einstein gravity value,  $(\ctt/a^*)^{\rm ECG}=12/\pi^3$. By doing so, we obtain
%\begin{align}\label{lin}
%\frac{S_{\infty}}{S_1}=1-\frac{1+\frac{\pi^3(\ctt/a^*)}{12}}{6\sqrt{\frac{\left[2(\ctt/a^*)+\frac{12}{\pi^3}\right]}{2\left[(\ctt/a^*)+\frac{12}{\pi^3}\right]}}} \simeq 
%\left[1-\frac{2}{3\sqrt{3}}\right]-\frac{5\pi^3 }{216\sqrt{3}} \left[(\ctt/a^*)-\frac{12}{\pi^3}\right]+\dots\, ,
%\end{align}
\begin{align}\label{lin}
	\frac{S_{\infty}}{S_1}=1-\frac{\pi^3  \left[(\ctt/a^*)+\frac{12}{\pi^3}\right]^{3/2}}{72  \left[(\ctt/a^*)+\frac{6}{\pi^3}\right]^{1/2}} \simeq 
	\left[1-\frac{2}{3\sqrt{3}}\right]-\frac{5\pi^3 }{216\sqrt{3}} \left[(\ctt/a^*)-\frac{12}{\pi^3}\right]+\dots\, ,
\end{align}
where the first omitted correction is quadratic in the expansion parameter. As it turns out, the linear approximation in \req{lin} fits the exact curve very well for most values of $\ctt/a^*$ --- see dashed line in Fig. \ref{figr2}. We suspect a similar phenomenon occurs for smaller values of $q$. 

In spite of this `pseudo-linearity', it seems clear that $S_q^{\rm ECG}$ does not have a simple dependence on universal CFT quantities. This fact, which agrees with the exact $d\geq 4$ results of \cite{HoloRen} for Gauss-Bonnet and QTG, was actually anticipated in that paper also for $d=3$, where $S_q$ was computed  at leading order in the gravitational coupling for a  bulk model consisting of Einstein gravity plus a  Weyl$^3$ correction.  

% If we restrict ourselves to the physical range \req{phiss}, the allowed values of $\ctt/a^*$ are
%\begin{equation}
%\frac{105}{107}\leq  \frac{\pi^3}{12}\frac{\ctt}{a^*} \leq 1\, .
%\end{equation}
%Then, we can rewrite \req{eq:x_q 2} and \req{reni}
%\begin{equation}\label{eq:x_q 22}
%x_q^2\left(3q^2x_q^2-q^2-2qx_q\right)=\frac{\left[(\ctt/a^*)-\frac{12}{\pi^3} \right]}{\left[(\ctt/a^*)+\frac{12}{\pi^3}\right] }\left(1-q^2 x_q^4 \right)	\ ,
%\end{equation}
%\begin{equation}\label{renii}
%S_q= a^* \frac{ V_{\mathbb{H}^2} \pi^3 q}{24(q-1)}  \left[(\ctt/a^*)+\frac{12}{\pi^3} \right]\left[1-x_q-\frac{x^2_q}{q}+x_q^3+\frac{1}{3}\frac{\left[(\ctt/a^*)-\frac{12}{\pi^3} %\right]}{\left[(\ctt/a^*)+\frac{12}{\pi^3}\right] } \left(\frac{3}{q^2 x_q}-3-\frac{1}{q^3}+x_q^3\right)\right]\ .
%\end{equation}

\subsection{Scaling dimension of twist operators}\label{twist}
Let us now turn to the scaling dimension of twist operators. In the context of computing R\'enyi entropies for some region $V$ using the replica trick, the boundary conditions which glue together the different copies of the replicated geometry at the entangling surface $\partial V$, can be alternatively implemented through the insertion of dimension-$(d-2)$ operators $\tau_q$ extending over $\partial V$ \cite{Calabrese:2004eu,HoloRen,Hung:2014npa,Swingle:2010jz}. The replicated-geometry construction is then replaced by a path integral over the symmetric product of $q$ copies of the theory on a single copy of the geometry, with the $\tau_q$ inserted. Given $V$, $\Tr \rho_V^q$ can be then obtained as the expectation value of these `twist operators', $\Tr \rho_V^q=\braket{\tau_q}_q$, computed in the $q$-fold symmetric product CFT. A natural notion of scaling dimension, $h_q$, can be defined for $\tau_q$ from the leading singularity appearing in the correlator $\braket{T_{\mu\nu}\tau_q}$, as the stress tensor is inserted close to $\partial V$. In particular \cite{HoloRen,Hung:2014npa},
\begin{equation}
\braket{T_{\mu\nu}\tau_q}_q=-\frac{h_q}{2\pi} \frac{b_{\mu\nu}}{y^d}\, ,
\end{equation}
where $b_{\mu\nu}$ is a fixed tensorial structure and $y$ is the separation between the stress-tensor insertion and $\partial V$.

 Our interest in the $h_q$ for ECG is mostly related to the use that we will make of them in the following subsection, so let us just reproduce the most relevant result needed to compute them for holographic CFTs \cite{HoloRen,Hung:2014npa}. This establishes that, given some higher-derivative bulk theory, $h_q$ can be obtained from the thermal entropy and temperature of the corresponding hyperbolic AdS black hole as 
\begin{equation}
h_q=\frac{2\pi R q}{(d-1)V_{\mathbb{H}^{d-1}}}\int_{x_q}^{1}T(x)S'(x)dx\ .
\end{equation}
%In order to evaluate this expression for ECG, it is convenient to use that 
%\begin{equation}
%\int_1^x T(x)S'(x)dx= I(x)\ ,
%\end{equation}
%where 
%\begin{equation}
%I(x)=-T_0 \frac{\tilde L^2 V_{\mathbb{H}^2}}{4G}\left[\frac{2x^3}{f_\infty}+\frac{2x\left(2-\frac{3x^2}{f_\infty}\right)\left(\frac{3x^2}{f_{\infty}}-1\right)}{3\left[1+\sqrt{1-\frac{3 f_{\infty}^2\mu}{x^4}\left(\frac{3x^2 }{f_{\infty}}-1\right)}\right]}+\frac{4x\left(\frac{3x^2 }{f_{\infty}}-1\right)^2}{3\left[1+\sqrt{1-\frac{3 f_{\infty}^2\mu}{x^4}\left(\frac{3x^2 }{f_{\infty}}-1\right)}\right]^2}\right] \ .
%\end{equation}
%We get
Then, using \req{sx}, we find, for the universal piece,
\begin{equation}\label{scalingdimension}
h^{\rm ECG}_q=-\frac{q\tilde L^2}{8G}\left[x_q^3-x_q-\mu f^2_{\infty}\left(x_q^3+\frac{2}{q^3}-\frac{3}{q^2x_q}\right)\right] \ ,
\end{equation}
which reduces to the Einstein gravity result \cite{HoloRen}
\begin{equation}
h_{q}^{\rm E}=\frac{q \tilde L^2}{8G}x_q\left(1-x_q^2\right)\, ,
\end{equation}
when $\mu=0$. It is easy to perform some checks of this result. In particular, we find
\begin{equation}
 \lim_{q\rightarrow 0}h^{\rm ECG}_q=-\frac{1}{12\pi^2 q^2}\cs^{\rm ECG}\, , \quad \partial_q h^{\rm ECG}_q|_{q=1} =\frac{\pi^3}{24}\ctt^{\rm ECG}\ ,
\end{equation}
as expected from the general identities found in \cite{Bueno3} and \cite{Hung:2014npa}, respectively.
Similarly, using \req{reni}, it is possible to verify that the general relations \cite{Bueno3}\footnote{For $j=1$, the second term is ignored.}
\begin{equation}
\left.\partial^j_q h_q\right|_{q=1}=\frac{1}{4\pi}\left[(j+1) \left. \partial^j_q S_q\right|_{q=1}+j^2 \left. \partial^{j-1}_q S_q\right|_{q=1} \right]\, ,
\end{equation}
hold for general $j$ and arbitrary values of $\mu$, as they should.

%Another check would be to take the limit $q\rightarrow 0$. For this, we use that 
%\begin{equation}
%x_q(q)=\frac{2f_\infty}{3 q}+\frac{1+\frac{27}{4}\mu}{2}q+\mathcal O(q^3)\ .
%\end{equation}
%Plugging this expansion into (\ref{scalingdimension}), we get 
%\begin{equation}\label{hq0}
%\lim_{q\rightarrow 0}h_q=-\frac{\tilde L^2}{27 G}\frac{f^2_{\infty}\left(1-\frac{27 }{4}\mu\right)}{q^2}=-\frac{\cs}{12\pi^2 q^2}\ ,
%\end{equation}
%where $\cs=\tfrac{4\pi^2\tilde L^2}{9G}f^2_{\infty}\left(1-\tfrac{27 }{4}\mu\right)$ is the thermal entropy charge.

\subsection{Stress tensor three-point function charge $t_4$}\label{t44}
For general CFTs in $d=3$, the stress tensor three-point function is a combination of fixed tensorial structures controlled by two theory-dependent quantities \cite{Osborn:1993cr}, which can be chosen to be $\ctt$ plus an additional parameter\footnote{In general, in $d=3$, there is also a parity-violating structure \cite{Giombi:2011rz,Maldacena:2011nz,Chowdhury:2017vel}, which is controlled by yet another parameter. Capturing this would require introducing another bulk density involving some contraction of curvature tensors with the Levi-Civita symbol --- see \eg \cite{Maldacena:2011nz}.}, $t_4$. The latter was originally introduced in  \cite{Hofman:2008ar}, where it was shown to appear in the general formula for the energy flux reaching null infinity in a given direction after inserting an operator of the form $\epsilon_{ij}T^{ij}$, where $\vec{\epsilon}$ is some symmetric polarization vector.  For any CFT$_3$, the result takes the general form
\begin{equation}\label{ees}
\braket{\mathcal{E}(\vec{n})}=\frac{E}{2\pi}\left[1+t_4 \left(\frac{|\epsilon_{ij}n^in^j|^2}{\epsilon_{ij}^*\epsilon_{ij}}-\frac{1}{8}\right)\right]\, ,
\end{equation}
where $E$ is the total energy, and $\vec{n}$ is the unit vector indicating the direction in which we are measuring the flux. Hence, the only theory-dependent quantity appearing in the above expression is $t_4$ which, along with $\ctt$, fully characterize $\braket{TTT}$ --- see \eg \cite{Buchel:2009sk,Bobev:2017asb} for the explicit connection.  For $d\geq 4$, there is an extra parity-preserving structure weighted by another theory-dependent constant, customarily denoted $t_2$.

Higher-dimensional versions of \req{ees} have been used to identify $t_4$ and $t_2$ for holographic theories dual to certain higher-order gravities in $d\geq 4$, such as Lovelock \cite{Buchel:2009sk,deBoer:2009gx} or QTG \cite{Myers:2010jv}. It is known that $t_4=0$ for general supersymmetric theories \cite{Hofman:2008ar,Kulaxizi:2009pz}, as well as for theories of the Lovelock class \cite{Buchel:2009sk,deBoer:2009gx,Camanho:2009hu,Camanho:2013pda}, including Einstein gravity in general dimensions. In fact, one of the original motivations for the construction of QTG in \cite{Quasi}, was to provide a nonperturbative holographic model with a non-vanishing $t_4$ in $d=4$. Here, we show that ECG provides an analogous model in $d=3$.

In order to determine $t_4$ for ECG, we will use the results in \cite{Chu:2016tps}, where it was shown that the scaling dimension of twist operators in holographic theories is related to the parameters controlling the stress-tensor three-point function. In particular, it was shown that the expression
\begin{equation}\label{minino}
\frac{h_q}{\ctt}=\frac{\pi^3}{24}(q-1)-\frac{\pi^3}{11520}(420+t_4)(q-1)^2+\mathcal{O}(q-1)^3\, ,
\end{equation}	
holds for general holographic higher-order gravities in $d=3$, at least at leading order in the couplings. Performing the corresponding expansion in the twist-operator scaling dimension \req{scalingdimension}, we find
\begin{equation}\label{t4ecg}
t^{\rm ECG}_4=\frac{-1260 \mu f_{\infty}^2}{(1-3\mu f_{\infty}^2)}\, ,
\end{equation}
which, as expected, vanishes for Einstein gravity. One may worry about the validity of \req{minino} beyond leading order, for which $t_4^{\rm ECG}= -1260 \mu +\mathcal{O}(\mu^2)$. However, we have good reasons to believe that \req{t4ecg} is correct for general values of $\mu$. First of all, observe that \req{t4ecg} singles out $\mu=4/27$ as a special value of the coupling, since $t^{\rm ECG}_4$ diverges in that case. Of course, this is nothing but the critical limit of the theory, for which some sort of bizarre behavior was to be expected. Secondly, in appendix \ref{ttt}, we use the results found in \cite{HoloRen} for the twist-operator scaling dimensions in $d$-dimensional holographic Gauss-Bonnet and $d=4$ QTG, and show that the ($d$-dimensional versions of) \req{minino} provide expressions for $t_2$ and $t_4$ which exactly agree with the fully nonperturbative ones found in \cite{Buchel:2009sk} and \cite{Myers:2010jv}. These observations strongly suggest that \req{t4ecg} is an exact expression.
%At leading order in $\mu$, it reads
%\begin{equation}
%t_4= -1260 \mu +\mathcal{O}(\mu^2)\, .
%\end{equation}

Now, in $d=3$, imposing the positivity of energy fluxes in arbitrary directions gives rise to the constraint $-4\leq t_4 \leq 4$, which is valid for general CFTs \cite{Buchel:2009sk}, as long as the additional parity-odd structure is absent, as in the case of ECG\footnote{Observe, in particular, that for a CFT$_3$ consisting of $n_{\rm s}$ real conformal scalars and $n_{\rm f}/2$ Dirac fermions, $t_4=4(n_{\rm s}-n_{\rm f})/(n_{\rm s}+n_{\rm f})$, which therefore covers the full space of allowed values of $t_4$ \cite{Buchel:2009sk}, the limiting values corresponding to an arbitrary number of fermions, and to an arbitrary number of scalars, respectively.}. 
When written in terms of the gravitational coupling for ECG, this constraint translates into
\begin{equation}\label{fifif}
\frac{312}{313}\leq f_{\infty} \leq \frac{318}{317}\, ,
\end{equation}
which, together with the previous constraint $1\leq f_{\infty} \leq 3/2$ becomes
\begin{equation}
1\leq f_{\infty} \leq \frac{318}{317}\simeq 1.00315\, .
\end{equation}
This can in turn be explicitly written in terms of $\mu$ as
\begin{equation}\label{phiss}
0\leq \mu \leq \frac{100489}{32157432}\simeq 0.00312491\, .
\end{equation}
This reduces the range of allowed values of $\mu$ quite considerably.
Observe that for $f_{\infty}=318/317$, $t_4=-4$, which is precisely the value corresponding to a free fermion. The other limiting value, $t_4=4$, corresponding to a free scalar, would imply a negative value of $\mu$, and is therefore excluded. Observe also that the bound is maximally violated at the critical value $\mu=4/27$.

%%%%%%%%%%%%%%%%%%%%%%%%%%%%%%%%%%%%%%%%%%%%%%%%%%%%%%%%%%%%%%%%%%%
%%%%%%%%%%%%%%%%%%%%%%%%%%%%%%%%%%%%%%%%%%%%%%%%%%%%%%%%%%%%%%%%%%%
%%%%%%%%%%%%%%%%%%%%%%%%%%%%%%%%%%%%%%%%%%%%%%%%%%%%%%%%%%%%%%%%%%%

\section{Holographic hydrodynamics}\labell{shear}
One of the paradigmatic applications of higher-order gravities in the AdS/CFT context has been the construction of counterexamples to the famous Kovtun-Son-Starinets (KSS) bound for the shear viscosity over entropy density bound \cite{Kovtun:2004de}. The latter was originally conjectured to satisfy $\eta/s \geq \frac{1}{4\pi}$ (in natural units) for any fluid in any number of dimensions, the saturation occurring for holographic plasmas dual to Einstein gravity AdS$_{(d+1)}$ black branes.  Violations of the bound --- generically produced by finite-$N$ effects from the gauge-theory side --- were argued to occur for holographic plasmas dual to black branes in several higher-order theories --- see, \eg \cite{Buchel:2004di,Kats:2007mq,Brigante:2007nu,Myers:2008yi,Cai:2008ph,Ge:2008ni} for some of the earliest works and \cite{Cremonini:2011iq} for a review. A thorough study of various consistency conditions --- such as subluminal propagation of excitations, energy positivity  or unitarity --- on some of the holographic theories for which the corresponding branes could be actually constructed --- hence allowing for fully nonperturbative calculations in the higher-curvature couplings --- suggested that the bound can be lowered down to  $\eta/s \sim 0.4\cdot \frac{1}{4\pi}$ for $d=4$ \cite{Myers:2010jv}, and arbitrarily close to zero for large enough $d$ \cite{Camanho:2010ru}. 
These results give rise to three possibilities for finite-$d$: (i) the parameter space which would permit violations of the KSS bound is in fact not allowed by some other unidentified physical conditions --- see below --- and the KSS bound is true after all; (ii) there exists some lower bound, but it is lower than the KSS one; (iii) there is no bound at all. 
It was shown later \cite{Camanho:2014apa} that higher-derivative theories with nonperturbative couplings are in fact generally acausal unless the spectrum is supplemented by higher-spin modes. While it is still unclear under what circumstances such additional degrees of freedom play a relevant role --- specially given the success of holographic higher-curvature models in other holographic applications --- the reliability of the aforementioned conclusions regarding the fate of the bound was
put in suspense by this result. The current belief seems to be that some non-trivial bound, lower than the KSS one, does exist for general $d$ --- see \eg \cite{Fouxon:2008pz}.

In this section we compute the shear viscosity to entropy density ratio for ECG, providing the first calculation of such a quantity for a holographic higher-curvature gravity in $d=3$ which is fully nonperturbative in the gravitational coupling. We will proceed along the lines of \cite{Myers:2010jv,Paulos:2009yk}.  Let us start considering the ECG planar black hole in \req{bhss}, \ie we set $k=0$ and $N^2=1/f_{\infty}$, 
\begin{equation}\label{planarbh}
ds^2=\frac{r^2}{L^2}\left[-\frac{f(r)}{f_{\infty}}dt^2+dx_1^2+dx_2^2\right]+\frac{L^2}{r^2f(r)}dr^2\, .
\end{equation}
Now, it is convenient to perform the change of coordinates $z=1-\rh^2/r^2$, so that the horizon corresponds to $z=0$, the asymptotic boundary being at $z=1$. The metric reads then
\begin{equation}\label{mm}
ds^2=\frac{\rh^2}{L^2(1-z)}\left(-\frac{f(z)}{f_{\infty}}dt^2+dx_1^2+dx_2^2 \right)+\frac{L^2}{4f(z)(1-z)^2}dz^2\, .
\end{equation}
On the other hand, the cubic equation that determines $f(r)$, \req{eqsf}, reads, in terms of $z$
\begin{equation}\label{reee}
1-f(z)+\mu \left[f^3-3 (1-z)^2f f'^2 -2(1-z)^3 f' (f'^2-3f f'')\right]=\left(1-\frac{27}{4}\mu\right)(1-z)^{3/2}\, ,
\end{equation}
where now $f'\equiv df/dz$, and so on. In order to determine the shear viscosity, we will need the near-horizon behavior of $f$, so let us perform a Taylor expansion of the form
\begin{equation}\label{ftay}
f(z)=f_0' z +\frac{1}{2}f_0'' z^2 + \frac{1}{6}f_0'''(z)z^3+\dots,
\end{equation}
The coefficients in this expansion can be of course written in terms of those in the $r$-expansion series \req{nH}, but it is easier to work directly with the variable $z$. Inserting \req{ftay} in \req{reee} and imposing it to hold order by order in $z$, one finds
\begin{equation}\label{f0}
f_0'=\frac{3}{2}\, , \quad f_0'''=\frac{-144 \mu  f_0''^2+4 (135 \mu +4) f_0''-81 \mu +12}{216 \mu }\, ,\quad f_0^{(4)}=\ldots,
\end{equation}
etc. All the coefficients are determined by $f_0''$, whose value is fixed by the asymptotic condition $\lim_{z\rightarrow 1}f(z)=f_{\infty}$. Analogously to the discussion in section \ref{fullsol}, there is a unique value of this parameter for which the desired boundary condition is achieved. This defines $f_0''$ as a function of $\mu$, which we denote $f_0''(\mu)$. We can compute this numerically with arbitrary precision, but let us also try an analytic computation using the following logic. Observe that, for $\mu=0$, the solution is simply $f(z)=1-(1-z)^{3/2}$, from where we read all the derivatives 
\begin{equation}
f_0^{(n)}(0)=(-1)^{n+1}\frac{\Gamma(5/2)}{\Gamma(5/2-n)}.
\end{equation}
 Now, since the solution for general $\mu$ should reduce to the Einstein gravity one when $\mu\rightarrow 0$, the derivatives \req{f0} should coincide with the previous ones in that limit. It turns out that we can use this condition to determine the derivatives of $f_0''(\mu)$ with respect to $\mu$ at $\mu=0$. Let us see how this works. Obviously, we have $\lim_{\mu\rightarrow 0}f_0''(\mu)\equiv f_0''(0)=-3/4$. Then, we should also have $\lim_{\mu\rightarrow 0}f_0'''(\mu)\equiv f_0'''(0)=-3/8$. If we take this limit in the second equation of \req{f0}, we get the condition
 \begin{equation}
 \lim_{\mu\rightarrow 0}\left[\frac{-144 f_0''(\mu)^2+540f_0''(\mu) -81}{216}+\frac{2}{27}\frac{f_0''(\mu)+3/4}{\mu}\right]=-\frac{3}{8}\, .
 \end{equation}
 The limit of the first term is finite and we can simply substitute $f_0''(0)=-3/4$. However, in the second term we have
 \begin{equation}
\lim_{\mu\rightarrow 0} \frac{f_0''(\mu)+3/4}{\mu}=\lim_{\mu\rightarrow 0} \frac{f_0''(\mu)-f_0''(0)}{\mu}\equiv\frac{df_0''(\mu)}{d\mu}\bigg|_{\mu=0}\, .
 \end{equation}
 Therefore, this equation is actually giving us the value of the derivative of $f_0''(\mu)$ at $\mu=0$, the result being $243/8$. The same process can be repeated at every order and we can obtain all derivatives of this function at $\mu=0$. Up to second order, we have
\begin{equation}
f_0''(\mu=0)=-\frac{3}{4}\, ,\quad \frac{df_0''(\mu)}{d\mu}\bigg|_{\mu=0}=\frac{243}{8}\, ,\quad \frac{d^2f_0''(\mu)}{d\mu^2}\bigg|_{\mu=0}=-\frac{115911}{16}\, .
\end{equation}
Now, if the function $f_0''(\mu)$ were analytic, we could in principle construct it as
\begin{equation}\label{f0ser}
f_0''(\mu)=\sum_{n=0}^{\infty}\frac{1}{n!}\frac{d^nf_0''(\mu)}{d\mu^n}\bigg|_{\mu=0}\mu^n\, .
\end{equation}
However, a convergence analysis, including many terms in the expansion, reveals that this series is actually divergent for every $\mu\neq 0$ --- in other words, the radius of convergence is 0. The fact that the series diverges is telling us that the function does not allow for a Taylor expansion around $\mu=0$. This is an example of a $\mathcal{C}^{\infty}$ function which is not analytic\footnote{See \eg \cite{Pasini:2015zlx} for  another explicit example in a different context.}. 
Nevertheless, the series can be used to provide an approximate result for small enough $\mu$ if we truncate it at certain $n$. For example, to quadratic order we obtain
\begin{equation}\label{f0approx}
f_0''(\mu)\approx-\frac{3}{4}+\frac{243}{8}\mu-\frac{115911}{32}\mu^2\, ,
\end{equation}
but the approximation is only good for rather small values of the coupling, \eg for $\mu=0.003$, the error  is $\sim 3\%$ (with respect to the numerical value) and the precision is not increased by the addition of further terms. 
Observe also that in the  critical limit, $\mu=4/27$, we have $f_{\rm cr}(z)=\frac{3}{2}z$, and hence $f_0''(4/27)=0$ in that case. 

After this dissertation, which we will use to get a grasp on the small-$\mu$ behavior of $\eta$, let us now turn to the actual computation. In order to do so,  we perturb the black hole metric \req{mm} by shifting
\begin{equation}
dx_1\rightarrow dx_1+\varepsilon e^{-i\omega t} dx_2\,,
\end{equation}
where $\varepsilon$ is a small parameter. Then, the shear viscosity can be obtained as\footnote{See \cite{Fan:2018qnt} for a recent alternative method.} \cite{Paulos:2009yk}
\begin{equation}
\eta=-8\pi T \lim_{\omega,\epsilon\rightarrow 0}\frac{{\rm Res}_{z=0}\mathcal{L} }{\omega^2 \epsilon^2}\, ,
\end{equation}
where $\mathcal{L}$ is the corresponding full gravitational Lagrangian (including the $\sqrt{|g|}$ term) in \req{ECG} evaluated on the perturbed metric.  Using \req{ftay}, we can evaluate this quantity, and the result reads
\begin{equation}
\eta^{\rm ECG}=\frac{3\rh^2}{64\pi G L^2 f_0' } \left[2+(21f_0'^2+36f_0''^2-114f_0'f_0''+36f_0' f_0''')\mu \right]\, .
\end{equation}
Then, using the values of $f_0'$ and $f_0'''$ in \req{f0}, we find
\begin{equation}
\eta^{\rm ECG}=\frac{\rh^2}{32\pi G L^2} \left[5+27\mu+(4-36\mu)f_0''(\mu)\right]\,.
\end{equation}
Finally, from \req{entropy} it follows that the shear viscosity over entropy density ratio reads
\begin{equation}\label{eta/s}
\left[\frac{\eta}{s}\right]^{\rm ECG}=\frac{5+27\mu+(4-36\mu)f_0''(\mu)}{8\pi\left(1-\frac{27}{4} \mu\right)}\,.
\end{equation}
Some comments are in order. First, note that this expression is very different from the rest of nonperturbative results for $\eta/s$ available in the literature for $d\geq 4$ theories, corresponding to Lovelock \cite{Brigante:2007nu,Ge:2009eh,Brustein:2008cg,Shu:2009ax} and QTG \cite{Myers:2010jv}. In those cases, it is found that $\eta/s$ depends on the gravitational couplings in a polynomial way\footnote{Note however that, \eg for Gauss-Bonnet gravity, some of the remaining second-order coefficients have a nonpolynomial dependence on the corresponding coupling \cite{Grozdanov:2015asa,Grozdanov:2016fkt}. } --- see also \cite{Parvizi:2017boc}.
On the contrary, the ECG result has a very nonpolynomial character, for two reasons. First, the presence of the function $f_0''(\mu)$, which is non-analytic, implies that $\eta/s$ cannot be Taylor-expanded around $\mu=0$. And second, the denominator `$(1-27/(4 \mu))$' in \req{eta/s} is also a new feature, which gives rise to a divergence in the critical limit. The appearance of such contribution in the denominator is rooted in the different way in which ECG modifies the result for the thermal entropy charge $\cs$ with respect to the other theories mentioned above --- see discussion in subsection  \ref{cssex}.

\begin{figure}[t]
	\centering 
	\includegraphics[scale=0.8]{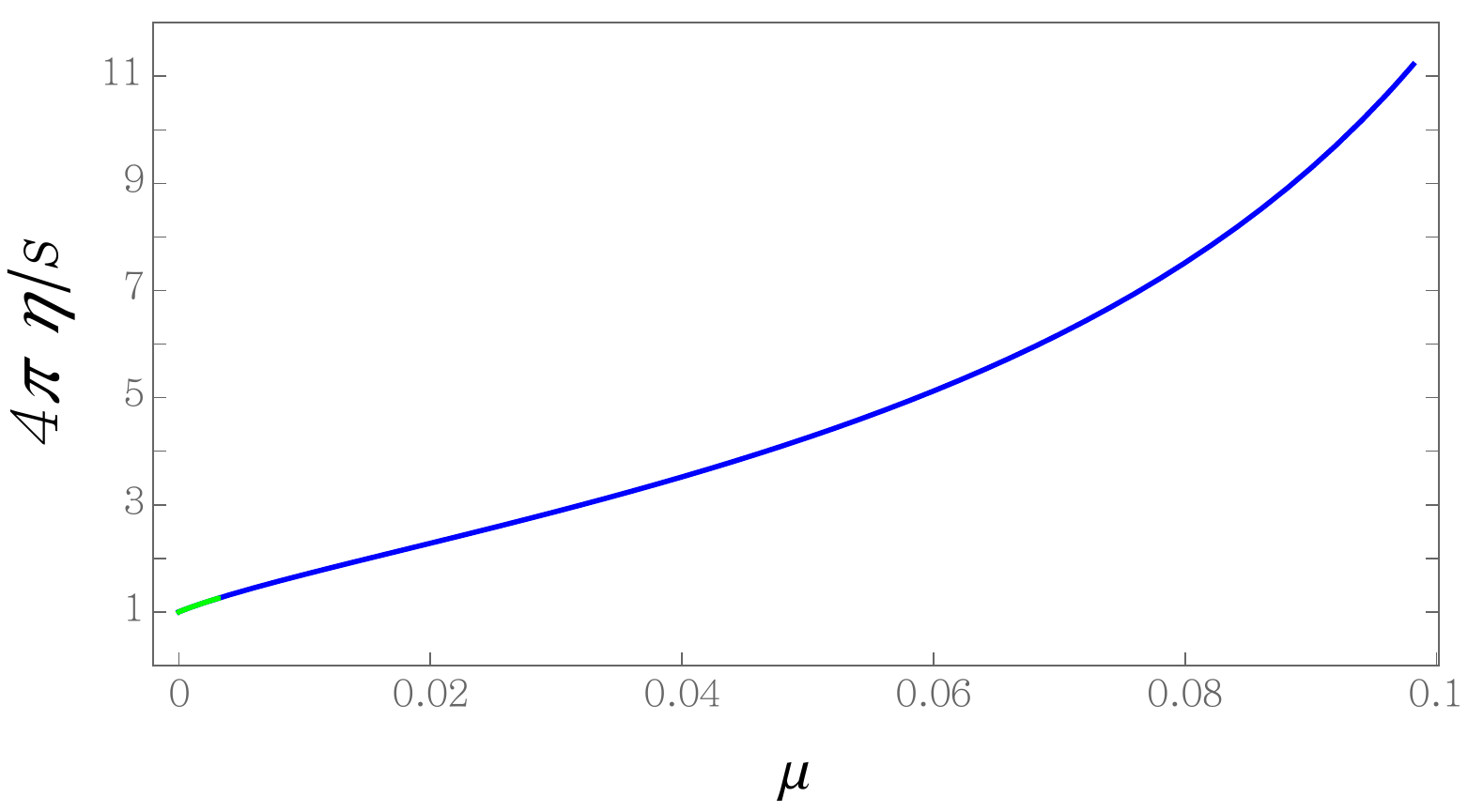}
	\caption{Shear viscosity to entropy density ratio as a function of $\mu$. The green line represents the region allowed by the constraint $t_4\ge -4$.}
	\label{ratio}
\end{figure}

Let us analyze the profile of $\eta/s$ as a function of $\mu$. When $\mu\ll 1$, we can use \req{f0approx} to obtain  
\begin{equation}
\left[\frac{\eta}{s}\right]^{\rm ECG}\approx \frac{1}{4\pi}\left(1+\frac{189 \mu }{2}-\frac{114453 \mu ^2}{16}\right)\, .
\end{equation}
Again, remember that, strictly speaking, this is not a Taylor expansion and it only provides a good approximation for very small $\mu$. In any case, note that the leading correction is positive, so $\eta/s$ is increasing with $\mu$.
On the other hand, in the critical limit, we have $f_0''(\mu\rightarrow 4/27)\rightarrow 0$, so the leading behavior of $\req{eta/s}$ can be captured analytically,
\begin{equation}
\left[\frac{\eta}{s}\right]^{\rm ECG}=\frac{9}{8\pi\left(1-\frac{27}{4} \mu\right)}+\mathcal{O}(1)\, ,\quad \text{for}\, \quad \mu\rightarrow\frac{4}{27}\, .
\end{equation}
Hence, this ratio takes arbitrarily high values as we approach the critical limit\footnote{Observe that, from this point of view, the critical limit of ECG is very different from that corresponding to its higher-dimensional cousins, such as Gauss-Bonnet. In that case, $\eta/s$ diverges for $\lambda_{\rm \ssc GB} \rightarrow -\infty$, while it stays finite for the critical value $\lambda_{\rm \ssc GB}=1/4$.}. The full profile of $\eta/s$ can be obtained with arbitrary precision from a numerical computation of $f_0''(\mu)$. The result is shown in Fig. \ref{ratio}.  
The curve is monotonically increasing, and blows up in the critical limit, $\mu=4/27$. Therefore, the KSS bound is not violated for any value of $\mu$ in the dynamically allowed region, $0\leq \mu\leq 4/27$, which is precisely a consequence of the nonexistence of $\mu<0$ solutions with positive energy. In that sense, as opposed to previously studied theories in higher dimensions, ECG simply does not allow for violations of the bound, not even in principle. It would be interesting to find out whether this phenomenon is common to the rest of $d=3$ theories constructed in \cite{PabloPablo4} and, more generally, to the new theories belonging to the Generalized QTG class \cite{Hennigar:2017ego,PabloPablo3,Ahmed:2017jod} in general dimensions.
% non-negative value of $\mu$ and in principle $\eta/s$ it could take any value above $1/(4\pi)$ . 

As we explained in subsection \ref{t44}, imposing the positivity of energy fluxes in the CFT, gives rise to the constraint  $0\le \mu\le 0.00312$ --- see green region in Fig.  \ref{ratio}. This would imply a maximum possible value for $\eta/s $ in ECG, given by
\begin{equation}
\left[\frac{\eta}{s}\right]_{\rm max.}^{\rm ECG}\simeq 1.253\times \frac{1}{4\pi}\, .
\end{equation}
 %We find the maximum value of the shear viscosity to entropy density ratio compatible with this bound is $\max \eta/s \approx 1.253/(4\pi)$
%The fact that we cannot violate the KSS bound seems to be related to the fact that we are forced to choose $\mu\ge 0$, since for negative $\mu$ ECG does not allow for positive energy black holes or black branes. This is an intriguing connection that requires of further analysis. The computation performed here for the shear viscosity to entropy density ratio in ECG can be extended elsewhere for any (or all) of the theories introduced in \cite{PabloPablo4}. An interesting question is whether the addition of additional higher-curvature terms with arbitrary couplings makes possible the violation of the KSS bound, or if it is a common property of all these theories that they respect the bound.\\

From the results here, we can extract some general lessons regarding calculations of $\eta/s$ in higher-curvature holographic CFTs. First, we have seen that the ECG result is highly nonperturbative in the gravitational coupling. There is in principle no reason to expect this to be different for more general theories. The results found for Lovelock and QTG, polynomial in the gravitational couplings, are probably less generic --- for those, the metric function $f(r)$ is determined by an algebraic equation, which is a highly exceptional property \cite{PabloPablo3}.  Besides, as we have seen, there may be regions of the parameter space for which the corresponding black branes do not exist, even for arbitrarily small values of the couplings. None of this is seen when working perturbatively in the gravitational couplings, which means that the results obtained in that way must be taken carefully. This is a lesson which extends to most calculations in higher-curvature gravities.

\section{Final comments}\label{discu}
In this paper we have studied various aspects of $d=3$ holographic ECG, which, as we have argued, is a toy model of a nonsupersymmetric CFT ($t_4\neq 0$) analogous to QTG in $d=4$.  A detailed summary of our findings can be found in subsection \ref{summ}. We close the paper with a few additional comments. 

Throughout the paper, we have computed several universal charges characterizing the dual CFT. In Table \ref{diction} we have collected some of them, as well as their $d=4$ QTG counterparts (with the Gauss-Bonnet coupling set to zero).
For Einstein gravity, these are all proportional to the only dimensionless quantity present in the Lagrangian, namely, $\tilde{L}^2/G$.  Just like for the higher-dimensional examples previously considered, including QTG, the introduction of the ECG coupling breaks this degeneracy, and all charges become independent from each other, in the sense that all possible ratios formed from them are $\mu$-dependent. As we can see, the stress-tensor parameters have remarkably similar numerical coefficients in both theories when expressed in terms of the Einstein-gravity charges. The expressions for $a^*$ are also similar, whereas the ones for $\cs$ are considerably different, due to the appearance of an extra factor which vanishes in the critical limit for ECG. The differences extend to many other quantities, as we have tried to illustrate throughout the text, which suggests that similar analyses for other members of the Generalized QTG family \cite{Hennigar:2017ego,PabloPablo3,Ahmed:2017jod,PabloPablo4} should be performed. In particular, it would be interesting to find out whether the highly-nonperturbative result for the shear viscosity over entropy density ratio extends to those theories, and whether violations of the KSS bound are also forbidden for them.

\begin{table*}[t] \hspace{-0.55cm}
	\begin{tabular}{|c|c|c|c|c|}
		\hline
		&   $\ctt$  & $\ctt\cdot t_4$   & $\cs $   & $a^* $\\
		\hline\hline
		Einstein &  $\frac{\Gamma(d+2)}{8(d-1)\Gamma(d/2)\pi^{(d+2)/2}}\frac{\tilde{L}^{d-1}}{G}$  & $0$ & $\frac{\Gamma(d+1)\pi^{(2d-1)/2}2^{d-3}}{\Gamma(\frac{d+1}{2})\Gamma(d/2)d^d}\frac{\tilde{L}^{d-1}}{G}$ &  $\frac{\pi^{(d-2)/2}}{8\Gamma(d/2)}\frac{\tilde{L}^{d-1}}{G}$ \\ \hline
		ECG $(d=3)$	& $(1-3\mu f_{\infty}^2)\ctte$   & -$1260 \mu f_{\infty}^2 \ctte$ & $\left(1-\frac{27}{4}\mu \right)f_{\infty}^2 \cse $ & $\left(1+3\mu f_{\infty}^2\right)a^{*{\rm \ssc E}}$ \\ \hline
		QTG $(d=4)$	& $(1-3\mu f_{\infty}^2)\ctte $   & $3780 \mu f_{\infty}^2 \ctte$ & $f_{\infty}^3\cse $ & $\left(1+9\mu f_{\infty}^2\right)a^{*{\rm \ssc E}}$ \\
		\hline
	\end{tabular}
	\caption{From left to right: stress-tensor two- and three-point function charges $\ctt$ and $\ctt \cdot t_4$, thermal entropy charge $\cs$, and universal contribution to the entanglement entropy across a spherical region, $a^*$, for holographic theories dual to Einstein gravity in $d$ dimensions, ECG ($d=3$) and Quasi-topological gravity (with vanishing Gauss-Bonnet coupling) in $d=4$ \cite{Myers:2010jv}. }
	\label{diction}
\end{table*}

\subsection{Generalized action, $a^*$ and holographic complexity?}
In section \ref{osa}, we proposed a new method for evaluating Euclidean on-shell actions for higher-order gravities whose linearized equations of motion on maximally symmetric backgrounds are second order. Throughout the paper, we have performed several successful and highly non-trivial checks of the proposal --- see appendix \ref{BTcheck} as well. It would be interesting to perform further studies of our generalized action \req{SEcomplete} for other theories, such as higher-order Lovelock theories, QTG and its higher-order generalizations and, more generally, for theories of the Generalized QTG type. One of the most striking aspects of \req{SEcomplete} is that it avoids the --- usually very challenging --- problem of determining the correct generalization of the Gibbons-Hawking-York boundary term. At the same time, and somewhat surprisingly, it involves the universal charge $a^*$ controlling the EE of spherical regions in the corresponding dual CFT. This acts as a weight that changes from one theory to another. 

As pointed out in \cite{Carmi:2016wjl}, one of the open questions in the context of holographic complexity, is to determine what kind of universal information (if any) is encoded in the results obtained using the `complexity$=$volume' \cite{Susskind:2014rva,Stanford:2014jda} and `complexity$=$action' \cite{Brown:2015bva,Brown:2015lvg} prescriptions. In order to do so, one possible venue would consist in studying the corresponding quantities for holographic higher-order gravities. For those, the charge-degeneracy inherent to Einstein gravity (where all charges are proportional to $\tilde{L}^2/G$) is broken, and one would hope to be able to identify the nature of possible universal quantities\footnote{Certain ambiguities related to the normalization of affine parameters along null boundaries seem to pollute the would-be universal terms in complexity$=$action calculations \cite{Carmi:2016wjl}. Dealing with those is an additional challenge.} appearing in holographic complexity. Our generalized action \req{SEcomplete} suggests that $a^*$ may appear universally in some of these terms. Observe that complexity$=$action calculations require the introduction of additional terms in the gravitational action when the boundary contains null pieces and joints --- see \cite{Lehner:2016vdi} and references therein. In that case one can only speculate on whether a similar mechanism could make $a^*$ --- or some other characteristic charge --- appear in the corresponding generalized terms for higher-order gravities.

\acknowledgments
We wish to thank Marco Baggio, Nikolay Bobev, Fridrik Gautason, Robie Hennigar, Diego Hofman, Carlos Hoyos, Robb Mann and Rob Myers for useful discussions. The work of PB was supported by a postdoctoral fellowship from the National Science Foundation of Belgium (FWO). The work of PAC is funded by Fundaci\'on la Caixa through a ``la Caixa - Severo Ochoa" International pre-doctoral grant. AR was supported by a ``Centro de Excelencia Internacional UAM/CSIC" FPI pre-doctoral grant and by a grant from the ``Residencia de Estudiantes". PAC and AR were further supported by the MINECO/FEDER, UE grant FPA2015-66793-P, and from the ``Centro de Excelencia Severo Ochoa" Program grant  SEV-2016-0597. PAC also thanks the Perimeter Institute ``Visiting Graduate Fellows" program. Research at Perimeter Institute is supported by the Government of Canada through the Department of Innovation, Science and Economic Development and by the Province of Ontario through the Ministry of Research, Innovation and Science.

\appendix

\section{$\braket{TTT}$ parameters from $h_q$}\label{ttt}
In this appendix we show that the formulas in \cite{Chu:2016tps} for the twist operator scaling dimensions $h_q$ around $q=1$ can be used to obtain the exact values of the parameters $t_2$ and $t_4$ for holographic Gauss-Bonnet in general dimensions, and for QTG in $d=4$. The general-$d$ version of \req{minino} reads  \cite{Chu:2016tps}
\begin{equation}\label{Miaoformula}
\frac{h_q}{\ctt}=2\pi^{\frac{d}{2}+1}\frac{\Gamma(d/2)}{\Gamma(d+2)}(q-1)+\frac{h''_q(1)}{2\ctt}(q-1)^2+\mathcal O(q-1)^3\ ,
\end{equation}
where 
\begin{equation}\label{Miaoformula2}
\begin{aligned}
\frac{h''_q(1)}{\ctt}=&-\frac{2\pi^{1+d/2}\Gamma(d/2)}{(d-1)^3d(d+1)\Gamma(d+3)}\Bigg[d\left(2d^5-9d^3+2d^2+7d-2\right)\\
&+(d-2)(d-3)(d+1)(d+2)(2d-1)t_2+(d-2)(7d^3-19d^2-8d+8)t _4\Bigg]\ .
\end{aligned}
\end{equation}
This expression is valid for general holographic higher-order gravities, at least at leading order in the gravitational couplings. 
%We want to show that, for some theories, this expression also holds  to all order in the couplings. In order to do so, we consider Gauss-Bonnet in arbitrary dimensions and quasi-topological gravity in $4+1$ dimensions.
\subsection{Gauss-Bonnet in arbitrary dimensions}
In this case, the expression for the scaling dimension of twist operators is given by \cite{HoloRen} 
\begin{equation}\label{eq:scaling_dimension_GB}
\frac{h_q}{\ctt}=\frac{\Gamma(d/2)}{4\Gamma(d+2)}\pi^{1+d/2} qx_q^{d-4}(x_q^2-1)\left[d-3-(d+1)x_q^2+(d-3)\frac{1-2\frac{d-1}{d-3}\lambda f_\infty}{1-2\lambda f_\infty}(x_q^2-1)\right]\, ,
\end{equation}
where $x_q$ satisfies the following quartic equation 
\begin{equation}
x_q^4 d-\frac{2}{q}x_q^3-(d-2)x_q^2+\lambda f_\infty \left[4\frac{x_q}{q}-x_q^4 d+d-4\right]=0\, .
\end{equation}
A Taylor expansion around $q=1$ gives 
\begin{equation}
x_q=1+\frac{1}{1-d}(q-1)+\frac{d}{(d-1)^3}\frac{-2d+3+\lambda f_\infty (4d-10)}{-2+4\lambda f_\infty}(q-1)^2+\mathcal O(q-1)^3 \, .
\end{equation}
Plugging this expansion into \req{eq:scaling_dimension_GB}, we find 
\begin{align}
&\frac{h_q}{\ctt}=\frac{2\Gamma(d/2)\pi^{1+d/2}}{\Gamma(d+2)}(q-1)\\ \notag
&-\frac{(d-1)\Gamma(d/2)\pi^{1+d/2}}{\Gamma(d+2)}\left[-1+4d-2d^2+\lambda f_\infty(6-16d+4d^2)\right](q-1)^2+\mathcal O(q-1)^3\, .
\end{align}
Comparing this with (\ref{Miaoformula}), we find that $t_2$ and $t_4$ should be given by
\begin{equation}\label{eq:t2_GB}
t_2=\frac{4d(d-1) \lambda f_\infty}{(d-2)(d-3) (1-2\lambda f_\infty)}\, , \quad t_4=0\, ,
\end{equation}
which matches the exact nonperturbative result \cite{Buchel:2009sk}.

\subsection{Quasi-topological gravity}
In this case, the scaling dimension $h_q$ was obtained in \cite{HoloRen} in terms of the charges $a$, $c$ and $t_4$ of the theory as
\begin{equation}\label{eq:hq_quasitopo}
h_q=\frac{aq}{4\pi x_q^2}(x_q^2-1)\left[x_q^4\left(1-5\frac{c}{a}-10\frac{c}{a}t_4\right)-x_q^2\left(1-\frac{c}{a}-8\frac{c}{a}t_4\right)+2\frac{c}{a}t_4\right]\ ,
\end{equation}
where 
\begin{eqnarray}\label{centralchargesquasitopological}
c&=&\pi^2\frac{\tilde L^3}{8\pi G}\left(1-2\lambda f_\infty -3\mu f_\infty^2\right)\ , \\
a&=&\pi^2\frac{\tilde L^3}{8\pi G}\left(1-6\lambda f_\infty +9\mu f_\infty^2\right)\ , \\
t_4&=&\frac{3780\mu f_\infty^2}{1-2\lambda f_\infty-3\mu f_\infty^2}\ ,
\end{eqnarray}
 and where $x_q$ satisfies the following quartic equation
\begin{equation}\label{eq:xq-quasitopo}
2x_q^6-\frac{x_q^5}{q}-x_q^4+2\lambda f_\infty x_q^3\left(\frac{1}{q}-x_q^3\right)+\mu f_\infty^2\left(-1+\frac{3x_q}{q}-2x_q^ 6\right)=0\ .
\end{equation}
Moreover, we have \cite{Myers:2010jv}
\begin{equation}
t_2=\frac{24f_\infty\left(\lambda-87\mu f_\infty\right)}{1-2\lambda f_\infty-3\mu f_\infty^2}\ ,
\end{equation}
which properly reduces to the Gauss-Bonnet formula (\ref{eq:t2_GB}) for $\mu=0$ and $d=4$. Before computing the Taylor expansion of $h_q$ around $q=1$, we invert (\ref{centralchargesquasitopological}) and find\footnote{There seems to be a small typo in eq. (2.58) of \cite{HoloRen}. Note also that our convention for $t_4$ differs by a factor of $1890$ with respect to that in  \cite{HoloRen}, but agrees with the one in \cite{Myers:2010jv}.}
\begin{eqnarray}
\frac{\tilde L^3}{8\pi G}&=&\frac{a}{2\pi^2}\left(3\frac{c}{a}\left(1+\frac{3t_4}{1890}\right)-1\right)\, , \\
\lambda f_\infty&=&\frac{1}{2}\frac{\frac{c}{a}\left(1+\frac{6t_4}{1890}\right)-1}{3\frac{c}{a}(1+\frac{3t_4}{1890})-1}\, , 
\label{eq:lambda}
\\
\mu f_\infty^2&=&\frac{\frac{c}{a}\frac{t_4}{1890}}{3\frac{c}{a}\left(1+\frac{3t_4}{1890}\right)-1}\, .
\end{eqnarray}
and rewrite (\ref{eq:xq-quasitopo}) in terms of $c/a$ and $t_4$. We get 
\begin{equation}
x_q(q)=1-\frac{q-1}{3}+\frac{4+\frac{8t_4}{1890}-\frac{2}{3}\frac{a}{c}}{9}(q-1)^2+\mathcal O(q-1)^3 \, ,
\end{equation}
and plugging it into (\ref{eq:hq_quasitopo}), we find 
\begin{equation}
\frac{h_q}{c}=\frac{2}{3\pi}(q-1)+ \frac{7 \frac{a}{c}-24-\frac{84 t_4}{1890}}{27\pi}(q-1)^2+\mathcal O(q-1)^3 \, .
\end{equation}
Comparing the leading term, we notice that $\ctt$ should be related to $c$ via $\tfrac{\ctt}{c}=\tfrac{40} {\pi^4}$, which is correct. Finally, using
\begin{equation}
\frac{a}{c}=1-\frac{t_2}{6}-\frac{4}{45}t_4\ ,
\end{equation}
we find
\begin{equation}
\frac{h''_q(1)}{2\ctt }=-\pi^3\frac{102+7t_2+4t_4}{6480}\ ,
\end{equation}
which exactly agrees with the general formula \req{Miaoformula2} when we particularize it to $d=4$.

\section{Generalized action for Gauss-Bonnet gravity}\label{BTcheck}
In this appendix we perform an additional check of the generalized action introduced in section \ref{osa}. In particular, we apply it here to a theory for which the exact generalization of the GHY term is known, namely, $D$-dimensional Gauss-Bonnet gravity \cite{Teitelboim:1987zz,Myers:1987yn}. The full Euclidean action of the theory reads
\begin{equation}
I_E^{\rm GB}=-\frac{1}{16\pi G}\int_{\mathcal{M}}d^Dx\sqrt{g}\left[\frac{(D-1)(D-2)}{L^2}+R+\frac{L^2\lambda}{(D-3)(D-4)}\mathcal{X}_4\right]+I^{\rm GB}_{\rm GHY}+I^{\rm GB}_{\rm CT}\, ,
\end{equation}
where the generalization of the GHY term reads
\begin{equation}
\begin{aligned}
I^{\rm GB}_{\rm GHY}&=-\frac{1}{8\pi G}\int_{\partial \mathcal{M}}d^{D-1}x\sqrt{h}\Bigg\{K+\frac{L^2\lambda}{(D-3)(D-4)}\delta^{a_1a_2a_3}_{b_1b_2b_3}K^{b_1}_{a_1}\left(\mathcal{R}^{b_2b_3}_{a_2a_3}-\frac{2}{3}K^{b_2}_{a_2}K^{b_3}_{a_3}\right)\Bigg\}\, ,
\end{aligned}
\end{equation}
and the counterterms can be chosen as \cite{Emparan:1999pm, Mann:1999pc, Balasubramanian:1999re, Brihaye:2008xu,Astefanesei:2008wz}
\begin{align}\notag
I^{\rm GB}_{\rm CT}&=\frac{1}{8\pi G}\int_{\partial \mathcal{M}}d^{D-1}x\sqrt{h}\Bigg\{\frac{(D-2)(f_{\infty}+2)}{3f_{\infty}^{1/2}L}+\frac{ L(3f_{\infty}-2)}{2f_{\infty}^{3/2}(D-3)}\mathcal{R}+\frac{ L^3\Theta[D-6]}{2f_{\infty}^{5/2}(D-3)^2(D-5)}\\
&\times\left[(2-f_{\infty})\left(\mathcal{R}_{ab}\mathcal{R}^{ab}-\frac{D-1}{4(D-2)}\mathcal{R}^2\right)-\frac{(D-3)(1-f_{\infty})}{D-4}\mathcal{X}_4(h)\right]+\ldots\Bigg\}\, .
\end{align}
With these boundary contributions, the Gauss-Bonnet action functional is differentiable and finite in AdS spaces. Since Gauss-Bonnet gravity has an Einstein-like spectrum in pure AdS$_D$ (in fact, in any background), our generalized GHY term should also be applicable to GB gravity, as long as the boundary consists only of asymptotically AdS pieces. The prescription in \req{SEcomplete} gives the following result when applied to the Gauss-Bonnet Lagrangian,
\begin{equation}
\begin{aligned}
I_{\rm GGHY}^{\rm GB}+I_{\rm GCT}^{\rm GB}&=-\frac{1-2\lambda f_{\infty}\frac{D-2}{D-4}}{8 \pi G}\int_{\partial \mathcal{M}}d^{D-1}x\sqrt{h}\bigg[K-\frac{D-2}{\tilde L}-\frac{\tilde L}{2(D-3)}\mathcal{R}\\
&-\frac{\tilde L^3\Theta[D-6]}{2(D-3)^2(D-5)}\left(\mathcal{R}_{ab}\mathcal{R}^{ab}-\frac{D-1}{4(D-2)}\mathcal{R}^2\right)+\ldots\bigg]\, ,
\end{aligned}
\end{equation}
where we included a set of counterterms valid up to $D=7$ and where $\tilde L=L/\sqrt{f_{\infty}}$. Recall that the coefficient in front of the integral is proportional to the universal constant $a^*$ appearing in the EE across a spherical region, which for GB gravity reads
\begin{equation}
a^*=\left(1-2\lambda f_{\infty}\frac{d-1}{d-3}\right)\frac{\tilde L^{d-1}\Omega_{(d-1)}}{16 \pi G}\, .
\end{equation}
In order to compare both boundary terms, let us consider a metric of the form
\begin{equation}
ds^2=f(r)d\tau^2+\frac{dr^2}{f(r)}+r^2d\Sigma_{k}^2\, ,
\end{equation}
where $d\Sigma_{k}^2$ is the metric of a maximally symmetric space of curvature $k=-1,0,1$ and $\tau$ has period $\beta$. For $f(r)=f_{\infty} r^2/L^2+k$, the previous metric reduces to pure Euclidean AdS$_D$, with the boundary at $r=+\infty$, which we regulate as $r\rightarrow L^2/\delta$. Let us now switch on arbitrary radial perturbations
\begin{equation}
f(r)=f_{\infty}\frac{r^2}{L^2}+k+\frac{f_1}{r}+\frac{f_2}{r^2}+\frac{f_3}{r^3}+\ldots\, .
\end{equation}
Evaluated at  $r\rightarrow L^2/\delta$, the boundary terms coming from both prescriptions yield, respectively
\begin{eqnarray*}
I_{\rm GHY+CT}^{\rm GB}&=&\frac{\beta \pi}{4 G}\left[\frac{(5 f_{\infty}-6) L^6}{\delta^4}
+\frac{f_1 (5 f_{\infty}-6) L^2}{f_{\infty}\delta}+\frac{(5 f_{\infty}-6) \left(4 f_2 f_{\infty}-3 L^2 k^2\right)}{8 f_{\infty}^2}+\mathcal{O}(\delta^2)\right]\, ,\\
I_{\rm GGHY+GCT}^{\rm GB}&=&\frac{\beta \pi}{4G}\left[\frac{(5 f_{\infty}-6) L^6}{\delta^4}
+\frac{f_1 (5 f_{\infty}-6) L^2}{f_{\infty}\delta}+\frac{(5 f_{\infty}-6) \left(4 f_2 f_{\infty}-3 L^2 k^2\right)}{8 f_{\infty}^2}+\mathcal{O}'(\delta^2)\right]\, .
\end{eqnarray*}
This is, all divergent and finite terms are equal! The difference only appears in the decaying terms, which of course give no contribution to the action. For the sake of simplicity, we evaluated the above expressions for $D=5$, but it is straightforward to check that the same phenomenon happens in higher dimensions (with the expressions above we have checked $D=6,7$). Therefore, at least from a practical point of view, our generalized boundary term is as good as the Gauss-Bonnet one when applied to asymptotically AdS spaces. We expect our method to work also for general Lovelock gravities, as well as for QTG, and the rest of theories belonging to the Einstein-like class in the classification of \cite{Aspects}.
%The advantage of the new boundary term is that it can be applied to arbitrary higher-order theories, providing that their linearized equations are Einstein-like. 

\section{Boundary terms in the two-point function}\label{2pbdy}
In this appendix we evaluate explicitly the boundary contribution in \req{EuclideanECG} for the metric perturbation considered in section \ref{tt}. The sum of all boundary contributions appearing in \req{ICT}, which includes the one coming from $I^{\rm ECG}_{E\, \rm Bulk}$ in \req{ttw}, as well as the generalized GHY term and the counterterms in  \req{EuclideanECG}, reads
\begin{equation}
I_{E\, \rm bdry}^{\rm ECG}=-\frac{1}{8\pi G}\int d^3x\left[\frac{1}{2}\Gamma_r+(1+3\mu f_{\infty}^2)\sqrt{h}\left(K-\frac{2\sqrt{f_{\infty}}}{L}-\frac{L}{2\sqrt{f_{\infty}}}\mathcal{R}\right)\right]\, ,
\end{equation}
where $\Gamma_r$ comes from integration by parts in the bulk action, and is given by
\begin{align}\label{gaga}
\Gamma_r=\frac{1}{\sqrt{f_{\infty}}}\bigg[&-\frac{2(4f_{\infty}-3)r^3}{L^4}+\frac{4f_{\infty}-3}{L^4}\left(2r^4\phi\partial_r\phi+r^3\phi^2\right)+\frac{(f_{\infty}-1)r^5}{L^4}(\partial_r\phi)^2\\
&+6\mu f_{\infty}^2\left(-\frac{r}{2}(\partial_{\tau}\phi)^2+r^2\partial_r\phi\partial_{\tau}^2\phi\right)\bigg]\, .
\end{align}
The rest are the boundary terms in the action \req{EuclideanECG}. The induced metric on a hypersurface of fixed $r$ is
\begin{equation}
^{(3)}ds^2=\frac{r^2}{L^2}\left(d{\tau}^2+dx^2+dy^2+2dxdy\phi(r,{\tau})\right)\, .
\end{equation}
At quadratic order in $\phi$ we have
\begin{equation}
\sqrt{h}=\frac{r^3}{L^3}\left(1-\frac{1}{2}\phi^2\right)\, ,\quad \mathcal{R}=\frac{L^2}{2 r^2}\left(3(\partial_{\tau}\phi)^2+4\phi\partial_{\tau}^2\phi\right)\, , \quad K=\frac{3\sqrt{f_{\infty}}}{L}-\frac{r\sqrt{f_{\infty}}}{L}\phi\partial_r\phi\, .
\end{equation}
Then, we obtain at that order
\begin{equation}
\begin{aligned}
I_{E\, \rm bdry}^{\rm ECG}=\frac{1}{8\pi G}\int d^3x\bigg[&\frac{3r^3}{L^4\sqrt{f_{\infty}}}(1-f_{\infty}+\mu f_{\infty}^3)\left(-1+\frac{\phi^2}{2}+r\phi\partial_r\phi\right)-\frac{3(f_{\infty}-1)r^5}{2\sqrt{f_{\infty}}L^4}(\partial_r\phi)^2\\
&+\frac{r}{\sqrt{f_{\infty}}}\left((1+3\mu f_{\infty}^2)\left(\frac{3}{4}(\partial_{\tau}\phi)^2+\phi\partial_{\tau}^2\phi\right)+\frac{3\mu}{2}f_{\infty}^2(\partial_{\tau}\phi)^2\right)\\
&-3\mu f_{\infty}^{3/2}r^2\partial_r\phi\partial_{\tau}^2\phi\bigg]\, .
\end{aligned}
\end{equation}
The first term vanishes  because $1-f_{\infty}+\mu f_{\infty}^3=0$ is precisely the AdS$_4$ embedding equation \req{roo}. Now it proves useful to perform the Fourier transformation of $\phi$:
\begin{equation}
\phi(r,{\tau})=\frac{1}{2\pi}\int dp \phi_0(p)e^{ip {\tau}}H_p(r)\, ,
\end{equation}
with $H_p(r)=e^{-\frac{L^2 |p|}{\sqrt{f_{\infty}}r}}\left(1+\frac{L^2 |p|}{\sqrt{f_{\infty}}r}\right)$. Then,
\begin{equation}
\begin{aligned}
I_{E\, \rm bdry}^{\rm ECG}=&\frac{V_{\mathbb{R}^2}}{16 \pi^2 G}\int  dpdq \delta(q+p)\phi_0(p)\phi_0(q)\bigg[-\frac{3(f_{\infty}-1)r^5}{2\sqrt{f_{\infty}}L^4}(\partial_r H_p)^2\\
&+\frac{r H_p^2}{\sqrt{f_{\infty}}}\left((1+3\mu f_{\infty}^2)\left(-\frac{3}{4}p q-q^2\right)-\frac{3\mu}{2}f_{\infty}^2pq\right)+3\mu q^2 f_{\infty}^{3/2}r^2H_q\partial_rH_p \bigg]\, .
\end{aligned}
\end{equation}
Now, since $\partial_rH_p\sim 1/r^3$ , the first and last terms vanish for $r\rightarrow\infty$. Then, we are left with the final result
\begin{equation}
\begin{aligned}
I_{E\, \rm bdry}^{\rm ECG}=-\frac{V_{\mathbb{R}^2}(1-3\mu f_{\infty}^2)}{64\pi^2 G\sqrt{f_{\infty}}}\int dpdq \delta(q+p)\phi_0(p)\phi_0(q) p^2 rH_p^2(r)\, ,
\end{aligned}
\end{equation}
which appears in the main text.

%On the other hand, eq.~(\ref{traceless}) becomes
%\be \label{traceless2}
%\frac{2}{3}\left[\bar{\nabla}_{\mu}\bar{\nabla}^{\rho}\hat{h}_{\rho\nu}-\frac{1}{4}\bar{g}_{\mu\nu}\bar{\nabla}^{\rho}\bar{\nabla}^{\sigma} \hat{h}_{\rho\sigma} \right]\left[1-36\lambda_1\right]- \frac{1}{2}\left[1-24\lambda_1\right]\left[\bar{\Box}+\frac{2}{L^2}\right]\hat h_{\mu\nu}=0\, .
%\ee

%\section{}\label{integrals}

%\renewcommand{\leftmark}{\MakeUppercase{Bibliography}}
%\phantomsection
\bibliography{Gravities}
\bibliographystyle{JHEP-2}
\label{biblio}

\end{document}